\documentclass[aps, prd, amsmath, floats, floatfix, twocolumn,
superscriptaddress, nofootinbib, showpacs]{revtex4}

\usepackage{graphicx}
\usepackage{amsmath,amssymb}
\usepackage{amsfonts}
\usepackage{xspace} % Sensible space treatment at end of simple macros
\usepackage[usenames,dvipsnames]{color}
\usepackage{dcolumn}% Align table columns on decimal point
\usepackage{bm}% bold math

\usepackage{ulem}
\newcommand{\Caltech}{\affiliation{Theoretical Astrophysics 130-33,
    California Institute of Technology, Pasadena, CA 91125}}
\newcommand{\Cornell}{\affiliation{Center for Radiophysics and Space
    Research, Cornell University, Ithaca, New York, 14853}}
\newcommand{\Maryland}{\affiliation{Maryland Center for Fundamental
    Physics, Department of Physics, University of Maryland, College
    Park, MD 20742}}

\newcommand{\PsiFourPhase}{\phi}
\newcommand{\hDotPhase}{\varphi}
\newcommand{\OrbitalPhase}{\Phi}
\newcommand{\PsiFourFreq}{\omega}
\newcommand{\hDotFreq}{\varpi}
\newcommand{\OrbitalFreq}{\Omega}

\definecolor {darkgreen}{rgb}{0.2,0.7,0.2}

\newcommand{\Real}{\mbox{Re}}
\newcommand{\Imag}{\mbox{Im}}
\newcommand{\pade}{Pad\'{e}\xspace}

\newcommand{\vp}{\mbox{\boldmath${p}$}}

\begin{document}

\title{High-accuracy numerical simulation of black-hole binaries:
Computation of the gravitational-wave energy flux and comparisons with
post-{N}ewtonian approximants}

\author{Michael Boyle} \Caltech %
\author{Alessandra Buonanno} \Maryland %
\author{Lawrence E. Kidder} \Cornell %
\author{Abdul H. Mrou\'e} \Cornell %
\author{Yi Pan} \Maryland %
\author{Harald P. Pfeiffer} \Caltech %
\author{Mark A. Scheel} \Caltech %

\begin{abstract} 
Expressions for the gravitational wave (GW) energy flux and
center-of-mass energy of a compact binary are integral building blocks
of post-Newtonian (PN) waveforms.  In this paper, we compute the GW
energy flux and GW frequency derivative from a highly accurate
numerical simulation of an equal-mass, non-spinning black hole
binary. We also estimate the (derivative of the) center-of-mass energy
from the simulation by assuming energy balance.  We compare these
quantities with the predictions of various PN approximants (adiabatic
Taylor and \pade models; non-adiabatic effective-one-body (EOB)
models).  We find that \pade summation of the energy flux does not
accelerate the convergence of the flux series; nevertheless, the \pade
flux is markedly closer to the numerical result for the whole range of
the simulation (about 30 GW cycles). Taylor and \pade models
overestimate the increase in flux and frequency derivative close to
merger, whereas EOB models reproduce more faithfully the shape of and
are closer to the numerical flux, frequency derivative and derivative
of energy.  We also compare the GW phase of the numerical simulation
with \pade and EOB models.  Matching numerical and untuned 3.5 PN
order waveforms, we find that the phase difference accumulated until
$M \omega = 0.1$ is -0.12 radians for \pade approximants, and 0.50
(0.45) radians for an EOB approximant with Keplerian (non-Keplerian)
flux.  We fit free parameters within the EOB models to minimize the
phase difference, and confirm the presence of degeneracies among these
parameters. By tuning the pseudo 4PN order coefficients in the radial
potential or in the flux, or, if present, the location of the pole in
the flux, we find that the accumulated phase difference at $M \omega =
0.1$ can be reduced---if desired---to much less than the estimated
numerical phase error (0.02 radians).
\end{abstract}

\date{\today}


\maketitle

\section{Introduction}
\label{sec:intro}

The first-generation interferometric gravitational wave (GW)
detectors, such as LIGO~\cite{BarishWeiss1999,Waldman2006},
GEO600~\cite{Hild2006} and
Virgo~\cite{AcerneseAmico2002,AcerneseAmico2006}, are now operating at
or near their design sensitivities.  One of the most promising sources
for these detectors is the inspiral and merger of binary black holes
(BBHs) with masses $m_1 \sim m_2 \sim
10$--$20\,M_\odot$~\cite{FlanaganHughes1998,DamourIyer2000}.  A detailed and
accurate understanding of the gravitational waves radiated as the
black holes spiral towards each other will be crucial not only for the
initial detection of such sources, but also for maximizing the
information that can be obtained from signals once they are observed.
Both the detection and subsequent analysis of gravitational waves from
compact binaries depends crucially on our ability to build an accurate
bank of templates, where each template is a theoretical model that
accurately represents the gravitational waveform from a binary that
has a certain set of parameters (e.g., masses and spins).  For
detection, the technique of matched filtering is applied to noisy data
to extract any signals that match members of the template bank.  For
analysis, the best-fit parameters are determined, most likely by an
iterative process that involves constructing further templates to zero
in on the best fit.

When the black holes are far apart and moving slowly, the
gravitational waveform (i.e., the template) can be accurately computed
using a post-Newtonian (PN) expansion.  As the holes approach each
other and their velocities increase, the post-Newtonian expansion is
expected to become less and less reliable. However, until recently
there has been no independent way to determine how close
comparable-mass holes must be before PN methods become inaccurate.
This has changed with recent advances in numerical relativity (NR),
which make it possible for the first time to quantify the disagreement
between PN predictions~\cite{Blanchet2006} and the true
waveform~\cite{BuonannoCook2007,BakerVanMeter2007,
HannamHusa2008,BoyleBrown2007,GopakumarHannam0712.3737,HannamHusa0712.3787}.  In a previous
paper~\cite{BoyleBrown2007}, some of us described numerical simulations of
15 orbits of an equal-mass non-spinning binary black hole system.
Gravitational waveforms from these simulations covering more than 30
GW cycles and ending about 1.5 GW cycles before merger, were compared
with those from quasi-circular PN formulas for several time-domain
Taylor approximants computed in the so-called {\it adiabatic}
approximation.  We found that there was excellent agreement (within
$0.05$ radians) in the GW phase between the numerical results and the
PN waveforms over the first $\sim 15$ cycles, thus validating the
numerical simulation and establishing a regime where PN theory is
accurate.  In the last 15 cycles to merger, however, {\em generic}
time-domain Taylor approximants build up phase differences of several
radians.  But, apparently by coincidence, one specific PN approximant,
TaylorT4 at 3.5PN order, agreed much better with the numerical
simulations, with accumulated phase differences of less than 0.05
radians over the 30-cycle waveform.  Simulations by Hannam et
al.~\cite{HannamHusa0712.3787} for equal-mass, non-precessing spinning
binaries confirm that this agreement in the non-spinning case is a
coincidence: they find the phase disagreement between TaylorT4 and the
numerical waveform can be a radian or more as the spins of the black
holes are increased.

To build a template bank to be used by ground-based GW detectors, one
possibility would be to run a separate numerical simulation for each
template. This is not currently possible, however, due to the large
computational cost per numerical waveform (on the order of a week for
a single waveform) and the large number of templates needed to cover
the parameter space, especially when spins are present.  A more
realistic possibility is to perform a small number of simulations and
develop an analytic template family (i.e., a fitting formula) which
interpolates the parameter space between the
simulations~\cite{PanBuonanno2008,BuonannoPan2007,AjithBabak2008,
DamourNagar2008,DamourNagar2008a,DamourNagar2008b}.

Before the NR breakthrough several analytic prescriptions were
proposed to address the loss of accuracy of the adiabatic Taylor
approximants.  Damour, Iyer and Sathyaprakash~\cite{DamourIyer1998}
introduced the \pade summation of the PN center-of-mass energy and
gravitational energy flux in order to produce a series of \pade
approximants for the waveforms in the adiabatic.  Buonanno and
Damour~\cite{BuonannoDamour1999,BuonannoDamour2000,DamourJaranowski2000,Damour2001}
introduced the effective-one-body (EOB) approach which gives an
analytic description of the motion and radiation beyond the adiabatic
approximation of the binary system through inspiral, merger, and
ringdown. The EOB approach also employs the \pade summation of the
energy flux and of some crucial ingredients, such as the radial
potential entering the conservative dynamics. So far, the EOB
waveforms have been compared with several numerical waveforms of
non-spinning binary black
holes~\cite{BuonannoCook2007,PanBuonanno2008,
BuonannoPan2007,DamourNagar2008,DamourNagar2008a,DamourNagar2008b}.
Buonanno et al.~\cite{BuonannoPan2007} showed that by using three
quasi-normal modes~\cite{BuonannoCook2007} and by tuning the pseudo
4PN order coefficient~\cite{DamourIyer2003} in the EOB radial
potential to a specific value, the phase difference accumulated by the
end of the ringdown phase can be reduced to $\sim 0.19 \mbox{--} 0.50$ radians,
%$\sim 3 \mbox{--}8 \times10^{-2}$ of a GW cycle, 
depending on the mass ratio and the number of
multipole moments included in the waveform. Those results were
obtained using waveforms with $5\mbox{--}16$ GW cycles and mass ratios
$1:4$, $1:2$, $2:3$ and $1:1$. In
Refs.~\cite{DamourNagar2008,DamourNagar2008a,DamourNagar2008b} the
authors introduced other improvements in the EOB
approach, in part obtained by tuning the test-mass limit
results~\cite{DamourNagar2007} --- for example \pade summation of the
PN amplitude corrections in the inspiral waveform; ringdown matching
over an interval instead of a point; inclusion of non-circular terms
in the tangential damping force; use of five quasi-normal modes. They
found that the phase differences accumulated by the end of the
inspiral (ringdown) can be reduced to 
$\pm 0.001$ ($\pm 0.03$) radians
%$\pm 2 \times 10^{-4}$ ($\pm 5 \times 10^{-3}$) of a GW cycle
for equal-mass binaries~\cite{DamourNagar2008,DamourNagar2008a} and to 
$\pm 0.05$ radians
%$\pm 8 \times 10^{-3}$ of a GW cycle 
for binaries with mass ratio
$1:2$~\cite{DamourNagar2008b}. Note that these phase differences
are smaller than the numerical errors in the simulations.

The energy flux and the center-of-mass energy are two fundamental
quantities of the binary dynamics and crucial ingredients in building
GW templates.  In this paper we extract these quantities, and compare
the results from our numerical inspiral
simulation~\cite{BoyleBrown2007} with PN results in both their
Taylor-expanded and summed (\pade and EOB) forms. The agreement
between the numerical and analytical results for the energy flux and
the center-of-mass energy is a further validation of the numerical
simulation.  It also allows us to study whether or not the agreement
of the phase evolution of PN and numerical waveforms is accidental.
In addition, we compute waveforms based on adiabatic \pade and
non-adiabatic EOB approximants in their {\it untuned} form (i.e.,
without introducing fitting coefficients) and study their agreement
with our numerical simulations.

We try to understand whether these approximants can reproduce features
of the numerical simulations that can be exploited to develop a
faithful analytic template family. By introducing unknown higher-order
PN coefficients into the dynamics and tuning them to the numerical
data, we investigate how to improve the agreement with the numerical
results.  Although our study only examines non-spinning, equal-mass
binary black holes, by combining it with other
studies~\cite{PanBuonanno2008,BuonannoPan2007,AjithBabak2008,
DamourNagar2008,DamourNagar2008a,DamourNagar2008b} one can already
pinpoint which parameters are degenerate and which have the largest
effect on the waveforms.  This is particularly relevant during the
last stages of inspiral and plunge.  The overall methodology can be
extended to a larger region of the parameter space.  We will defer to
a future paper a complete study of the flexibility of the EOB approach
with the extension of our numerical waveform through merger and
ringdown.

This paper is organized as follows:
Section~\ref{sec:NumericalEnergyFlux} gives a quick review of the
numerical simulations presented in~\cite{BoyleBrown2007}, and then
presents the computation of the GW energy flux from the simulation.
In Sec.~\ref{sec:PNapproximants} we summarize the PN approximants that
will be compared to the numerical simulation.  In
Sec.~\ref{sec:CompareFlux}, we compare the GW energy flux for the
various PN approximants with numerical results and explore the
possibility of improving the agreement with the numerical flux by
adding phenomenological
parameters~\cite{PanBuonanno2008,BuonannoPan2007,DamourNagar2008,DamourNagar2008a,DamourNagar2008b}.
In Sec.~\ref{sec:CompareEnergy}, we examine the evolution of the
center-of-mass energy for the various PN approximants and compare to
the numerical results assuming balance between the change in the
center-of-mass energy and the energy carried from the system by the
gravitational waves.  In Sec.~\ref{sec:CompareWaveforms} we compare
waveforms constructed from the \pade and EOB approximants with our
numerical results, and study how to improve the agreement by
exploiting the flexibility of the EOB model (i.e., by fitting free
parameters of the EOB model).  Finally, we present some concluding
remarks in Sec.~\ref{sec:Conclusions}.  In the Appendix we review the
performance of the \pade summation of the Taylor series of the energy
flux in the test particle limit.

\section{Computation of the numerical gravitational-wave energy flux}
\label{sec:NumericalEnergyFlux}

\subsection{Overview and Definitions}

\begin{figure}
  %15OrbitOverview
  \includegraphics[width=\linewidth]{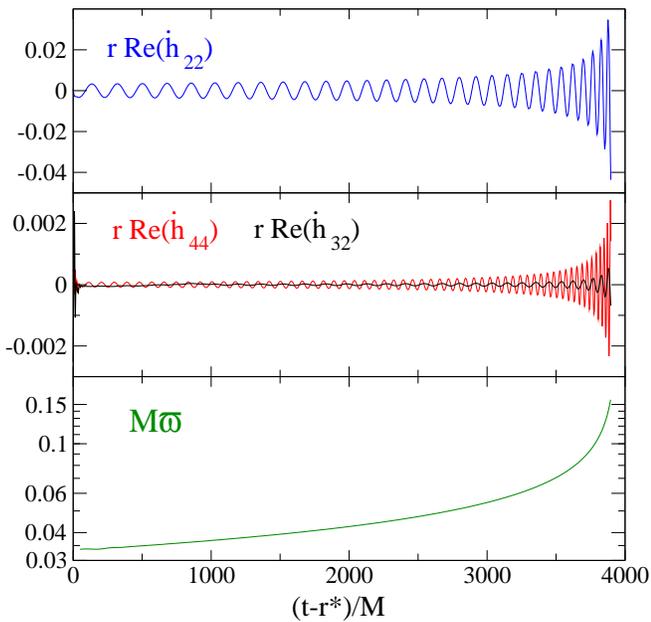}
  \caption{\label{fig:15OrbitOverview}Some aspects of the numerical
    simulation.  From top panel to bottom: the leading mode
    $\dot{h}_{22}$; the two next largest modes, $\dot h_{44}$ and
    $\dot h_{32}$ (smallest); the frequency of $\dot{h}_{22}$ [see
    Eq.~(\ref{eq:Def-hDotQuants})]. }
\end{figure}

The data used in this paper is the same as that described in Sec. II
of Boyle et al.~\cite{BoyleBrown2007}.  The simulation is a 16-orbit
inspiral, with very low spin and eccentricity.
Figure~\ref{fig:15OrbitOverview} presents a view of some relevant
quantities of that simulation.

The Newman-Penrose scalar $\Psi_4$, defined using a coordinate-based
tetrad, is extracted from the simulation at several extraction radii
and expanded in spin-weighted spherical harmonics,
\begin{equation}
  \label{eq:Psi4Decomp}
  \Psi_{4}(t,r,\theta,\phi) = \sum_{l,m}\, \Psi_{4}^{lm}(t,r)\,
  _{-2}Y_{lm}(\theta, \phi)\,.
\end{equation}
Then $\Psi_{4}^{lm}(t,r)$ is extrapolated to infinite extraction
radius using an $n$-th order polynomial in $1/r$, where typically
$n=3$.  This results in the asymptotic field
$r\Psi_{4}^{lm}(t-r^\ast)$ as function of retarded time\footnote{See
Sec. II F of Ref.~\cite{BoyleBrown2007} for a precise definition of
$r^\ast$ and a description of the extrapolation.}  $t-r^\ast$.

Gravitational radiation may also be expressed via the standard
metric-perturbation quantities $h_{+}$ and $h_{\times}$, which we
similarly write in terms of spin-weighted spherical harmonic
components,
\begin{equation}
  \label{eq:hDef}
  h \equiv h_{+}-ih_{\times} = \sum_{l,m}\, h_{lm}\, _{-2}Y_{lm}\,.
\end{equation}
For linear perturbations around Minkowski space,
$\Psi_{4}^{lm}(t-r^{\ast}) = \ddot{h}_{lm}(t-r^{\ast})$.  In
particular, this relation should be true for the waveforms we have
extrapolated to infinity.

However, to compute the energy flux we do not need to determine $h$;
we need only its time derivative $\dot{h}$.  The energy flux depends
on the spin-weighted spherical harmonic coefficients of the time
derivative $\dot{h}$ via
\begin{equation}
  \label{eq:Flux}
  {F} = \frac{1}{16\pi} \sum_{l=2}^\infty \sum_{m=-l}^l |r\, \dot
  h_{lm}|^2 \,.
\end{equation}
We obtain $\dot{h}_{lm}$ by time-integration of $\Psi_4^{lm}$, as
discussed in detail below.

Finally, we define gravitational wave phase and frequency in two
ways---one based on $\Psi_4^{22}$, and one based on $\dot{h}_{22}$:
\begin{equation}
  \label{eq:Def-PsiFourQuants}
  \PsiFourPhase = -\arg ( \Psi_{4}^{22} )\,, \qquad
  \PsiFourFreq = \frac{d}{dt} \PsiFourPhase \,,
\end{equation}
\begin{equation}
  \label{eq:Def-hDotQuants}
  \hDotPhase = -\arg \left( \dot{h}_{22} \right)\,, \qquad
  \hDotFreq = \frac{d}{dt} \hDotPhase \,.
\end{equation}
In both cases, we define the $\arg$ function to be the usual function,
with discontinuities of $2\pi$ removed.  Many PN formulae (see
Sec.~\ref{sec:PNapproximants}) involve yet another frequency and
phase: the {\em orbital} phase $\OrbitalPhase$ and {\em orbital}
frequency $\OrbitalFreq$.  Although the three frequencies satisfy
$\PsiFourFreq\approx \hDotFreq\approx 2\OrbitalFreq$, the slight
differences between different frequencies are significant at the level
of precision of our comparison (see Fig.~\ref{fig:CompFNorm_vs_omegas}
below), so it is important to distinguish carefully between them.

When discussing our numerical solution, we write all dimensionful
quantities in terms of the mass scale $M$, which we choose to be the
sum of the irreducible masses of the two black holes.\footnote{This
quantity was denoted by $m$ in Ref.~\cite{BoyleBrown2007}.}

\subsection{Calculation of \boldmath$\dot{h}$}

The energy flux depends on the spin-weighted spherical harmonic
coefficients of $\dot h$ via Eq.~(\ref{eq:Flux}).  We therefore need
to perform one time integration on $\Psi_{4}^{lm}$:
\begin{equation}\label{eq:hdot-integration}
  \dot{h}_{lm}(t) = \int_{t_0}^t \Psi_{4}^{lm}(t')\,dt' + H_{lm}.
\end{equation}
This integration is performed for each mode $(l,m)$ separately and
requires the choice of two integration constants, which are contained
in the complex number $H_{lm}$.  Ideally, $H_{lm}$ should be chosen
such that $\dot{h}_{lm}\to 0$ for $t\to -\infty$.  Because our
numerical simulations do not extend into the distant past, this
prescription cannot be implemented.  Rather, we make use of the
approximation that the real and imaginary parts of $\dot h_{lm}$
should oscillate symmetrically around zero.

Let us consider a pure sine/cosine wave, with constant amplitude and
phase:
\begin{align}
  \Psi_4^{\rm ex}&=A[\cos(\omega t)+i\sin(\omega t)],\\
  \dot h^{\rm ex}&=\frac{A}{\omega}[\sin(\omega t)-i\cos(\omega
  t)]+H^{\rm ex},
\end{align}
where the superscript `ex' stands for example. The amplitude is given
by
\begin{equation}
  |\dot{h}^{\rm ex}|^2=\frac{A^2}{\omega^2}+2\frac{A}{\omega}[\Real
  H^{\rm ex}\sin(\omega t)-\Imag H^{\rm ex}\cos(\omega t)]+|H^{\rm
    ex}|^2.
\end{equation}
Only for the correct choice of integration constants, $H^{\rm ex}=0$,
is the amplitude $|\dot h^{\rm ex}|$ {\em constant}.

Therefore, we propose to determine the integration constants $H_{lm}$
in Eq.~(\ref{eq:hdot-integration}) by minimizing the time derivative
of the amplitude over the entire waveform. In particular we minimize
\begin{equation}\label{eq:Functional}
  {\cal I}_{lm}\equiv \int_{t_1}^{t_2} \left(\frac{d}{dt}
    |\dot{h}_{lm}|^2\right)^2\,dt .
\end{equation}
From this minimization principle it follows that $H_{lm}$ is
determined by the linear system
\begin{subequations}
  \label{eq:MinimizeAmpVariations}
  \begin{align}
    \Real H&\int\!\! (\Real\Psi_4)^2 dt + \Imag H \int\!\! \Real
    \Psi_4\Imag\Psi_4 dt \nonumber \\ & = -\!
    \int\!\left[(\Real\Psi_4)^2
      \Real \dot{h}_{0} + \Real\Psi_4\Imag\Psi_4\Imag\dot{h}_{0}\right]dt,\\
    \Real H&\int\!\!\Real\Psi_4\Imag\Psi_4dt + \Imag H \int\!\!
    (\Imag\Psi_4)^2 dt \nonumber \\ & = -\!
    \int\!\left[(\Imag\Psi_4)^2\Imag\dot{h}_0 +
      \Real\Psi_4\Imag\Psi_4\Real\dot{h}_0\right]dt.
  \end{align}
\end{subequations}
Here, we have suppressed the indices $lm$ for clarity, all integrals
are definite integrals from $t_1$ to $t_2$, and $\dot{h}_0(t)\equiv
\int_{t_0}^t \Psi_4(t')\,dt'$.  For a given integration interval
$[t_1, t_2]$, Eqs.~(\ref{eq:MinimizeAmpVariations}) provide a
deterministic procedure to determine the integration constants
$H_{lm}$. We note that there have been several earlier proposals to
fix integration constants~\cite{PfeifferBrown2007, BertiCardoso2007,
PollneyReisswig2007, DamourNagar2008b, SchnittmanBuonanno2008}.  While
we have not tested those proposals, we point out that
Eqs.~(\ref{eq:MinimizeAmpVariations}) allow for very accurate
determination of the integration constants and one can easily obtain
an error estimate, as we discuss in the next subsection.

\subsection{Uncertainties in numerical quantities}

Because the amplitude and frequency of the waveform are not constant,
this procedure is imperfect, and the result depends somewhat on the
chosen values of $t_{1}$ and $t_{2}$.  To estimate the residual
uncertainty in $H$ due to this choice, we select nine different values
for $t_1$ and eleven values for $t_{2}$: $t_1=200M, 220M, \ldots,
360M$; $t_2=2000M, 2100M, \ldots, 3000M$.  The values of $t_1$ vary
over roughly one GW cycle and test the sensitivity to the GW phase at
the start of the integration interval; the values of $t_2$ are
designed to test the dependence on the amplitude at the end of the
integration interval.  For $t_2>3000M$ we find that the errors in our
procedure rapidly increase for several reasons: (a) the minimization
principle is based on the approximation that the amplitude is
constant; this approximation becomes worse toward merger; (b) ${\cal
I}_{lm}$ in Eq.~(\ref{eq:Functional}) weights absolute changes in
$|\dot h|$, not relative ones; close to merger, the amplitude becomes
so large that it dominates ${\cal I}_{lm}$; and (c) the integration
constants shift the waveform $\dot h_{lm}$ vertically, and we are
trying to determine the particular vertical shift such that $\dot
h_{lm}$ is centered around zero.  Determination of such an offset is
most accurate in a regime where the oscillations are {\em small},
i.e., at early times.

For each of these 99 integration intervals, we compute integration
constants using Eqs.~(\ref{eq:MinimizeAmpVariations}) for the three
dominant modes, $\dot{h}_{22}$, $\dot h_{44}$ and $\dot h_{32}$, and
we compute $F(t)$ from Eq.~(\ref{eq:Flux}) using only these modes and
we compute $\hDotFreq(t)$.  (We will show below that the contributions
of other modes are far below our numerical errors on the flux.)  We
average the 99 functions $F(t)$ and $\hDotFreq(t)$ and then use a
parametric plot of $F(t)$ versus $\hDotFreq(t)$ in our comparisons
presented below.  The variation in these 99 values yields an
uncertainty in $F$ due to the choice of integration constants.

\begin{figure}
  %NumericalFlux
  \includegraphics[width=\linewidth]{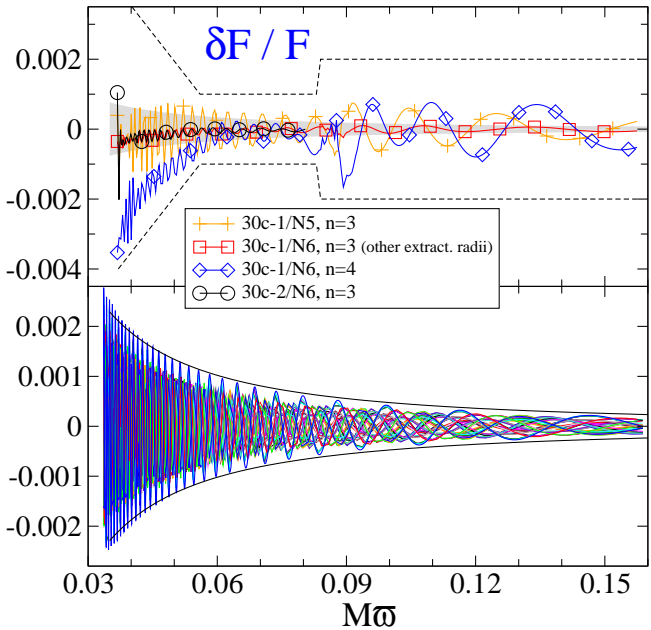}
  \caption{\label{fig:NumericalFlux} {\bf Lower panel:} Relative
    difference between flux $F(\hDotFreq)$ computed with 99 different
    intervals $[t_1,t_2]$ and the average of these.  {\bf Upper
    panel:} Relative change in the flux $F(\hDotFreq)$ under various
    changes to the numerical simulation.  The grey area in the upper
    panel indicates the uncertainty due to the choice of integration
    constants, which is always dominated by numerical error.  The
    dashed line in the upper panel is our final error estimate, which
    we plot in later figures.  }
\end{figure}

The lower panel of Fig.~\ref{fig:NumericalFlux} shows the variation in
flux from the 99 different integration intervals.  We find that the
{\em maximum} deviation can be well approximated by $\max|\delta F|/F=
1.5\times 10^{-5}(M\hDotFreq)^{-3/2}$ (see the solid line in lower
panel of Fig.~\ref{fig:NumericalFlux}).  The {\em average} $F$
computed from all 99 intervals $[t_1, t_2]$ will have a smaller error.
Inspection of the lower panel of Fig.~\ref{fig:NumericalFlux} reveals
that the $\delta F/F$ curves fall into 11 groups, corresponding to the
11 values of $t_2$.  Assuming that $\delta F$ between these groups is
randomly distributed, the error of the average will be reduced by a
factor $\sqrt{11}$, i.e., $\delta F/F = 5\times
10^{-6}(M\hDotFreq)^{-3/2}$.  This error is indicated as the grey
shaded area in the upper panel of Fig.~\ref{fig:NumericalFlux}.

\begin{figure}
  %NumericalFlux_Modes
  \includegraphics[width=0.97\linewidth]{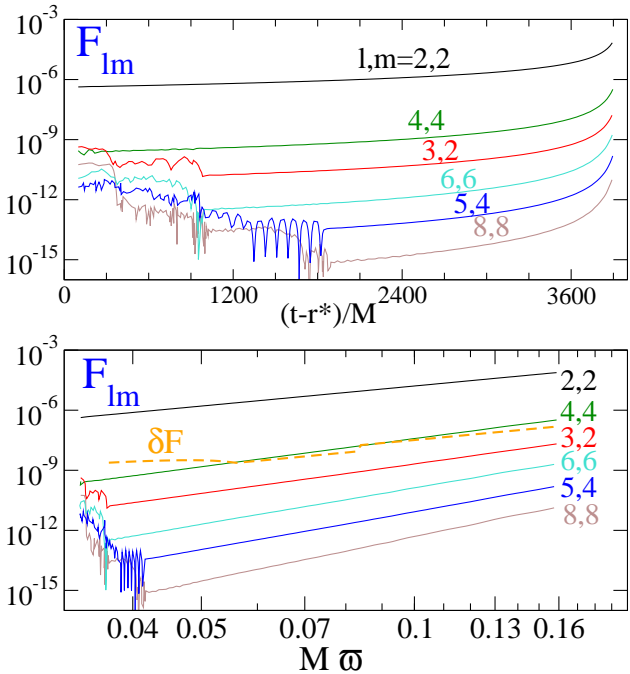}
  \caption{\label{fig:FluxModes} Contributions of various
    $(l,m)$-modes to the total numerical gravitational wave flux.
    {\bf Upper panel:} plotted as a function of time. {\bf Lower
    panel:} Plotted as a function of frequency $M\hDotFreq$. The lower
    panel also contains the error estimate derived in
    Fig.~\ref{fig:NumericalFlux}. }
\end{figure}

The upper panel of Fig.~\ref{fig:NumericalFlux} plots the relative
change in $F(\hDotFreq)$ for several changes in our numerical
simulation: (a) Computing the flux from a run with lower resolution
(0030c/N5 in the language of Boyle et al.~\cite{BoyleBrown2007}); (b) using
a different set of extraction radii for the extraction of the
gravitational wave; (c) increasing the polynomial order of
extrapolation of $\Psi_4$ to infinite extraction radius from $n=3$ to
$n=4$; and (d) computing the flux from a separate evolution with a
different outer boundary radius (0030c-2/N6).  At low frequencies, the
error is dominated by extrapolation to infinite radius and is a few
tenths of a percent; at intermediate frequencies, $0.055\lesssim
M\hDotFreq<0.083$, all errors are smaller than 0.1 percent.  At
frequency $M\hDotFreq\approx 0.084$ we change the gauge conditions in
the evolutions to allow wave-escorting; this introduces high-frequency
features, which are small when extrapolation order $n=3$ is used, but
which dominate for $n=4$ extrapolation.  The numerical data we use in
the PN comparisons below is extrapolated with $n=3$, for which the
features due to change of gauge are small, but nevertheless we will
use conservative error bars encompassing the $n=4$ extrapolation as
indicated in Fig.~\ref{fig:NumericalFlux}, i.e. a relative error of
0.2 per cent for $M\hDotFreq>0.083$.  We find that the uncertainty in
the flux due to numerical error in determining $\Psi_4$ is always
larger than the uncertainty due to the choice of integration
constants.

\begin{figure}
  %NumericalDotOmega2
  \includegraphics[width=\linewidth]{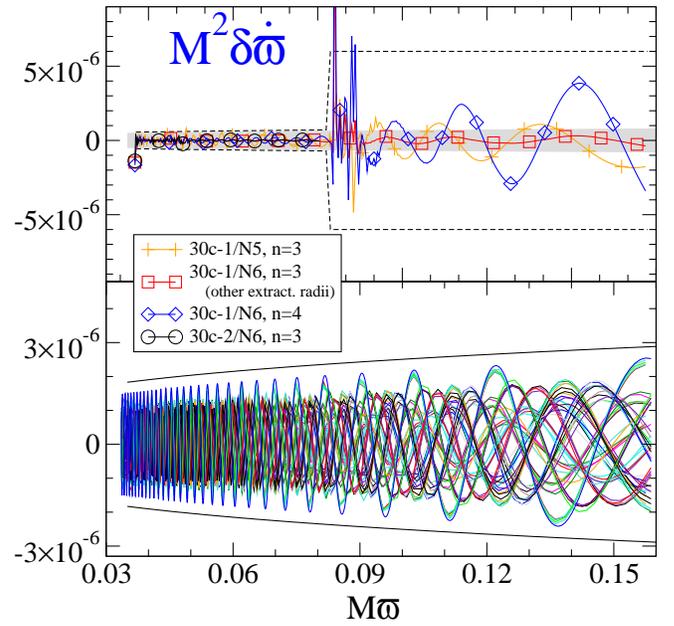}
  \caption{\label{fig:Numericalomega} {\bf Lower panel:} Difference
    between frequency derivative $\dot{\hDotFreq}$ computed with 99
    different intervals $[t_1, t_2]$ and the average of these. {\bf
    Upper panel:} Change in the frequency derivative $\dot{\hDotFreq}$
    under various changes to the numerical simulation.  The grey area
    in the upper panel indicates the uncertainty due to choice of
    integration constants, which dominates the overall uncertainty for
    low frequencies.  The dashed line in the upper panel is our final
    error estimate, which we plot in later figures. }
\end{figure}

The contributions of the various $(l,m)$-modes to the total flux [see
Eq.~(\ref{eq:Flux})] are plotted in Fig.~\ref{fig:FluxModes}.  The top
panel plots the flux as a function of time; the lower panel as a
function of frequency $M\hDotFreq$.  The dashed line in the lower
panel corresponds to the error estimate of
Fig.~\ref{fig:NumericalFlux}.  Because the modes $(5,4)$, $(6,6)$, and
$(8,8)$ are significantly smaller than our error estimate, we do not
include them in the present analysis.

To estimate the uncertainty in $\dot{\hDotFreq}$, we proceed in a
similar fashion.  Each one of the 99 different integration intervals
yields an $\dot h_{22}$ from which we determine $\dot{\hDotFreq}$.  We
average these to obtain the final $\dot{\hDotFreq}$ to be used in the
post-Newtonian comparisons.  The lower panel of
Fig.~\ref{fig:Numericalomega} shows the variation in $\dot{\hDotFreq}$
between the 99 different integration intervals.  We find that the {\em
maximum} deviation can be well approximated by $\max|M^{2}\delta
\dot{\hDotFreq}|= 5\times 10^{-6}(M\hDotFreq)^{-0.3}$ (see the solid
line in lower panel of Fig.~\ref{fig:Numericalomega}).  The {\em
average} $\dot{\hDotFreq}$ computed from all 99 intervals $[t_1, t_2]$
will have a smaller error.  Inspection of the lower panel of
Fig.~\ref{fig:Numericalomega} reveals that the $\delta\dot{\hDotFreq}$
curves fall into 11 groups, corresponding to the 11 values of $t_2$.
As for the case of $\delta F$, if we assume that $\delta
\dot{\hDotFreq}$ between these groups is randomly distributed, then
the error of the average will be reduced by a factor $\sqrt{11}$,
i.e., $M^{2}\delta\dot{\hDotFreq} = 1.5\times
10^{-6}(M\hDotFreq)^{-0.3}$.  This error is indicated as the grey
shaded area in the upper panel of Fig.~\ref{fig:Numericalomega}.

The upper panel of Fig.~\ref{fig:Numericalomega} plots also the change
in $\dot{\hDotFreq}(\hDotFreq)$ for the same changes in our numerical
simulation already discussed above.  We find that at
$M\hDotFreq<0.083$, the uncertainty in $\dot{\hDotFreq}$ is dominated
by the choice of integration constants, whereas at higher frequencies
the uncertainty is dominated by the numerical errors in the
calculation of $\Psi_4$. {As discussed above, at frequency
$M\hDotFreq\approx 0.084$ we change the gauge conditions in the
evolutions to allow wave-escorting; this introduces high-frequency
features leading to more conservative error estimates.

Note that $\dot{\hDotFreq}$ is a very steep function of $\hDotFreq$.
While the absolute errors in $\dot{\hDotFreq}$ are roughly constant
for our simulation, the {\em relative} errors change significantly:
$\delta\dot{\hDotFreq}/\dot{\hDotFreq}$ drops from about 10 per cent
early in the run to about 0.2 percent at late times.

We also point out that the first $1000M$ of our simulation are
contaminated by noise due to a pulse of ``junk-radiation'' at the
start of the simulation.  While this contamination is not apparent on
a plot of the waveform as in Fig.~\ref{fig:15OrbitOverview}, it
nevertheless limits accurate PN-NR comparisons to the region,
$t-r^\ast\gtrsim 1000M$, i.e., $M\hDotFreq\gtrsim 0.037$.

\section{Post-Newtonian approximants}
\label{sec:PNapproximants}

In this paper we will compare the numerical simulation to various
approximants based on the PN expansion.  The PN expansion is a
slow-motion, weak-field approximation to general relativity with an
expansion parameter $\epsilon \sim (v/c)^2 \sim (GM/rc^2)$.  For a
binary system of two point masses $m_1$ and $m_2$, $v$ is the
magnitude of the relative velocity, $M$ is the total mass, and $r$ is
the separation.  For a review of the PN expansion applied to
gravitational radiation from inspiralling compact binaries, see
Ref.~\cite{Blanchet2006}.

\begin{table}
  \begin{tabular}{|c|c|c|c|c|}
    \hline {\rm approximant} & {\rm notation} & {\rm see Eqs.} & {\rm
    adiabatic} & {\rm Keplerian} \\ \hline {\rm Taylor (T-)} &
    $F_n/E_p$ & (\ref{flux})/(\ref{energy}) & {\rm yes} & {\rm yes} \\
    {\rm \pade (P-)} & $F_n^m/E_p^q$ &
    (\ref{eq:pade-flux})/(\ref{eq:pade-energy})& {\rm yes} & {\rm yes}
    \\ {\rm EOB (E-)} & $F_n^m/H_{p}$ & (\ref{fluxPK})/(\ref{himpr})&
    {\rm no} & {\rm yes} \\ {\rm EOB (E-)} & ${}^{\rm nK}F_n^m/H_{p}$
    & (\ref{fluxPnK})/(\ref{himpr}) & {\rm no} & {\rm no} \\ {\rm EOB
    (E-)} & $F_n/H_{p}$ & (\ref{fluxTK})/(\ref{himpr}) & {\rm no} &
    {\rm yes} \\ {\rm EOB (E-)} & ${}^{\rm nK}F_n/H_{p}$ &
    (\ref{fluxTnK})/(\ref{himpr}) & {\rm no} & {\rm no} \\ \hline
    \end{tabular}
  \caption{\label{tab:approximants} Summary of PN-approximants. The
    T-approximants are always Taylor T4~\cite{BoyleBrown2007} except in
    Fig.~\ref{fig:dphase-eobandpade}. The P-approximant in the second
    row was introduced in Refs.~\cite{DamourIyer1998,DamourJaranowski2000,BuonannoChen2003}
    and the {\it original} E-approximant in third row was introduced
    in Refs.~\cite{BuonannoDamour1999,BuonannoDamour2000,DamourJaranowski2000}.  The last three rows
    refer to three possible variations of E-approximants introduced in
    Refs.~\cite{BuonannoChen2003,BuonannoChen2006}. In a few tests aimed at
    improving the closeness between numerical data and E-approximants,
    we vary $v_{\rm pole}$ and treat the logarithms as constants when
    \pade summation to the flux is applied~\cite{DamourNagar2008}.  We
    shall denote this flux by $\overline{F}_n^m$.  Finally, when using
    tuned PN-approximants with pseudo 4PN order terms in the flux,
    energy, or Hamiltonian, we denote the latter as $\small{p}F$,
    $\small{p}E$ and $\small{p}H$. Note that if known test-mass limit
    coefficients in the flux are used, the latter is still denoted as
    $F$ even at PN orders larger than 3.5PN.  Finally, the values of
    $v_{\rm pole}$ and $v_{\rm lso}$ used in the P-approximants
    $F_n^m$ and ${}^{\rm nK}F_n^m$ are $v^{\rm 2PN}_{\rm pole}=0.6907$
    and$v_{\rm lso}^{\rm 2PN}= 0.4456$.}
\end{table}

In Table~\ref{tab:approximants} we summarize the PN-approximants that
we use, and our notation.  We shall use the PN approximants in the
so-called adiabatic approximation, both in the standard
Taylor-expanded form (reviewed in Sec.~\ref{sec:TaylorApproximants})
and in a form based on \pade summation (reviewed in
Sec.~\ref{sec:PadeApproximants}).  In addition we shall use the
non-adiabatic EOB model (reviewed in Sec.~\ref{sec:EOBapproximants})
in its original form~\cite{BuonannoDamour1999,BuonannoDamour2000,DamourJaranowski2000}, as well as
several variations that differ in the form of the radiation-reaction
force~\cite{BuonannoChen2003,BuonannoChen2006,DamourGopakumar2006}.  After
summarizing the various PN approximants in
Secs.~\ref{sec:TaylorApproximants}, \ref{sec:PadeApproximants}, and
\ref{sec:EOBapproximants}, we describe how we construct the waveform
for these approximants in Sec.~\ref{sec:PNwaveforms}.

In the adiabatic approximation the inspiral is modeled as a
quasi-stationary sequence of circular orbits.  The evolution of the
inspiral (and in particular of the orbital phase $\OrbitalPhase$) is
completely determined by the \emph{energy-balance
  equation}~\cite{Blanchet2006}
\begin{equation}
  \frac{d {E}(v_\OrbitalFreq)}{d t} = - {F}(v_\OrbitalFreq)\,.
  \label{be}
\end{equation}
This equation relates the time derivative of the center-of-mass energy
${E}(v_\OrbitalFreq)$ (which is conserved in absence of radiation
reaction) to the gravitational wave energy flux
${F}(v_\OrbitalFreq)$. Both functions are known for quasicircular
orbits as a PN expansion in the invariantly defined velocity
\begin{equation}\label{eq:PN-velocity}
  v_\OrbitalFreq= \left( M \OrbitalFreq \right)^{1/3}\,,
\end{equation}
where $\OrbitalFreq = \dot{\OrbitalPhase} $ is the orbital frequency
(we use units such that $G=c=1$).\footnote{In Ref.~\cite{BoyleBrown2007} we
  used $x = v_\OrbitalFreq^2$ as the expansion parameter.}  We will
denote the Taylor-expanded flux (energy) by $F_k$ ($E_k$) where $k$
denotes the maximum power of $v_\OrbitalFreq$ retained in the series.
(Recall that $k=2N$ for an $N$th order PN expansion.)  We will denote
the \pade-expanded flux (energy) by $F^m_n$ ($E^m_n$) where $m+n=k$,
with $m$ and $n$ denoting the order of the polynomial in the numerator
and denominator, respectively.

\subsection{Adiabatic Taylor approximants}
\label{sec:TaylorApproximants}

For generic values of the symmetric mass ratio $\nu = m_1 m_2 / M^2$,
the center-of-mass energy is known through 3PN
order~\cite{JaranowskiSchafer1998,AndradeBlanchet2001, DamourJaranowski2000a, BlanchetFaye2001,DamourJaranowski2001}. For
circular orbits the Taylor PN-approximants (henceforth,
T-approximants) to the energy are given by
\begin{equation}
  \label{energy}
  {E}_{2k}(v_\OrbitalFreq) = -\frac{M \nu}{2}\,v_\OrbitalFreq^2\,
  \sum_{i=0}^k {\cal E}_{2i}(\nu)\,v_\OrbitalFreq^{2i}\,,
\end{equation}
where the known coefficients are
\begin{eqnarray}
  {\cal E}_0(\nu) &=& 1\,,\\
  {\cal E}_2(\nu) &=& - \frac{3}{4} - \frac{\nu}{12}\,,\\
  {\cal E}_4(\nu) &=& -\frac{27}{8} + \frac{19}{8}\,\nu -
  \frac{1}{24}\,\nu^2\,,\\
  {\cal E}_6(\nu) &=& - \frac{675}{64}+\left ( \frac{34445}{576} -
    \frac{205}{96}\pi^2 \right )\,\nu - \frac{155}{96}\,\nu^2
  \nonumber \\ && - \frac{35}{5184}\,\nu^3\,.
\end{eqnarray}

The GW energy flux for arbitrary masses has been computed through
3.5PN order~\cite{BlanchetFaye2002,BlanchetDamour2004}:
\begin{equation}
  \label{flux}
  {F}_k(v_\OrbitalFreq) = \frac{32}{5}\,\nu^2\,v_\OrbitalFreq^{10}\,
  \sum_{i=0}^{k} {\cal F}_i(\nu)\,v_\OrbitalFreq^{i}\,,
\end{equation}
where
\begin{eqnarray}
  {\cal F}_0(\nu) &=& 1\,, \\
  {\cal F}_1(\nu) &=&0 \,, \\
  {\cal F}_2(\nu) &=& -\frac{1247}{336} - \frac{35}{12}\nu \,, \\
  {\cal F}_3(\nu) &=& 4\pi\,, \\
  {\cal F}_4(\nu) &=& -\frac{44711}{9072} + \frac{9271}{504}\nu +
  \frac{65}{18}\nu^2\,, \\
  {\cal F}_5(\nu) &=& - \left( \frac{8191}{672} +
    \frac{583}{24}\nu \right) \pi\,, \\
  {\cal F}_6(\nu) &=& \frac{6643739519}{69854400} +
  \frac{16}{3}\,\pi^2 - \frac{1712}{105}\,\gamma_E \nonumber \\ && -
  \frac{856}{105}\,\log (16 v_\OrbitalFreq^2) + \left
    (-\frac{134543}{7776} + \frac{41}{48}\,\pi^2\right )\nu \nonumber \\ 
  && \label{f6} -\frac{94403}{3024}\,\nu^2 - \frac{775}{324}\,\nu^3 \,,\\
  {\cal F}_7(\nu) &=& \left (-\frac{16285}{504} +
    \frac{214745}{1728}\nu + \frac{193385}{3024}\nu^2 \right )\,\pi\,,
\end{eqnarray}
where $\gamma_E$ is Euler's constant. Notice that starting at 3PN
order ($k=6$) logarithms enter the flux.

\subsection{Adiabatic \pade approximants}
\label{sec:PadeApproximants}

\subsubsection{Center-of-mass energy}

Damour, Iyer and Sathyaprakash~\cite{DamourIyer1998} (henceforth DIS)
proposed a new class of approximate waveforms constructed by
introducing new energy and flux functions and by applying \pade
summation~\cite{BenderOrszag} to build successive approximants to
these two functions (henceforth P-approximants).  Their motivation for
introducing these new functions and using their P-approximants came
from an examination of the behavior of the standard PN-expansion and
the new P-approximants in the test-mass limit in which the exact
gravitational energy flux is known
numerically~\cite{Poisson1995}, the PN expansion of the flux
is known through 5.5PN order~\cite{TanakaTagoshi1996}, and the center-of-mass
energy is known analytically as
\begin{equation}
  \frac{E(v_\OrbitalFreq; \nu=0)}{\mu} =
  \frac{1-2v^2_\OrbitalFreq}{\sqrt{1-3v^2_\OrbitalFreq}} - 1\,,
\end{equation}
where $\mu = M \nu$ is the reduced mass.

DIS first observed that in the quantum two-body problem the symmetric
quantity
\begin{equation}
  \epsilon \equiv \frac{E^2_{\rm tot} -m_1^2 -m_2^2}{2 m_1\,m_2}\,,
  \label{eq:epsdef}
\end{equation}
(where the total relativistic energy ${E}_\mathrm{tot} = E + M$), is
the best energy function when treating the two-body problem as an
effective one-body problem in an external field.  Because in the
test-mass limit
\begin{equation}
  \epsilon(v_\OrbitalFreq;\nu=0) = \frac{1 - 2
    v^2_\OrbitalFreq}{\sqrt{1 - 3 v^2_\OrbitalFreq}}\,,
\end{equation}
DIS defined the new energy function as
\begin{equation}
  e(v_\OrbitalFreq) \equiv \epsilon^2 - 1\,,
  \label{eq:edef}
\end{equation}
as this function has a simple pole singularity on the real axis in the
test-mass limit, and DIS conjectured that such a pole would continue
to exist in the comparable mass case.\footnote{A motivation for having
  using Eq.~(\ref{eq:edef}) instead of Eq.~(\ref{eq:epsdef}) as a
  basic quantity is that the former (unlike the latter) is amenable to
  \pade summation in the test mass limit.} The energy function
$E(v_\OrbitalFreq)$ entering the balance equation (\ref{be}) can be
expressed in terms of $e(v_\OrbitalFreq)$ as
\begin{equation}
  E(v_\OrbitalFreq) = \left \{ M^2 + 2 \nu M^2 \left [ \sqrt{1 +
        e(v_\OrbitalFreq)} -1 \right ] \right \}^{1/2} - M\,.
  \label{emap}
\end{equation}
by combining Eqs.~(\ref{eq:epsdef}) and (\ref{eq:edef}).  [Note that
  the map between the adiabatic functions $e$ and $E$ given by
  Eq.~(\ref{emap}) is the same map found in the EOB model between the
  effective Hamiltonian $H^{\rm eff}$ and the real Hamiltonian $H^{\rm
    real}$, as given by Eq.~(\ref{himpr}).]

Finally, DIS proposed as approximants to the energy function
$e(v_\OrbitalFreq)$ the diagonal or subdiagonal P-approximants,
depending on whether the PN order is even or odd.\footnote{As the
  energy is only a function of even powers of $v_\OrbitalFreq$, the
  choice of using diagonal or subdiagonal (superdiagonal) is based on
  the order of $v^2_\OrbitalFreq$ that is retained.  For notational
  consistency, the indices on all approximants will refer to the power
  of $v_\OrbitalFreq$.  Other references define the indices on the
  energy approximants with respect to $v^2_\OrbitalFreq$.}
Investigating the behavior of the P-approximants under variations of
an (at the time) unknown coefficient in the 3PN center-of-mass energy,
Damour, Jaranowski and Sch{\"a}fer~\cite{DamourJaranowski2000} found it more robust
to use the superdiagonal P-approximant instead of the subdiagonal
P-approximant at 3PN order.\footnote{Subdiagonal P-approximants were
  extended to 3PN order in Ref.~\cite{DamourIyer2002}, and LAL~\cite{LAL}
  software uses those P-approximants for the energy function.} This
suggestion was also adopted in Ref.~\cite{BuonannoChen2003} and will be
used here; that is, we use subdiagonal P-approximants for 1PN,
diagonal for 2PN, and superdiagonal for 3PN.

The P-approximants for the center-of-mass energy are defined as
\begin{equation}
  \label{eq:pade-energy}
  E_p^q(v_\OrbitalFreq) = \left \{ M^2 + 2 \nu M^2\left [ \sqrt{1 +
        e_p^q(v_\OrbitalFreq)} -1 \right ] \right \}^{1/2} -M\,,
\end{equation}
where at 2PN order~\cite{DamourIyer1998}
\begin{equation}
  \label{eP2}
  e_2^2(v_\OrbitalFreq) = -v_\OrbitalFreq^2\,\frac{1+\frac{1}{3}\nu-\left (
      4-\frac{9}{4}\nu+\frac{1}{9}\nu^2 \right )\,v_\OrbitalFreq^2}
  {1+\frac{1}{3}\nu-\left ( 3-\frac{35}{12}\nu \right )\,v_\OrbitalFreq^2}\,, 
\end{equation}
and at 3PN order~\cite{DamourJaranowski2000}
\begin{eqnarray}
  \label{eP3}
  e^4_2(v_\OrbitalFreq) &=&
  -v_\OrbitalFreq^2\,\frac{1}{1-w_3(\nu)\,v_\OrbitalFreq^2} \left[
    1-\left (1+ \frac{1}{3}\nu +w_3(\nu) \right)\,v_\OrbitalFreq^2
    \right. \nonumber \\ && \left. - \left (3-\frac{35}{12}\nu-\left (
    1+\frac{1}{3}\nu\right )\, w_3(\nu)\right )\,v_\OrbitalFreq^4
    \right] \,,
\end{eqnarray}
where
\begin{eqnarray}
  w_3(\nu)&=&\frac{40}{36-35\nu}\,\left [\frac{27}{10}+\frac{1}{16}
    \left (\frac{41}{4}\pi^2-\frac{4309}{15}\right )\nu
  \right. \nonumber \\ && \left. +\frac{103}{120}\nu^2
-\frac{1}{270}\nu^3\right ]\,.
\end{eqnarray}

\subsubsection{Gravitational wave energy flux}

As originally pointed out in Refs.~\cite{Poisson1993,CutlerFinn1993}, the flux
function in the test-mass limit has a simple pole at the light-ring
position (i.e., the last unstable circular orbit of a
photon). Motivated by this, DIS introduced a new flux-type function
\begin{equation}
  f_k(v_\OrbitalFreq) = \left ( 1 - \frac{v_\OrbitalFreq}{v_{\rm
        pole}(\nu)} \right ) \,F_k(v_\OrbitalFreq;\nu)\,,
\end{equation}
with the suggestion that $v_{\rm pole}$ be chosen to be at the light
ring (pole singularity) of the new energy function.

In order to construct well behaved approximants, DIS proposed to
normalize the velocity $v_\OrbitalFreq$ entering the logarithms in
Eq.~(\ref{f6}) to some relevant scale which they chose to be $v_{\rm
  lso}(\nu)$, where the last stable orbit (LSO) is defined as the
minimum of the energy.  Also, they factored out the logarithms
yielding
\begin{eqnarray}
  f_k(v_\OrbitalFreq) &=& \frac{32}{5} \,\nu^2\,v_\OrbitalFreq^{10}\,
  \left [1 + \log \frac{v_\OrbitalFreq}{v_{\rm lso}(\nu)}\,
    \left ( \sum_{i \geq 6}^k \ell_i\,\,v_\OrbitalFreq^i \right ) \right]
  \nonumber \\ && \times 
  \left (1 - \frac{v_\OrbitalFreq}{v_{\rm pole}(\nu)} \right )
  \,\sum_{i=0}^k {\cal F}^{\text{log-fac}}_i\, v_\OrbitalFreq^i\,,
  \label{logs}
\end{eqnarray}
where $\ell_i$ and ${\cal F}^{\text{log-fac}}_i$ are functions of
${\cal F}_i$.  Through 3.5PN order, $\ell_6 = -1712/105$, $\ell_7=0$,
and ${\cal F}^{\text{log-fac}}_i = {\cal F}_i$ with the replacement of
$v_\OrbitalFreq \to v_{\rm lso}$ in ${\cal F}_6$ [see Eq.~(\ref{f6})].

Finally, DIS proposed to define the P-approximant of the GW energy
flux as
\begin{equation}
  \label{eq:pade-flux}
  {F}_n^m(v_\OrbitalFreq) = \frac{1}{1-v_\OrbitalFreq/v_{\rm pole}(\nu)}
  \,f_n^m(v_\OrbitalFreq)\,.
\end{equation}
where
\begin{eqnarray}
  f_n^m(v_\OrbitalFreq) &=& \frac{32}{5}
  \,\nu^2\,v_\OrbitalFreq^{10}\,\left [1 + \log
    \frac{v_\OrbitalFreq}{v_{\rm lso}(\nu)}\, \left ( \sum_{i \geq
      6}^k \ell_i\,\,v_\OrbitalFreq^i \right ) \right] \nonumber \\ &\times& 
   {\rm P}^m_n \left[ \left (1 -
    \frac{v_\OrbitalFreq}{v_{\rm pole}(\nu)} \right ) \,\sum_{i=0}^k
         {\cal F}^{\text{log-fac}}_i\, v_\OrbitalFreq^i\, \right],
\label{eq:pade-flux2}
\end{eqnarray}
where ${\rm P}^m_n[x]$ denotes \pade summation of the series $x$.  DIS
proposed to use the diagonal or subdiagonal P-approximants, depending
on whether $k = n + m$ is even or odd. Furthermore, DIS proposed to
use $v_{\rm lso}(\nu)$ and $v_{\rm pole}(\nu)$ as the minimum and pole
of the center-of-mass energy P-approximant of the same PN order.  At
2PN (the order to which the PN expansion was known by DIS) $v_{\rm
  pole}$ is determined from the pole of the \pade energy function
$e_2^2$, yielding
\begin{equation}
  v^{\rm 2PN}_{\rm pole}(\nu)=\frac{1}{\sqrt{3}}\,\sqrt{\frac{1+
      \frac{1}{3}\nu}{1-\frac{35}{36}\nu}}\,.
  \label{vpoleDIS}
\end{equation} 
When the PN expansion was extended to 3PN order, it was found that
none of the 3PN P-approximants have a physical pole.  Therefore,
somewhat arbitrarily, we will follow previous analyses and use the
value (\ref{vpoleDIS}) also at 3PN order. We denote the P-approximants
defined by Eqs.~(\ref{eq:pade-flux}) and (\ref{eq:pade-energy}) as
$F_n^m/E_p^q$.

The denominator in the \pade summation of the GW energy flux can have
zeros.  They are called {\it extraneous poles} of the
P-approximant~\cite{BenderOrszag}.  It is desirable that these poles
be located at high frequency (i.e., beyond the transition from
inspiral to plunge). We shall see that depending on the PN order and
also the mass ratio, extraneous poles can be present at low
frequencies.  This could indicate poor convergence of the \pade
summation.

In Secs.~\ref{sec:FittingFlux}, \ref{sec:PadeWaveforms} and
\ref{sec:EOBWaveforms} we shall investigate how to improve the
closeness of the PN-approximants to the numerical data by varying 
$a_5$~\cite{DamourIyer2003,BuonannoPan2007,DamourNagar2008}, 
$v_{\rm pole}$~\cite{DamourIyer2003,DamourNagar2008} 
and also by introducing higher-order PN coefficients in
the flux function.  When varying $v_{\rm pole}$ in the P-approximant
at 3.5PN order, extraneous poles appear at low values of
$v_\OrbitalFreq$. Therefore, in order to push these poles to very high
frequency, we follow the suggestion of Ref.~\cite{DamourNagar2008}, and
use P-approximants at 4PN order, where the 4PN coefficient is set to
its known value in the test-mass limit.  This cure may fail for
different mass ratios if new extraneous poles appear at low frequency.
Furthermore the logarithm in the flux is not factored out as in
Eq.~(\ref{logs}), but treated as a constant when \pade summation is
done. In this case the flux function is denoted $\overline{F}_n^m$.

We notice that DIS motivated the introduction of the P-approximants
first in the test-mass limit case by observing much faster and
monotonic convergence of the \pade energy, flux and waveforms with
respect to Taylor energy, flux and waveforms.  Quantitative tests of
the convergence were done only for the \pade waveforms (see e.g.,
Tables III and IV in Ref.~\cite{DamourIyer1998}), while for the flux and the
energy conclusions were drawn qualitatively from Figs. 3 and 4 of
Ref.~\cite{DamourIyer1998}. DIS then conjectured that the comparable mass
case is a smooth deformation of the test-mass limit case, and proposed
to use close-to-diagonal P-approximants for the flux and the energy
when $\nu \neq 0$. In the Appendix we perform a few convergence tests
of the P-approximants of the flux function in the test-mass limit
case, and conclude that whereas the P-approximants provide a better
fit to the numerical flux at 5.5PN order, they do not accelerate the
convergence of the Taylor series expansion of the energy flux.

\subsection{Non-adiabatic effective-one-body approximants}
\label{sec:EOBapproximants}

The EOB model goes beyond the adiabatic approximation and can
incorporate deviations from the Keplerian law when the radial
separation become smaller than the last stable circular orbit.

Here we briefly review the main equations defining the EOB dynamics
and refer the reader to previous papers for more
details~\cite{BuonannoDamour2000,BuonannoDamour1999,DamourJaranowski2000,BuonannoChen2006,PanBuonanno2008,BuonannoPan2007,
  DamourNagar2008,DamourNagar2008a}.  The non-spinning EOB effective Hamiltonian
is~\cite{BuonannoDamour1999,DamourJaranowski2000}:
\begin{eqnarray}
  \label{eq:genexp}
  H^{\rm eff}(\mathbf{r},\vp) &=& \mu\,\widehat{H}^{\rm eff}({\mathbf r},
  {\mathbf p}) \nonumber \\ 
  &=& \mu\,\left\{ A (r) \left[ 1 + {\mathbf p}^{2} +
      \left( \frac{A(r)}{D(r)} - 1 \right) ({\mathbf n} \cdot {\mathbf p})^2
    \right. \right. \nonumber \\ && \left. \left. 
      + \frac{1}{r^{2}} 2(4-3\nu)\,\nu\,({\mathbf n} 
      \cdot {\mathbf p})^4 \right] \right\}^{1/2}  \,,
\end{eqnarray}
with $\mathbf{r}$ and $\mathbf{p}$ being the reduced dimensionless
variables; $\mathbf{n} = \mathbf{r}/r$ where we set $r =
|{\mathbf{r}}|$. In absence of spins the motion is constrained to a
plane. Introducing polar coordinates $(r,\OrbitalPhase,
p_r,p_\OrbitalPhase)$, the EOB effective metric reads
\begin{eqnarray}
  \label{eq:EOBmetric}
  ds_{\rm eff}^2 \equiv g_{\mu \nu}^{\rm eff}\,dx^\mu\, dx^\nu &=&
  -A(r)\,c^2dt^2 + \frac{D(r)}{A(r)}\,dr^2 \nonumber \\ && +
  r^2\,(d\theta^2+\sin^2\theta\,d\phi^2) \,.  
\end{eqnarray}
The EOB real Hamiltonian is
\begin{equation}
  \label{himpr}
  H^{\rm real} = M\,\sqrt{1 + 2\nu\,\left ( \frac{H^{\rm eff} -
        \mu}{\mu}\right )} -M\,,
\end{equation}
and we define $\hat{H}^{\rm real}= {H}^{\rm real}/\mu$. The
T-approximants to the coefficients $A(r)$ and $D(r)$ in
Eqs.~(\ref{eq:genexp}) and (\ref{eq:EOBmetric})
read~\cite{BuonannoDamour1999,DamourJaranowski2000}
\begin{eqnarray}
  \label{coeffA}
  A_{k}(r) &=& \sum_{i=0}^{k+1} \frac{a_i}{r^i}\,,\\
  \label{coeffD}
  D_{k}(r) &=& \sum_{i=0}^k \frac{d_i}{r^i}\,,
\end{eqnarray}
where
\begin{eqnarray}
  \label{a0a3}
  && a_0=1\,, \quad a_1 =2\,, \quad a_2 =0\,, \quad a_3(\nu) =2 \nu \,, \quad 
  \nonumber \\ &&
  \label{a4a5} 
  a_4(\nu) = \left (\frac{94}{3}-\frac{41}{32}\pi^2\right )\,\nu \,,\\
  \label{d0d2}  
  && d_0=1\,, \quad d_1 =0\,, \quad d_2(\nu) = 6\,\nu\,, \quad \nonumber \\ && 
  \label{d3d4}  
  d_3(\nu) = 2\, (3\nu-26)\,\nu\,.
\end{eqnarray}

In Sec.~\ref{sec:EOBWaveforms}, we will explore the flexibility of the
EOB model by tuning the pseudo 4PN order coefficients $a_5(\nu)$ which
we will take to have the following functional form\footnote{Note that
  what we denote $a_5$ in this paper was denoted $\lambda$ in
  Ref.~\cite{BuonannoPan2007}.}
\begin{equation}\label{eq:a5}
a_5(\nu) = a_5\,\nu\,.
\end{equation}

In order to assure the presence of an horizon in the effective metric,
we need to factor out a zero of $A(r)$. This is obtained by applying
the \pade summation~\cite{DamourJaranowski2000}. Thus, the coefficients $A_{k}(r)$
and $D_{k}(r)$ are replaced by the \pade approximants~\cite{DamourJaranowski2000}
\begin{equation}
  A_2^1(r) = \frac{r\,(-4+2r+\nu)}{2r^{2}+2\nu+r\,\nu}\,,
  \label{coeffPA2}
\end{equation}
at 2PN order, and
\begin{equation}
  A_{3}^1(r) = \frac{{\rm Num}(A_3^1)}{{\rm Den}(A_3^1)}\,,
\end{equation}
with
\begin{equation}
{\rm Num}(A_3^1) = r^{2}\,[(a_4(\nu)+8\nu-16) + r\,(8-2\nu)]\,,
\end{equation}
and
\begin{eqnarray}
{\rm Den}(A_3^1) &=&
      r^{3}\,(8-2\nu)+r^{2}\,[a_4(\nu)+4\nu]
      \nonumber \\ &&+r\,[2a_4(\nu)+8\nu] 
      +4[\nu^2+a_4(\nu)]\,,
\end{eqnarray}
at 3PN order.  When exploring the flexibility of the EOB model, we
use the following \pade approximant at 4 PN order~\cite{DamourIyer2003,BuonannoPan2007}:
\begin{equation}
  A_{4}^1(r) = \frac{{\rm Num}(A_4^1)}{{\rm Den}(A_4^1)}\,,
\end{equation}
with
\begin{eqnarray}
  {\rm Num}(A_4^1) &=& r^3\,[32 -24 \nu - 4 a_4(\nu) -
  a_5(\nu)] \nonumber \\ &&
  + r^4 [a_4(\nu)  - 16 +8 \nu]\,,
\end{eqnarray} 
and
\begin{eqnarray} {\rm Den}(A_4^1) &=& -a_4^2(\nu) - 8a_5(\nu) - 8
  a_4(\nu) \nu + 2 a_5(\nu) \nu - 16 \nu^2 \nonumber \\ && + r\,[-8
  a_4(\nu) -4 a_5(\nu)
  -2 a_4(\nu)  \nu -  16 \nu^2] \nonumber \\
  && + r^2\,[-4a_4(\nu) - 2a_5(\nu) -16 \nu] \nonumber \\ && +r^3\,[-2
  a_4(\nu) - a_5(\nu) - 8 \nu] \nonumber \\ && + r^4\, [-16 + a_4(\nu)
  + 8 \nu]\,.
\end{eqnarray}
For the coefficient $D(r)$, the P-approximant used at
2PN, 3PN, and 4PN order respectively are~\cite{DamourJaranowski2000,DamourIyer2003,BuonannoPan2007}:
\begin{eqnarray}
  \label{D2PN}
  D_{2}^{0}(r)&=&1-\frac{6\nu}{r^2}\,,\\
  D_{3}^0(r)&=&\frac{r^3}{r^3+6\nu r+2 \nu(26-3\nu)}\,,\\
  \label{D4PN}
  D_{4}^0(r)&=& \frac{r^4}{r^4+6\nu r^2+2 \nu(26-3\nu)r - d_4(\nu) +36\nu^2}\,,
  \nonumber \\
\end{eqnarray}
and we choose somewhat arbitrarily $d_4(\nu) = 36\nu^2$, so that
$D_{4}^0=D_{3}^0$. (We note that the value of $d_4$ does not affect
much the EOB evolution~\cite{BuonannoPan2007}.)  The EOB Hamilton
equations written in terms of the reduced quantities $\widehat{H}^{\rm
  real}$ and $\widehat{t} = t/M$, $\widehat{\OrbitalFreq} =
\OrbitalFreq\,M$~\cite{BuonannoDamour2000}, are
\begin{eqnarray}
  \frac{dr}{d \widehat{t}} &=& 
  \frac{\partial \widehat{H}^{\rm real}}{\partial p_r}(r,p_r,p_\OrbitalPhase)\,, 
  \label{eq:eobhamone} \\
  \frac{d \OrbitalPhase}{d \widehat{t}} &\equiv& \widehat{\OrbitalFreq} = 
  \frac{\partial \widehat{H}^{\rm real}}
  {\partial p_\OrbitalPhase}(r,p_r,p_\OrbitalPhase)\,, \\
  \frac{d p_r}{d \widehat{t}} &=& - 
  \frac{\partial \widehat{H}^{\rm real}}{\partial r}(r,p_r,p_\OrbitalPhase)\,, \\
  \frac{d p_\OrbitalPhase}{d \widehat{t}} &=&
  \widehat{\cal F}[\widehat{\OrbitalFreq} (r,p_r,p_{\OrbitalPhase})]\,, 
  \label{eq:eobhamfour}
\end{eqnarray}
where for the $\OrbitalPhase$ component of the radiation-reaction
force a few approximants are available. Originally,
Ref.~\cite{BuonannoDamour2000} suggested the following Keplerian
P-approximants to the flux
  \begin{equation}
    \label{fluxPK}
    {}^{\rm K}\widehat{\cal F}_n^m \equiv - 
    \frac{1}{\nu v_{\OrbitalFreq}^3}\,{F}_{n}^m(v_{\OrbitalFreq};\nu, v_{\rm pole}) \,,
  \end{equation}
where ${F}_{n}^m$ is given by the \pade flux in
Eqs.~(\ref{eq:pade-flux}) and (\ref{eq:pade-flux2}).  Here by
Keplerian we mean that in the flux the tangential velocity
$V_\OrbitalPhase = \dot{\OrbitalPhase}\,r$ is set to $V_\OrbitalPhase
\equiv v_{\OrbitalFreq} = \dot{\OrbitalPhase}^{1/3}$, having assumed
the Keplerian relation $\dot{\OrbitalPhase}^2\,r^3 = 1$. It was then
pointed out in Ref.~\cite{DamourGopakumar2006} that the Keplerian relation
becomes less and less accurate once the binary passes through the last
stable orbit. A more appropriate approximant to the flux would be
\begin{equation}
    \label{fluxPnK}
    {}^{\rm nK} \widehat{\cal F}_n^m \equiv  - 
    \frac{v^3_\Omega}{\nu V_\OrbitalPhase^6}\,
    {F}_{n}^m(V_\OrbitalPhase;\nu, v_{\rm pole})\,,
\end{equation}
where $V_\OrbitalPhase \equiv \dot{\OrbitalPhase}\,r_\OrbitalFreq$.
Notice that because the EOB Hamiltonian is a deformation of the
Schwarzschild Hamiltonian, the exact Keplerian relation is
$\dot{\OrbitalPhase}^2\,r_\OrbitalFreq^3 = 1$ with $r_\OrbitalFreq
\equiv r\,[\psi(r,p_\OrbitalPhase)]^{1/3}$ and $\psi$ is defined following the argument presented around Eq. (19) to (22) in Ref.~\cite{DamourGopakumar2006}:
\begin{eqnarray}
\frac{1}{\psi r^3}&\equiv&\omega_{\rm circ}^2=\left (\frac{\partial\mathcal{H}(r,p_r=0,p_\phi)}{\partial p_\phi}\right )^2 \nonumber\\
&=&\frac{1}{r^3}\frac{p_\phi^2A(r)}{\left(1+\frac{p_\phi^2}{r^2}\right) r\left(1+2\eta\left(\sqrt{w(r,p_\phi)}-1\right)\right)} \nonumber\\
\end{eqnarray}
where $w(r,p_\phi)=A(r)\left(1+\frac{p_\phi^2}{r^2}\right)$.
The value of $p_\phi$ of circular orbits are obtained by minimizing with respect to $r$ the circular orbit Hamiltonian $\mathcal{H}(r,p_r=0,p_\phi)$ and it yields the following relation between $r$ and $p_\phi$
\begin{equation}\label{circpp}
\frac{2p_\phi^2A(r)}{r^3}=\left(1+\frac{p_\phi^2}{r^2}\right)\,\frac{dA(r)}{dr}\,.
\end{equation}
By inserting Eq.~\eqref{circpp} in the definition of $\psi$, and replacing all $p_\phi$ except 
the one which implicitly appears in $w(r,p_\phi)$ we obtain 
\begin{equation}
\psi=\frac{1+2\eta(\sqrt{w(r,p_\phi)}-1)}{r^2\,dA(r)/dr/2}\,.
\end{equation}

Finally, Refs.~\cite{BuonannoChen2003,BuonannoChen2006} introduced another possible
variation of the EOB flux approximants which use T-approximants for
the flux given by Eq.~(\ref{flux}), in either the Keplerian or
non-Keplerian form, i.e.
\begin{eqnarray}
    \label{fluxTK}
    {}^{\rm K}\widehat{\cal F}_n &=&  - \frac{1}{\nu v_{\OrbitalFreq}^3}\,
    F_n(v_{\OrbitalFreq})\,,
\end{eqnarray}
and
\begin{equation}
    \label{fluxTnK}
    {}^{\rm nK} \widehat{\cal F}_n = -  
    \frac{v^3_\Omega}{\nu V_\OrbitalPhase^6}\,F_n(V_\OrbitalPhase)\,.
\end{equation}
Note that the flux for the non-Keplerian EOB models are not simply
functions of the orbital frequency $\OrbitalFreq$.  We denote the
original E-approximants~\cite{BuonannoDamour1999,BuonannoDamour2000,DamourJaranowski2000} which use
the \pade flux (\ref{eq:pade-flux2}) as $F_n^m/H_p$ where $H_p$ is
$H^{\rm real}$ computed from $A_p^1$ and $D_p^0$.  Other
E-approximants used in this paper are summarized in
Table~\ref{tab:approximants}.  The initial conditions for
Eqs.~(\ref{eq:eobhamone})--(\ref{eq:eobhamfour}) are obtained
following Ref.~\cite{BuonannoDamour2000} and starting the evolution far apart
to reduce the eccentricity to negligible values.

\subsection{Waveforms}
\label{sec:PNwaveforms}

The PN waveforms are obtained by substituting the orbital phase and
frequency into the spherical harmonic mode (2,2) with amplitude
corrections through 3PN order~\cite{Kidder2008,ArunBlanchet2004}
\begin{eqnarray}
\label{eq:22-mode}
h_{22} &=& -8 \sqrt{\frac{\pi }{5}} \frac{\nu M}{R} e^{-2 i \OrbitalPhase } 
v_\OrbitalFreq^2
\left\{ 1 -  v_\OrbitalFreq^2 \left( \frac{107}{42} - \frac{55}{42} \nu \right)
\right. \nonumber \\* &+&
  \left. 2 \pi v_\OrbitalFreq^3 
-  v_\OrbitalFreq^4 \left( \frac{2173}{1512} + \frac{1069}{216} \nu - 
\frac{2047}{1512} \nu^2 \right)
\right. \nonumber \\* &-&
 \left. v_\OrbitalFreq^5
\left[ \left(\frac{107}{21} - \frac{34}{21} \nu \right) \pi + 24 i \nu \right]
\right. \nonumber \\* &+& \left.
 v_\OrbitalFreq^6 \left[ \frac{27027409}{646800} - \frac{856}{105} \gamma_E 
+ \frac{2}{3} \pi^2 - \frac{1712}{105} \ln{2} 
\right. \right. \nonumber \\* &-& \left. \left. 
 \frac{856}{105} \ln{v_\OrbitalFreq} 
- \left( \frac{278185}{33264} - \frac{41}{96} \pi^2 \right) \nu
- \frac{20261}{2772} \nu^2 \right. \right. \nonumber \\* &+& \left. \left.
\frac{114635}{99792} \nu^3 + \frac{428 i}{105} \pi
\right] + ~ O(\epsilon^{7/2}) \right\}.
\end{eqnarray}

For the adiabatic models, the orbital phase is obtained by rewriting
the energy balance equation~(\ref{be}) as
\begin{equation}
  \label{eq:dOrbFreq}
  \frac{d\OrbitalFreq}{dt} = -\frac{F}{dE/d\OrbitalFreq}\,.
\end{equation}
and integrating this equation along with $d\OrbitalPhase/dt =
\OrbitalFreq$.  The Taylor approximants are formed first by
substituting $F=F_n$ and $E=E_n$ into Eq.~(\ref{eq:dOrbFreq}).  The
P-approximant waveform is formed similarly by substituting $F=F^m_n$
and $E=E^m_n$ into Eq.~(\ref{eq:dOrbFreq}).  The TaylorT1 and \pade
approximants then numerically integrate Eq.~(\ref{eq:dOrbFreq}).  The
TaylorT4 approximant is formed by first re-expanding the right side of
Eq.~(\ref{eq:dOrbFreq}) as a single Taylor expansion truncated at the
appropriate order, and then numerically integrating the resulting
equation.  The TaylorT2 and TaylorT3 approximants perform the
integration analytically.  The various Taylor approximants are
reviewed in Sec. IIIE of Ref.~\cite{BoyleBrown2007}.

For the non-adiabatic EOB models, the orbital phase is determined by
solving Hamilton's
equations~(\ref{eq:eobhamone})-(\ref{eq:eobhamfour}).

After computing $h_{22}$, the appropriate time derivatives are taken
to form $\dot h_{22}$ and $\Psi_4^{22}$.

\section{Comparison with post-Newtonian approximants: energy flux}
\label{sec:CompareFlux}

We now compare the numerical GW energy flux with predictions from PN
theory.  In Sec.~\ref{sec:FluxComparison} we present comparisons with
T-, P- and E-approximants, and in Sec.~\ref{sec:FittingFlux} we
explore ways of fitting the numerical flux by introducing higher-order
PN coefficients and varying the value of $v_{\rm pole}$ away from
$v^{\rm 2PN}_{\rm pole}$ [Eq. (\ref{vpoleDIS})].

The PN flux is derived as a function of frequency, so it is natural to
perform this comparison as a function of frequency.  One alternative,
comparison as a function of time, would require computation of the PN
phase as a function of time.  This depends on the PN energy, so that a
comparison with respect to time would mix effects due to flux and
energy.  Furthermore, comparisons with respect to time are sensitive
to (and likely dominated by) secularly accumulating phase
differences~\cite{BakerMcWilliams2007}.

The PN flux is given in terms of the {\em orbital} frequency
$\OrbitalFreq$---see Eqs.~(\ref{flux}) and~(\ref{eq:PN-velocity})---so
at first glance, it might seem natural to compare PN and NR energy
fluxes at particular values of $\OrbitalFreq$. However, the orbital
frequency is gauge-dependent, and there is no simple relation between
the NR orbital frequency and the PN orbital frequency. Nor is there a
simple relation between the NR orbital frequency and any quantity
measured at infinity (where the energy flux is defined).  In
particular, it is very difficult to determine the NR orbital frequency
as a function of retarded time.  In contrast, the frequency
$\hDotFreq$ (see Eq.~(\ref{eq:Def-hDotQuants})) of the GWs at infinity
is an observable quantity, and is easily obtained from both PN
formulae and from the NR simulation. Therefore, to achieve a
meaningful comparison, we compare the PN and NR energy flux at
particular values of $\hDotFreq$.

\begin{figure}
  %omegas
  \includegraphics[width=.98\linewidth]{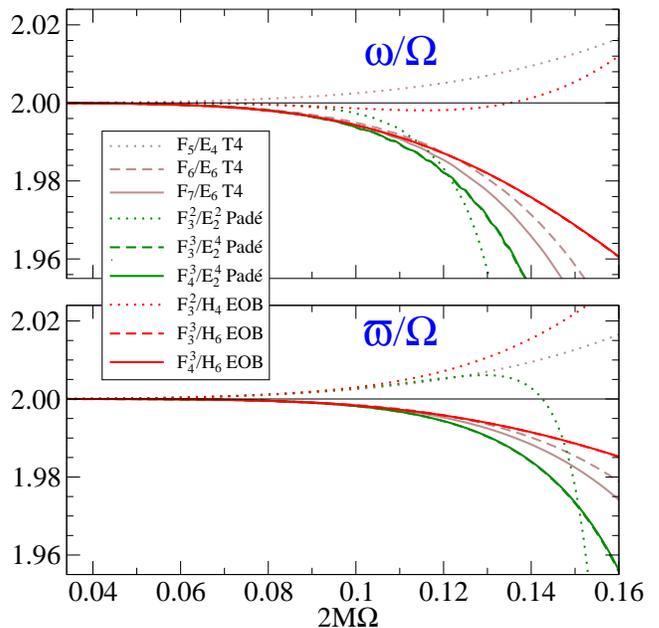}
\caption{Ratio of GW frequencies $\PsiFourFreq$ and $\hDotFreq$ to
  orbital frequency, as a function of (twice) the orbital frequency,
  for different PN models.  The GW frequencies $\PsiFourFreq$ and
  $\hDotFreq$ are defined in Eqs.~(\ref{eq:Def-PsiFourQuants})
  and~(\ref{eq:Def-hDotQuants}).  Solid lines correspond to 3.5PN,
  dashed and dotted lines to 3PN and 2.5PN, respectively.
\label{fig:GwVersusOrbitalFrequency}}
\end{figure}

In order to compute the PN flux as a function of $\hDotFreq$, we need
to find the mapping $\hDotFreq_{\rm PN} : \OrbitalFreq \to \hDotFreq$.
In order to find this mapping, we must build a PN waveform as a
function of $\OrbitalFreq$ and compute $\hDotFreq$ as defined by
Eq.~(\ref{eq:Def-hDotQuants}).  We construct the waveforms as
described in Sec.~\ref{sec:PNwaveforms}.  For the T-approximant of the
flux, we will use the TaylorT4 waveform.  In
Fig.~\ref{fig:GwVersusOrbitalFrequency} we plot both GW frequencies
(defined in Eqs.~(\ref{eq:Def-PsiFourQuants}) and
(\ref{eq:Def-hDotQuants})).  We then invert the mapping to obtain
$\OrbitalFreq_{\rm PN} = \hDotFreq_{\rm PN}^{-1} : \hDotFreq \to
\OrbitalFreq$.  So, given the PN flux $F(\OrbitalFreq)$ from
Sec.~\ref{sec:PNapproximants}, the flux as a function of the GW
frequency is given by $F(\hDotFreq)=F(\OrbitalFreq_{\rm
PN}(\hDotFreq))$.  The relation $\OrbitalFreq_{\rm PN}(\hDotFreq)$
depends on the instantaneous evolution of the PN model around
frequency $\OrbitalFreq$, and is therefore (unfortunately) dependent
on the PN model, in particular the choice of PN energy.  This
dependence, however, is local and will not lead to secularly
accumulating differences.

\begin{figure}
  %CompFNorm_vs_omegas
  \includegraphics[width=\linewidth]{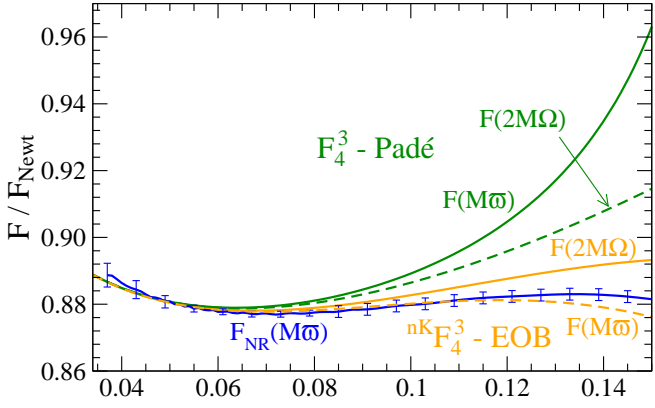}
\caption{Effect of choice of frequency.  Shown are the PN fluxes for
  two representative PN-approximants, plotted (correctly) as function
  of $\hDotFreq$ and (incorrectly) as function of $2\OrbitalFreq$.
  Plotting as a function of $2\OrbitalFreq$ changes the PN fluxes
  significantly relative to the numerical flux $F_{\rm NR}$.
\label{fig:CompFNorm_vs_omegas}}
\end{figure}

Notice from Fig.~\ref{fig:GwVersusOrbitalFrequency} that the orbital
frequency and the GW frequency differ by $\sim 1\% \mbox{--} 3\%$ at
large frequencies, depending on the PN model and the PN order, and the
difference in $\hDotFreq$ between different PN models is about $5\%$.
Because the energy flux is roughly proportional to $\hDotFreq^{10/3}$
(more precisely, $d\log F/d\log (M\hDotFreq)$ increases to $\sim$ 3.6
at $M \hDotFreq=0.15$), the difference in the flux caused by using GW
frequency from different PN models is about three to four times the
difference in GW frequencies.  Fig.~\ref{fig:CompFNorm_vs_omegas}
illustrates this effect by intentionally plotting the PN flux versus
the incorrect frequency $\OrbitalFreq$. Because changing the PN model
has a significant effect on the flux, we consider flux comparisons for
several different PN models below.

Note that for the flux comparison (and the comparisons of the
derivative of the energy in Sec. \ref{sec:CompareEnergy}), the PN
waveforms are used only to define the mapping between $\OrbitalFreq$
and $\hDotFreq$.  The PN flux is taken directly from the PN flux
expressions, e.g., Eq.~(\ref{flux}), and {\em not} computed by
applying Eq.~(\ref{eq:Flux}) to PN waveforms $h(t)$.
Equation~(\ref{eq:Flux}) is used only to compute the numerical flux.

\subsection{Flux comparison}
\label{sec:FluxComparison}

\begin{figure}
  %CompFomega_All
  \includegraphics[width=.98\linewidth]{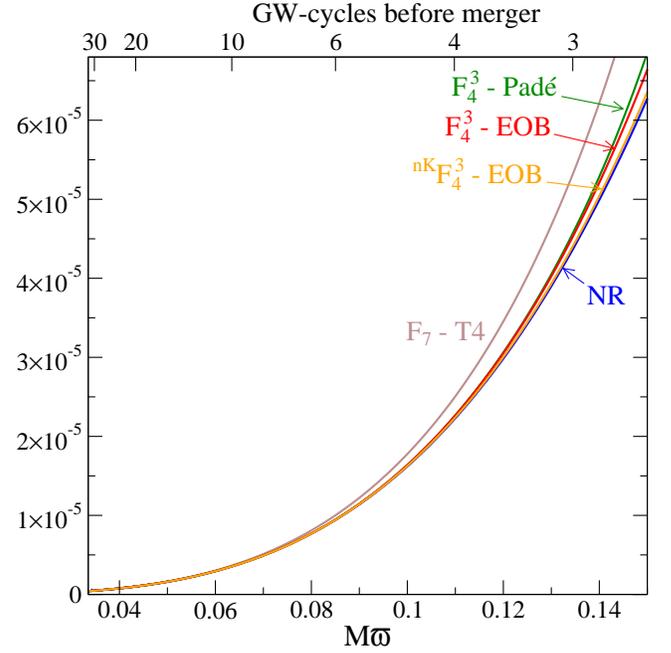}
 \caption{Comparison between the numerical energy flux and several PN
   approximants at 3.5PN order versus GW frequency $\hDotFreq$
   extracted from $\dot{h}_{22}$ in the equal-mass case.
\label{CompFomega1}}
\end{figure}

\begin{figure}
  %CompFomega_LF_diff
  \includegraphics[width=\linewidth]{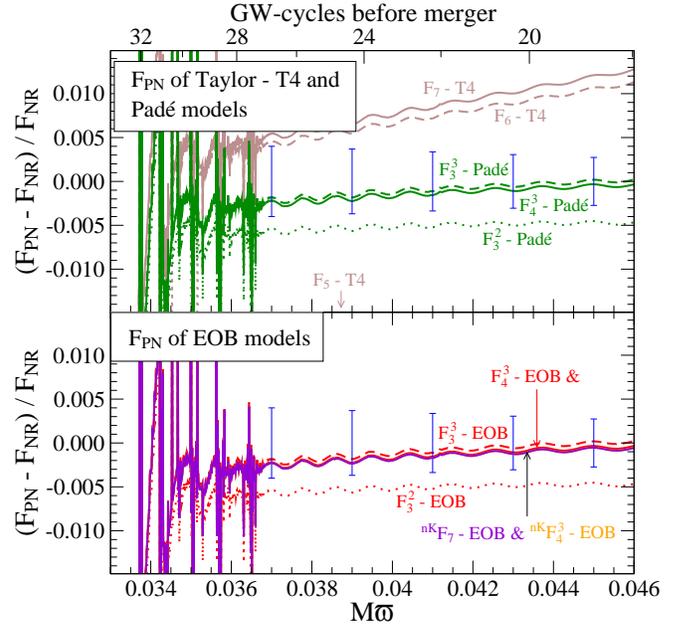}
\caption{Comparison between the numerical energy flux and several PN
  approximants versus GW frequency $\hDotFreq$ extracted from
  $\dot{h}_{22}$ in the equal-mass case.  We show the relative
  difference between numerical flux and PN flux, as well as the
  estimated error of the numerical flux (blue bars, see
  Fig.~\ref{fig:NumericalFlux}).  Solid lines represent 3.5PN models
  and NR; dashed and dotted lines correspond to 3PN and 2.5PN models,
  respectively. For notation see Table~\ref{tab:approximants} and
  caption therein.
\label{CompFomega2}}
\end{figure}

Figure~\ref{CompFomega1} plots the NR flux and the fluxes for the
\mbox{T-,} \mbox{P-,} and E-approximants at 3.5PN order as a function
of the GW frequency $\hDotFreq$ computed from $\dot{h}_{22}$. The
T-approximant is TaylorT4~\cite{BoyleBrown2007}.  Along the top of this
figure (as in several figures below) we indicate the number of
gravitational wave cycles up to merger, where we define ``merger'' as
the maximum of $|\Psi_4^{22}|$.  Figure~\ref{CompFomega2} zooms over
the first 15 GW cycles. We notice that during the first 15 GW cycles
the numerical data are fit best by the P- and E-approximants at 3PN
and 3.5PN order.  At these low frequencies the NR flux is best 
matched by the Keplerian and non-Keplerian EOB models and the \pade model.

\begin{figure}
  %CompFNorm_vs_omega
  \includegraphics[width=\linewidth]{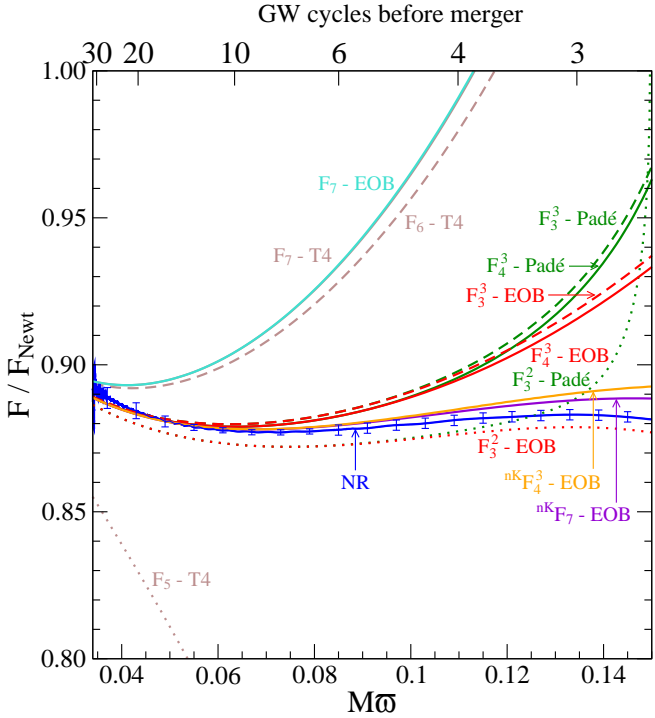}
\caption{Comparison of normalized energy flux $F/F_{\rm Newt}$ [see
    Eq.~(\ref{normflux})] for the equal-mass case.  Solid lines
  represent 3.5PN models and NR; dashed and dotted lines correspond to
  3PN and 2.5PN models, respectively.  For notation see
  Table~\ref{tab:approximants} and caption therein.
  \label{CompFomegaNorm}}
\end{figure}

\begin{table*}
  \begin{tabular}{c|cc|cc|cc|cc|cc}
    PN order
    & \multicolumn{2}{c|}{\footnotesize $v_\Omega\!=\!0.1$, 
      $2M\Omega\!=\!0.002$}
    & \multicolumn{2}{c|}{\footnotesize $v_\Omega\!=\!0.25$, 
      $2M\Omega\!=\!0.031$}
    & \multicolumn{2}{c|}{\footnotesize $v_\Omega\!=\!0.3$, 
      $2M\Omega\!=\!0.054$}
    & \multicolumn{2}{c|}{\footnotesize $v_\Omega\!=\!0.35$,
      $2M\Omega\!=\!0.086$}
    & \multicolumn{2}{c}{\footnotesize $v_\Omega\!=\!0.4$,
      $2M\Omega\!=\!0.128$}\\[2pt]
    (n+m)/2
    & $\displaystyle{\frac{{F}_{n+m}}{F_{\rm Newt}}}$ 
    & $\displaystyle{\frac{F_n^m}{F_{\rm Newt}}}$ 
    & $\displaystyle{\frac{{F}_{n+m}}{F_{\rm Newt}}}$ 
    & $\displaystyle{\frac{F_n^m}{F_{\rm Newt}}}$ 
    & $\displaystyle{\frac{{F}_{n+m}}{F_{\rm Newt}}}$ 
    & $\displaystyle{\frac{F_n^m}{F_{\rm Newt}}}$ 
    & $\displaystyle{\frac{{F}_{n+m}}{F_{\rm Newt}}}$ 
    & $\displaystyle{\frac{F_n^m}{F_{\rm Newt}}}$ 
    & $\displaystyle{\frac{{F}_{n+m}}{F_{\rm Newt}}}$ 
    & $\displaystyle{\frac{F_n^m}{F_{\rm Newt}}}$ 
    \\[5pt]
    \hline
    0.0&	        {\bf} 1.0000000&   {\bf }1.1692906 & {\bf }1.0000&   {\bf }1.5673   & 1.000&  1.7678   & 1.000 & 2.027 & 1.000 & 2.376	
    \\
    0.5&		{\bf} 1.0000000&   {\bf }1.0214102 & {\bf }1.0000&   {\bf }1.1507   & 1.000&  1.2325   & 1.000 & 1.345 & 1.000 & 1.505	
    \\
    1.0&		{\bf 0.9}555952&   {\bf 0.9}251084 & {\bf 0.}7225&   {\bf }-0.8648  & 0.939&  -7.8434  & 0.456 & 16.01& 1.091 & 8.443		
    \\
    1.5&		{\bf 0.96}81616&   {\bf 0.96}86094 & {\bf 0.}9188&   {\bf 0.}9074   & 0.940&   0.9069  & 0.995 & 0.924& 1.094 & 0.967 		
    \\
    2.0&		{\bf 0.96}81512&   {\bf 0.967}6191 & {\bf 0.}9184&   {\bf 0.8}850   & 0.939&   {\bf 0.8}671  & 0.993 & {\bf 0.8}60& 1.091 & 0.867	
    \\
    2.5&		{\bf 0.967}5775&   {\bf 0.967}6981 & {\bf 0.8}624&   {\bf 0.8}890   & 0.799&   {\bf 0.8}754 & 0.692 & {\bf 0.8}75 & 0.504& 0.893 	
    \\
    3.0&		{\bf 0.96772}65&   {\bf 0.96772}47 & {\bf 0.895}1&   {\bf 0.891}4   & {\bf 0.89}5&   {\bf 0.8}804 & {\bf 0.9}28 & {\bf 0.88}3& {\bf 1.0}22 & {\bf 0.90}3	
    \\
    3.5&		{\bf 0.96772}74&   {\bf 0.96772}33 & {\bf 0.895}7&   {\bf 0.891}2   & {\bf 0.89}7&   {\bf 0.8}798 & {\bf 0.9}34 & {\bf 0.88}2& {\bf 1.0}36 & {\bf 0.90}0
  \end{tabular}
  \caption{\label{tab:equalmass} Normalized energy flux $F/F_{\rm
      Newt}$ for the T- and P-approximants at subsequent PN orders for
    select velocities $v_\Omega$.  $v_\Omega=0.25$ corresponds to the
    start of the numerical simulation.  The P-approximant flux is
    given by Eq.~(\ref{eq:pade-flux}).  Note that the P-approximant
    has an extraneous pole at 1PN order at {$v_\Omega = 0.326$}.  We
    use $v_{\rm lso}=v^{\rm 2PN}_{\rm lso}=0.4456$ and $v_{\rm pole} =
    v^{\rm 2PN}_{\rm pole}=0.6907$.  We use boldface to indicate the
    range of significant figures that do not change with increasing PN
    order.}
\end{table*}

To more clearly show the behavior of the PN approximants, we plot in
Fig.~\ref{CompFomegaNorm} the energy flux normalized by the Newtonian
flux. The normalized flux is computed as
\begin{equation}
  \frac{F(\hDotFreq)}{F_{\rm Newt}(\hDotFreq)} \equiv 
  \frac{F(\hDotFreq)}{\frac{32}{5}\nu^2\left(\frac{M \hDotFreq}{2}\right)^{10/3}}\,,
  \label{normflux}
\end{equation}
where for the same reason mentioned above, the Newtonian flux is
expressed in terms of the GW frequency.  Notice that the
P-approximants and some of the E-approximants use the same \pade flux,
but they start differing at $M \hDotFreq\sim 0.12$ due to their
different GW frequencies (obtained from an adiabatic and non-adiabatic
evolution, respectively).  The E-approximants with Keplerian and
non-Keplerian flux increase less abruptly at high frequency than the
P- and T-approximants. This is a consequence of non-adiabatic effects
captured by the EOB model. Quite remarkably, the E-approximants with
non-Keplerian fluxes are rather close to the NR result for the entire
range of frequency spanned by the simulation.\footnote{We notice that
whereas the Keplerian \pade-based (or Taylor-based) approximants to
the flux differ from each other only when expressed in terms of the GW
frequency, the non-Keplerian \pade-based (or Taylor-based)
approximants to the flux differs from the others because their
functional dependence on the frequency is different (e.g., compare
Eq.~(\ref{fluxPnK}) with Eq.~(\ref{fluxPK})).}  We observe that
somewhat accidentally the PN-approximants at 2.5PN order are
also close to the numerical flux.

The normalized NR flux starts to decrease at $M \hDotFreq\sim
0.13$. We notice that this behavior is rather different from the
behavior of the normalized flux in the test-mass limit (see
Fig.~\ref{flux-tml} in the Appendix).  The E-approximants with
non-Keplerian \pade or Taylor flux show a similar decreasing behavior
at high frequency.

\begin{figure}
  %CauchyConvergence
  \includegraphics[width=0.9\linewidth]{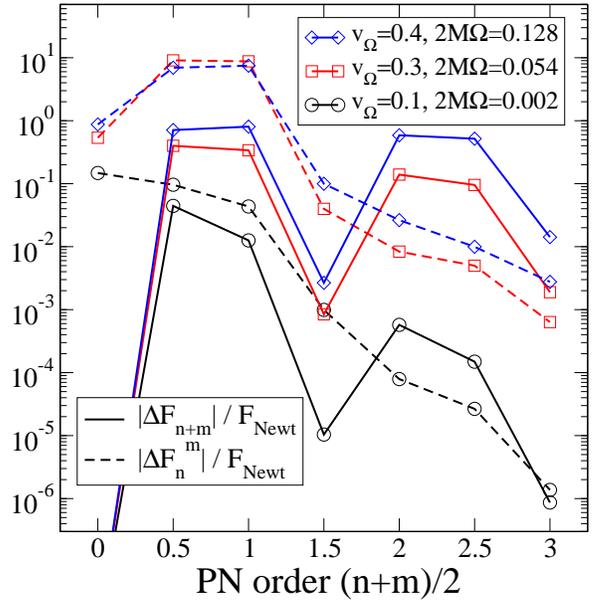}
  \caption{Cauchy convergence test of $F/F_{\rm Newt}$ for T- and
    P-approximants. We plot $\Delta{F}_{n+m} \equiv {F}_{n+m+1} -
    {F}_{n+m}$, and $\Delta{F}_n^m \equiv {F}_{n+1}^m - {F}_n^m$ for
    different values of $v_\Omega$.  The T- and P-approximants are
    given by Eqs.~(\ref{flux}) and (\ref{eq:pade-flux}), respectively.
    Note that the P-approximant has an extraneous pole at 1PN order at
    $v_\Omega = 0.326$.  We use $v_{\rm lso} = v_{\rm lso}^{\rm 2PN}$,
    and $v_{\rm pole}=v_{\rm pole}^{\rm
      2PN}$. \label{fig:CauchyConvergence}}
\end{figure}

Both Figs.~\ref{CompFomega2} and~\ref{CompFomegaNorm} show that in the
equal-mass case P-approximants fit the numerical results better than
T-approximants. In numerical analysis, however, \pade summation is
often used as a technique to accelerate the convergence of a
slowly-converging Taylor series (e.g., see Tables 8.9 and 8.12 in
Ref.~\cite{BenderOrszag}); hence it is natural to ask in the PN case
whether \pade summation indeed accelerates the convergence of the
series. In Table~\ref{tab:equalmass} we list the T- and P-approximants
of $F/F_{\rm Newt}$ computed at subsequent PN orders and for several
values of $v_\OrbitalFreq$ [from left to right $v_\OrbitalFreq = 0.1,
  0.25$ (i.e., beginning of the numerical simulation), $0.3, 0.35$,
  and $0.4$.] In Fig.~\ref{fig:CauchyConvergence} we perform a Cauchy
convergence test and compute the difference between T- and
P-approximants at subsequent PN orders. The figures do not suggest an
acceleration of the convergence. We notice that in the equal-mass case
P-approximants are converging more systematically than T-approximants.
However, this fact seems to depend on the mass ratio, as can be seen
by comparing Fig.~\ref{fig:CauchyConvergence} with
Table~\ref{tab:testmass} and Fig.~\ref{fig:testmassCC} in the Appendix
which are obtained in the test-mass limit.

\subsection{On the fitting of the numerical relativity energy flux}
\label{sec:FittingFlux}

In view of building accurate analytical templates that can interpolate
the NR waveforms during inspiral, merger and ringdown, we explore here
the possibility of improving the PN-approximants to the energy flux by
introducing {\it phenomenological} higher-order PN coefficients and/or
by varying the value of $v_{\rm pole}$.  This study should be
considered a first exploration of the problem, demonstrating only the
\emph{flexibility} of the PN models.  None of the \emph{quantities}
derived here should be used as the basis for further work.

We will minimize the difference between the PN flux and the numerical
flux by varying particular coefficients in the PN model.  Ideally, the
PN and numerical fluxes should be expressed as functions of
$\hDotFreq$ before taking this difference, so that the fluxes are
compared in a physically meaningful way.  Unfortunately, the
calculation of $\hDotFreq$ for the PN models is time-consuming,
because for each trial value of the phenomenological coefficient it is
necessary to compute a full waveform to determine the mapping between
$\hDotFreq$ and $\OrbitalFreq$.  So instead, in this section we simply
compare PN and numerical fluxes as functions of $\OrbitalFreq$, where
we define the numerical orbital frequency as $\OrbitalFreq \equiv
\hDotFreq/2$.  In Fig.~\ref{fig:CompFNorm_vs_omegas}, we can see that
the error introduced by the discrepancy between $\OrbitalFreq$ and
$\hDotFreq/2$ will be significant.  As we will show in
Sec.~\ref{sec:PadeWaveforms}, the waveforms produced using these
``tuned'' flux functions will improve agreement with the numerical
waveform at a significant level.  Nevertheless, the values derived in
this section may not be optimal.  Thus, we emphasize that the results
of this section constitute merely an exercise demonstrating the
feasibility of adjusting the PN parameters to optimize the agreement
of the PN flux function with numerical data.

\begin{figure}
  %fitfluxEM
  \includegraphics[width=\linewidth]{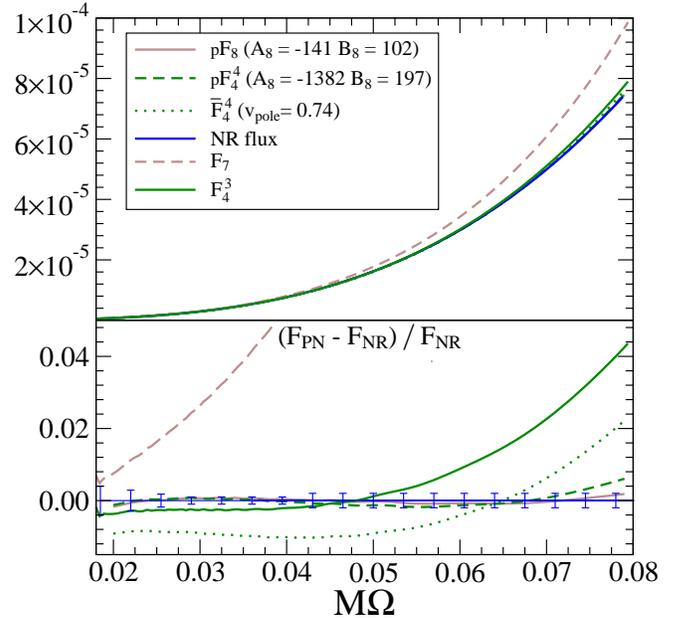}
\caption{Fitting several PN-approximants to the numerical flux. The
  $x$-axis denote the orbital frequency $\Omega$.  Because the
  numerical flux is computed as function of the GW frequency, we use
  for the numerical flux $\OrbitalFreq \equiv \hDotFreq/2$.  The blue
  bars indicate estimated errors on the numerical flux, see
  Fig.~\ref{fig:NumericalFlux}. For notation see
  Table~\ref{tab:approximants} and caption therein.
\label{fitflux}}
\end{figure}

The least-squares fits are done on $F(\hDotFreq)/F_{\rm
  Newt}(\hDotFreq)$ [see Eq.~(\ref{normflux})].  In the case of
T-approximants, we fit for the unknown 4PN-order coefficient in
Eq.~(\ref{flux}) for the equal-mass case.  We perform a least-squares
fit of the 4PN-order function ${\cal F}_{8}(\nu = 1/4) = A_{8} + B_{8}
\log v_\OrbitalFreq $ over the orbital-frequency range $M \OrbitalFreq = 0.02
\mbox{--} 0.08$ which starts after the first 9 GW cycles. We obtain
$A_8 = -141\,, B_8 = 102$. We notice that when we perform the fit over
the first 15 (or 20) GW cycles, spanning the frequency region $M
\OrbitalFreq =0.0168 \mbox{--} 0.0235$ ($M \OrbitalFreq =0.0168
\mbox{--} 0.0283$), the agreement becomes worse. The resulting flux is
shown in Fig.~\ref{fitflux}. The relative difference with the
numerical flux is at most $\sim 0.8\%$.

We repeat this analysis in the case of P-approximants. Because the
latter also depend upon $v_{\rm pole}$, we perform two least-squares
fits.  In the first fit, we fix $v_{\rm pole}$ to the value given by
Eq.~(\ref{vpoleDIS}) and apply the least-squares fit to ${\cal
  F}_{8}(\nu = 1/4)$ obtaining $A_8 = -1382\,, B_8 = 197$.

In the second fit, we vary $v_{\rm pole}$.  When varying $v_{\rm
  pole}$ in the P-approximant at 3.5PN order, extraneous poles appear
at low values of $v_\OrbitalFreq$. Therefore, in order to push these
poles to very high frequency, we follow the suggestion of
Ref.~\cite{DamourNagar2008}, and use P-approximants at 4PN order, where
the 4PN coefficient is set to its known value in the test-mass limit.
Furthermore the logarithm in the flux is not factored out, but treated
as a constant when \pade summation is done.  This cure may fail for
different mass ratios if new extraneous poles appear at low frequency.
The least-squares fit gives $v_{\rm pole}= 0.74$.  All the results for
the P-approximants are displayed in Fig.~\ref{fitflux}, where we also
show the T- and P-approximants at 3.5PN order without any fit.

Figure~\ref{fitflux} might suggest that by introducing higher-order PN
coefficients in the flux, the numerical flux can be fit better by
T-approximants than by P-approximants. However, this result can depend
on the use of orbital frequency instead of GW frequency. In
Sec.~\ref{sec:EOBWaveforms} (see Fig. ~\ref{fig:dphase-eob-time}) we
employ the fit values obtained in this study and show phase
differences between NR and tuned EOB models.

Finally, we attempted to extract PN coefficients higher than 3.5PN
order from the numerical flux, as was done at 2PN, 2.5PN and 3PN order
in Ref.~\cite{CutlerFinn1993} in the test-mass limit.  Unfortunately, the
differences between numerical flux and T-approximants are so large
---even at the beginning of the numerical waveform---that we were not
able to extract even known PN coefficients, like the ones at 3PN and
3.5PN order.  Thus, to fit unknown PN coefficients would require a
numerical simulation with more cycles starting at lower frequency.

\section{Estimation of the (derivative of the) center-of-mass energy}
\label{sec:CompareEnergy}

\begin{figure}
  %dOmOrb
  \includegraphics[width=\linewidth]{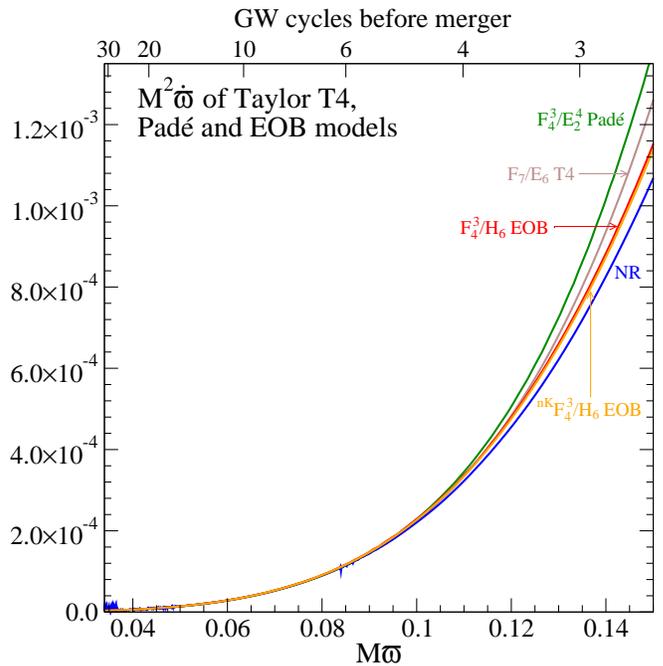} 
\caption{GW frequency derivative $\dot\hDotFreq$ for the numerical
  relativity simulation and various PN approximants at 3.5PN order.
  For notation see Table~\ref{tab:approximants} and caption therein.
\label{omegadot}}
\end{figure}

In the previous section, we analyzed and compared PN and numerical
energy fluxes.  The energy of the binary is the second fundamental
ingredient in the construction of adiabatic PN-approximants.
Unfortunately, there is no way to extract the energy for the numerical
simulation as a function of a gauge-invariant quantity such as the GW
frequency, so that it is impossible to compare PN and NR energies
directly.  The frequency derivative, $\dot{\hDotFreq}$, however, is
easily accessible in the numerical data, and, in the adiabatic
approximation is intimately related to the energy, as can be seen by
rewriting the energy balance, Eq.~(\ref{be}), in the form

\begin{equation}
\label{eq:BalanceEquationRearranged}
\frac{d\hDotFreq}{dt}=-\frac{F}{dE/d\hDotFreq}.
\end{equation}

Therefore, we begin this section with a comparison between numerical
$\dot\hDotFreq$ and the predictions of various PN-approximants.  For
the PN-approximants, we compute $h_{22}$ as usual (i.e., using energy
balance to compute the orbital frequency derivative
$\dot\OrbitalFreq$), and take a time derivative to obtain $\dot
h_{22}$ and extract $\dot\hDotFreq$ from it.  The waveform $h_{22}$
for the E-approximants is computed using Eqs.~(\ref{eq:genexp}),
(\ref{himpr}), (\ref{coeffA}) and (\ref{coeffD}) in
Sec.~\ref{sec:EOBapproximants}.  Figure~\ref{omegadot} plots the
numerical $\dot\hDotFreq$ and its value for T-, P- and also
E-approximants at 3.5PN order.

\begin{figure}
  %dOmOrbNorm
  \includegraphics[width=\linewidth]{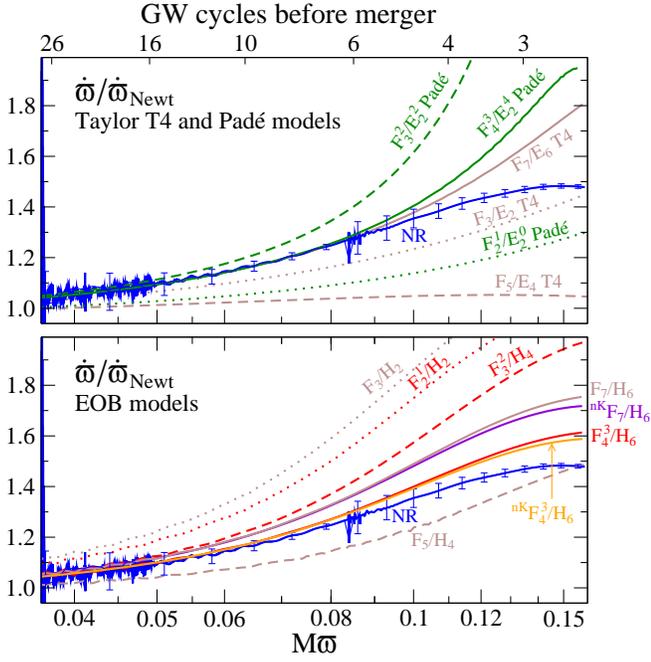}
  \caption{Comparison of $\dot{\hDotFreq}$ for the numerical results
    and various PN approximants.  Dotted, dashed and solid lines
    correspond to 1.5PN, 2.5PN and 3.5PN models, respectively.  For
    notation see Table~\ref{tab:approximants} and caption therein.
    \label{omegadotnorm}}
\end{figure}

In order to emphasize differences between the different
$\dot\hDotFreq$, we normalize the data in Fig.~\ref{omegadot} by the
Newtonian value of $\dot\hDotFreq$,
\begin{equation}
  \frac{\dot{\hDotFreq}}{\dot{\hDotFreq}_{\rm Newt}} \equiv
  \frac{\dot{\hDotFreq}} {\frac{192}{5} \frac{\nu}{M^2} \left(\frac{M
  \hDotFreq}{2}\right)^{11/3}}\,.
  \label{domeganorm}
\end{equation}
The normalization is used only to eliminate the leading-order behavior
of the various curves in Fig.~\ref{omegadot}; therefore, 
to compute the denominator of Eq.~(\ref{domeganorm}) we have
simply substituted $\hDotFreq/2$ rather than $\Omega$
into the Newtonian formula for the frequency derivative.

The normalized frequency derivatives are shown in
Fig.~\ref{omegadotnorm}.  At low frequencies, $\dot\hDotFreq$ is very
challenging to compute in numerical simulations, resulting in
comparatively large numerical uncertainties.  Therefore, for
frequencies $M\hDotFreq\lesssim 0.045$ we can merely conclude that PN
and NR are consistent with each other (i.e., are within the numerical
error bars of about 10 per cent).

The 3.5PN Taylor T4 model (labeled $F_7/E_6 T4$) agrees very well with
the numerical simulation up to $M\hDotFreq\approx 0.1$; this
observation is consistent with the excellent agreement between
TaylorT4 (3.5PN) and the numerical simulation observed in Boyle et
al.~\cite{BoyleBrown2007}, who compared up to this frequency.  Beyond
$m\hDotFreq=0.1$, however, $\dot\hDotFreq/\dot\hDotFreq_{\rm Newt}$
for Taylor T4 continues to increase (as for all other Taylor and \pade
models considered here), whereas for the numerical simulation,
$\dot\hDotFreq/\dot\hDotFreq_{\rm Newt}$ flattens (this behavior was
also observed in Ref.~\cite{DamourNagar2008}.)  Only the E-approximants at
3.5PN order reproduce the flattening of
$\dot\hDotFreq/\dot\hDotFreq_{\rm Newt}$ at high frequencies, with the
closest being the one which uses the non-Keplerian \pade flux ($^{\rm
  nK}F_4^3$).  Because the frequency derivative is the relevant
quantity that determines the phase evolution, the turning over of
$\dot\hDotFreq/\dot\hDotFreq_{\rm Newt}$ for the non-adiabatic models
in Fig.~\ref{omegadotnorm} suggests that, at high frequency,
non-adiabatic analytical models might be superior to adiabatic models.

\begin{figure}
 %dOmOrbNormSmoothed
  \includegraphics[width=\linewidth]{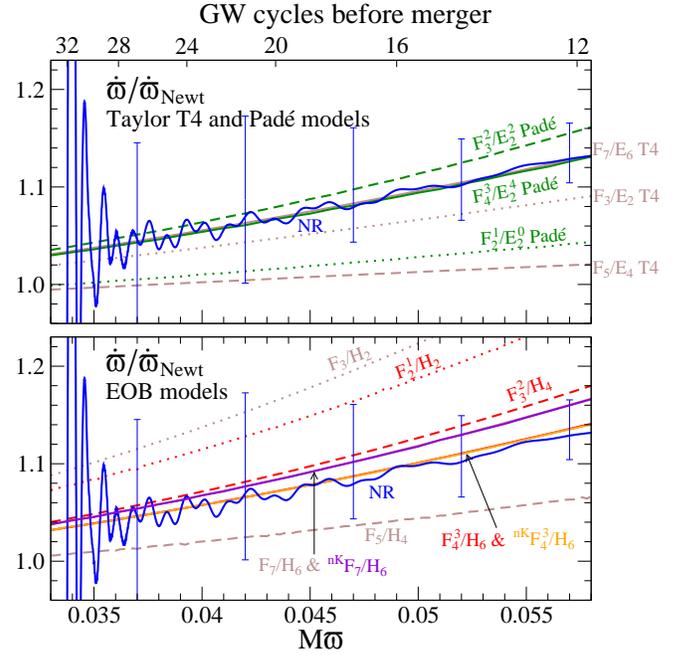}
\caption{Comparison of PN $\dot{\hDotFreq}$ with a heavily smoothed
  version of the numerical $\dot{\hDotFreq}$.  Solid lines represent
  3.5PN models and NR; dashed and dotted lines correspond to 3PN and
  2.5PN models, respectively.  For notation see
  Table~\ref{tab:approximants} and caption therein.
\label{fig:dOmOrbNormSmoothed}}
\end{figure}

If sufficient smoothing is applied to the numerical $\dot\hDotFreq$ it
becomes a smooth curve even at low frequencies.
Figure~\ref{fig:dOmOrbNormSmoothed} presents a comparison between such
a heavily smoothed numerical curve and the PN-approximants.  As
already pointed out, all PN approximants are consistent to within our
estimated numerical errors at low frequencies.  However, the NR result
in Fig.~\ref{fig:dOmOrbNormSmoothed} is notably closer to the 3.5PN
approximants than to lower order PN approximants.  This good agreement
provides a further validation of the numerical code used in Boyle et
al.~\cite{BoyleBrown2007}.  It also indicates that our error analysis in
Sec.~\ref{sec:NumericalEnergyFlux} may be overly conservative.

Our comparisons of $\dot\hDotFreq$ reveal a lot of information about
the PN approximants.  However, $\dot\hDotFreq$ depends on both flux
and energy (see Eq.~(\ref{eq:BalanceEquationRearranged})), and so
these comparisons do not yield information about flux or energy
separately.  To isolate effects due to the PN energy, we rearrange
Eq.~(\ref{eq:BalanceEquationRearranged}) further, such that it yields
in the adiabatic approximation the derivative of the center-of-mass
energy for the numerical simulation:
\begin{equation}
  \label{eq:NREnergyDerivative}
  \left [\frac{d {E}}{d \hDotFreq} \right ]_{\rm NR} = - \frac{F_{\rm
      NR}}{[d \hDotFreq/dt]_{\rm NR}}\,.
\end{equation}
The relative error in $ \left [d {E}/d \hDotFreq \right ]_{\rm NR}$ is
obtained as the root-square-sum of the relative errors of flux and
frequency derivative (see Figs.~\ref{fig:NumericalFlux}
and~\ref{fig:Numericalomega}).  In Fig.~\ref{dE} we compare the latter
with T-, P- and E-approximants. For adiabatic T4 and \pade models, we
compute $dE/d\hDotFreq$ by taking derivatives of $E(\OrbitalFreq)$ in
Eq.~(\ref{energy}) with respect to $\OrbitalFreq$ and then expressing
the derivative in terms of $\hDotFreq(\OrbitalFreq)$. For
non-adiabatic EOB models, we compute $dE/d\hDotFreq$ from the ratio of
$F_{\rm PN}$ and $[d \hDotFreq/dt]_{\rm PN}$ as obtained from
Figs.~\ref{CompFomega1} and \ref{omegadot}.  The closeness between the
numerical result and adiabatic PN-approximants is expected only in the
range of frequencies over which the balance equation and the adiabatic
approximation are valid. The upper panel of Fig.~\ref{dE} shows the
Taylor and \pade adiabatic models. The plot suggests that around $M
\hDotFreq \sim 0.08$ non-adiabatic effects are no longer negligible.
At lower frequencies, both 3.5PN order adiabatic approximants (\pade
and Taylor T4) match the numerical result very well.  Taylor T4 at
2.5PN matches well, too, although its frequency derivative
$\dot\hDotFreq$ and flux differ significantly from NR (see
Figs.~\ref{omegadotnorm} and~\ref{CompFomegaNorm}).  The T-approximant
at 3.5PN order is closest to the numerical result.  The lower panel of
Fig.~\ref{dE} shows the non-adiabatic E-approximants.  We notice that
the non-adiabatic models, especially at 3.5PN order, follow quite
nicely the behavior of the numerical derivative of the center of mass
energy. The E-approximant with non-Keplerian flux is closest to the
numerical result.  This analysis emphasizes again the relevance of
including non-adiabatic effects in the analytical
model~\cite{BuonannoDamour2000}.

\begin{figure}
  %diffdEOmNRPN_twopanel
  \includegraphics[width=\linewidth]{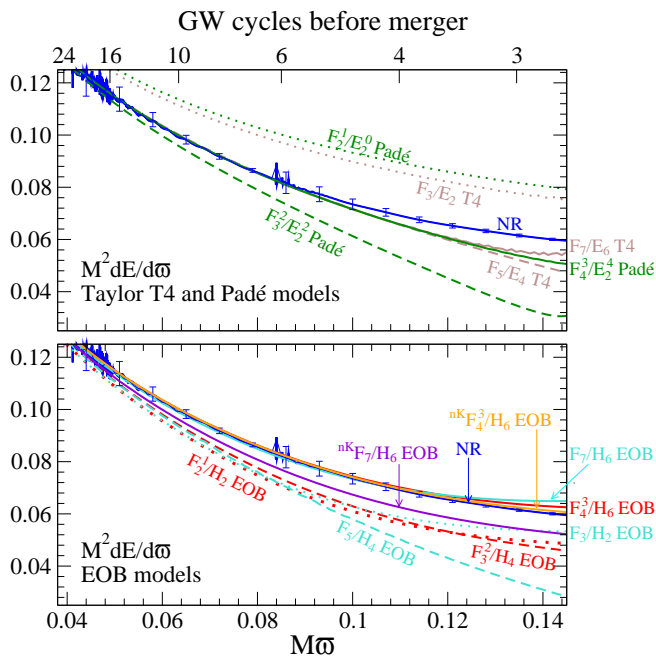}
  \caption{We compare $dE/d\hDotFreq$ versus GW frequency $\hDotFreq$
    for numerical relativity [see Eq.~(\ref{eq:NREnergyDerivative})]
    and PN approximants. Solid lines represent 3.5PN models and NR;
    dashed and dotted lines correspond to 3PN and 2.5PN models,
    respectively. For notation see Table~\ref{tab:approximants} and
    caption therein.
\label{dE}}
\end{figure}

\section{Comparing waveforms}
\label{sec:CompareWaveforms}

Here we compare the numerical waveform to various PN waveforms,
basically extending the analysis of Boyle et al.~\cite{BoyleBrown2007} to
include \pade and EOB waveforms.  Because the $(2,2)$ mode dominates
the waveform for an equal-mass non-spinning binary, we restrict the
comparison to only this mode.  As in~\cite{BoyleBrown2007}, we use
$\Psi_{4}^{22}$ and the GW phase and frequency $\PsiFourFreq$ defined
by Eq.~(\ref{eq:Def-PsiFourQuants}) when comparing waveforms.

For the comparisons presented in this section, the uncertainty in the
phase of the numerical waveform is roughly 0.02 radians.  This number
includes numerical errors (e.g. due to convergence and extrapolation
of the waveform to infinite extraction radius), as well as modelling
errors due to slightly nonzero eccentricity and spin of the numerical
simulation; see Ref.~\cite{BoyleBrown2007} Sec. V. for details.  We
note that the modelling errors have decreased since the analysis in
Ref.~\cite{BoyleBrown2007} because the new matching procedure
reduces the impact of eccentricity, and because the more
sophisticated spin-diagnostics presented in
Ref.~\cite{LovelaceOwen2008}) resulted in a smaller bound on the
residual spin.

\subsection{Matching procedure}
\label{sec:matching-procedure}

Each PN waveform has an arbitrary time offset, $t_{0}$, and phase
offset, $\PsiFourPhase_{0}$ with respect to the NR waveform.  The
procedure used by Boyle et al.~\cite{BoyleBrown2007}---as well as in
various other papers before it, such
as~\cite{BakerVanMeter2007,HannamHusa2008}---sets these constants by ensuring
that the GW phase and frequency match at a fiducial time.
Unfortunately, when matching at low frequency this method is 
sensitive to noise and to residual eccentricity in the numerical
waveform, and does not easily translate into a robust and automatic
algorithm.  Since we want to match as early as possible (where
we expect the PN approximants to be valid), we propose to use,
instead, a matching procedure which achieves the same goal, but
extends over a range of data. This procedure is similar to the one
proposed by Ajith et al.~\cite{AjithBabak2008}, but
whereas we match only the GW phase, Ajith et al. match the entire
gravitational waveform---including the amplitude---and include an
overall amplitude scaling. This method can be easily implemented as a
fairly automatic algorithm, robust against noise and residual
eccentricity.

Using the phase of the numerical and PN waveforms, we define the
quantity
\begin{equation}
  \label{eq:MatchingChiSquared}
  \Xi(\Delta t, \Delta \PsiFourPhase) = \int_{t_{1}}^{t_{2}}\,
  \left[ \PsiFourPhase_{\text{NR}}(t) - \PsiFourPhase_{\text{PN}}(t -
    \Delta t) - \Delta \PsiFourPhase \right]^{2}\, dt \,.
\end{equation}
Here, $t_{1}$ and $t_{2}$ represent the chosen range over which to
compare.  Minimizing this quantity by varying the time and phase
offsets $\Delta t$ and $\Delta \PsiFourPhase$ produces the optimal
values for these quantities in a least-squares sense.  Then to compare
PN and NR waveforms, we compare the (unchanged) NR waveform with an
offset PN waveform defined by
\begin{equation}
  \label{eq:MatchedWaveform}
  \Psi_{4,\text{PN}}(t) = A_{\text{PN}}(t+\Delta t)\,
  \text{e}^{-\text{i} \left[ \PsiFourPhase_{\text{PN}}(t + \Delta t) +
      \Delta \PsiFourPhase \right]} \,.
\end{equation}

With reasonable first guesses for $\Delta t$ and $\Delta
\PsiFourPhase$, the function $\Xi$ is quite nicely paraboloidal.
Thus, even simple minimization routines work well.  However, in cases
where speed is an issue, the problem can be reduced to one dimension.
For a given value of $\Delta t$, the optimization over $\Delta
\PsiFourPhase$ may be done analytically by setting
\begin{equation}
  \label{eq:MatchedDeltaPhi}
  \Delta \PsiFourPhase ( \Delta t) = \frac{\int_{t_{1}}^{t_{2}}\,
    \left[ \PsiFourPhase_{\text{NR}}(t) - \PsiFourPhase_{\text{PN}}(t -
      \Delta t) \right]\, dt} {t_{2}-t_{1}} \,.
\end{equation}
Using this value of $\Delta \PsiFourPhase$ for a given value of
$\Delta t$ decreases the number of function evaluations needed to find
the minimum. This can be very useful for large data sets, or
situations where many such matches need to be done.

The choice of $t_{1}$ and $t_{2}$ involves some degree of judgment.
Preferably, $t_{1}$ should be as early as possible, while not being
contaminated by junk radiation.  We choose $t_{1}=1100M$,
corresponding to $M\PsiFourFreq=0.037$. Similarly, $t_{2}$ should be
as early as possible, but far enough from $t_{1}$ so that the
integration averages over the noise.  In addition, the effects of the
small but nonzero orbital eccentricity show up as oscillations in the
phase, as can be seen, for example, in the range $t\in [1100,1900]M$
in Fig.~\ref{fig:dphase-pade-pole-f8}. We would like $t_2$ to be large
enough so that the integration averages over several cycles of this
oscillation, thus resulting in less bias due to eccentricity.  Here we
use $t_{2} = 1900M$, corresponding to $M\PsiFourFreq=0.042$.  We have
checked that changing the values of $t_{1}$ and $t_{2}$ by $\pm 100M$
changes the resulting phases by less than a few thousandths of a
radian through the end of the numerical waveform.

This method is quite similar to the one suggested in
Ref.~\cite{AjithBabak2008}. However, here we consider
only the phase and not the amplitude of the waveform.  Because we
restrict the analysis only to the $(2,2)$ waveform mode of an
equal-mass binary and compare only the phase and not the amplitude, we
think it is reasonable to have neglected the amplitude in the matching
procedure.

\subsection{Pad\'e waveforms}
\label{sec:PadeWaveforms}

\begin{figure}
  %EOBandPade-PNvsNR_byOrder
  \includegraphics[width=0.98\linewidth]{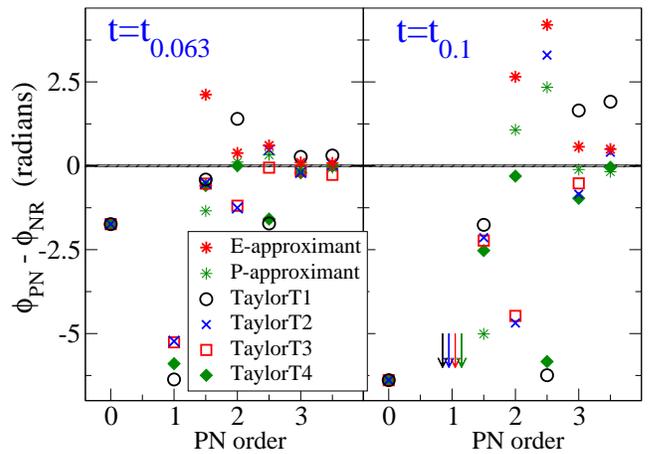}
  \caption{Phase differences between the numerical waveform, and
    untuned, original EOB, untuned \pade, and Taylor waveforms, at two
    selected times close to merger. The E-approximants are
    $F_n^m/H_p$, while the P-approximants are ${F}^m_n/E^q_p$ (see
    Table~\ref{tab:approximants} and caption therein).  Waveforms are
    matched with the procedure described in
    Sec.~\ref{sec:matching-procedure} and phase differences are
    computed at the time when the numerical simulation reaches $M
    \PsiFourFreq=0.063$ (left panel) and $M \PsiFourFreq=0.1$ (right
    panel).  Differences are plotted versus PN order. Note that at 1PN
    order the \pade flux has an extraneous pole at $v = 0.326$ causing
    a very large phase difference.  The thick black line indicates the 
uncertainty of the comparison as discussed in
    Sec.~\ref{sec:CompareWaveforms}, $|\Phi_{\rm PN}-\Phi_{\rm NR}|\leq 0.02$ radians.
    \label{fig:dphase-eobandpade}}
\end{figure}

In Fig.~\ref{fig:dphase-eobandpade} we plot the phase difference
between the numerical, T- and P-approximants~\cite{DamourIyer1998,DamourJaranowski2000,
  BuonannoChen2003} at the times when the numerical waveform reaches GW
frequencies $M \PsiFourFreq=0.063$ and $M \PsiFourFreq=0.1$.  The
phase differences are plotted versus the PN order. The phase
difference at $M \PsiFourFreq=0.1$ of the P-approximant at 3.5PN order
is $-0.12$ radians.  When comparing with generic Taylor approximants,
we notice that the phase differences of the P-approximants are less
scattered as the PN order is increased.  This might be due to the fact
that P-approximants of the energy flux are closer to the NR flux,
especially for lower $v_\OrbitalFreq$ where the phase accumulates the
most. Figure~\ref{fig:dphase-eobandpade} could be contrasted with
Tables III and IV of Ref.~\cite{DamourIyer1998} which show the overlaps
between the numerical waveform and P-approximants at subsequent PN
orders, in the test-mass limit case.  The behavior of the
P-approximants in Fig.~\ref{fig:dphase-eobandpade} are consistent with
the behavior of $\dot\hDotFreq$ seen in Fig.~\ref{omegadotnorm}: At
1.5PN, \pade has smaller $\dot\hDotFreq$ than the numerical
simulation, at 2.5PN, \pade has larger $\dot\hDotFreq$.  Consequently,
$\Phi_{\rm PN}-\Phi_{\rm NR}$ is negative at 1.5PN order and positive
at 2.5PN order.  For 3.5PN order, the P-approximant in
Fig.~\ref{omegadotnorm} agrees very well with the numerical simulation
(at least for $M\hDotFreq\lesssim 0.1$), which translates into
excellent agreement in Fig.~\ref{fig:dphase-eobandpade}.

\begin{figure}
  %diffPphases_new
  \includegraphics[width=\linewidth]{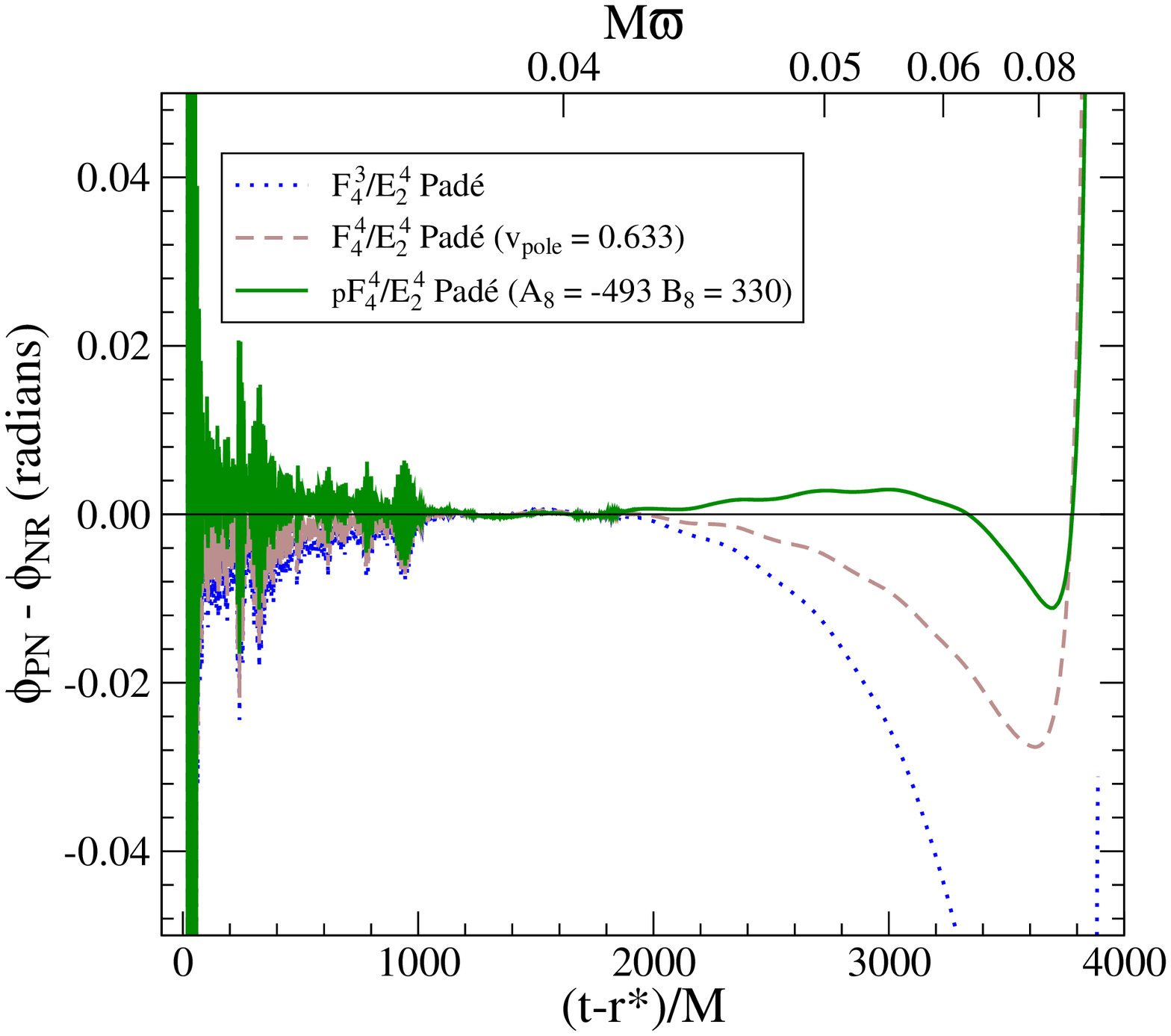}
  \caption{Phase differences between untuned and tuned P-approximants
    and NR.  The untuned P-approximant is ${F}^3_4/E^4_2$ ($v_{\rm
      lso}=v^{\rm 2PN}_{\rm lso}$, $v_{\rm pole}=v^{\rm 2PN}_{\rm
      pole}$). The tuned P-approximants are ${F}^4_4/E^4_2$ and
    tunable $v_{\rm pole}$ ($v_{\rm lso}=v^{\rm 2PN}_{\rm lso}$) and
    ${\small{p}F}^4_4/E^4_2$ ($v_{\rm lso}=v_{\rm lso}^{\rm 2PN}$,
    $v_{\rm pole}=v_{\rm pole}^{\rm 2PN}$) with tunable $A_8$ and
    $B_8$.  In all cases, waveforms are matched over $t-r^{\ast} \in
    [1100,1900]M$.
    \label{fig:dphase-pade-pole-f8}}
\end{figure}

In Fig.~\ref{fig:dphase-pade-pole-f8} we explore the possibility of
reducing the phase differences between the numerical waveform and
P-approximants: By (i) varying $v_{\rm pole}$ or (ii) introducing the
pseudo 4PN order coefficient ${\cal F}_8(\nu=1/4) = A_8 + B_8 \log
v_\OrbitalFreq$ in the energy flux.  We tune the coefficients by
minimizing the sum of the squares of the phase difference at
$t_{0.063}$ and $t_{0.1}$.  We find that if $v_{\rm pole}=0.633$, the
P-approximant $F_4^4/E_2^4$ has a maximum phase difference before $M
\PsiFourFreq=0.1$ smaller than the numerical error in the
simulation. A similar result is obtained for the the P-approximant
$\small{p}F_4^4/E_2^4$ if we use $v_{\rm pole} = v^{\rm 2PN}_{\rm
pole}=0.6907$, and tune $A_8 = -493$, $B_8 = 330$.

\subsection{Effective-one-body waveforms}
\label{sec:EOBWaveforms}

In Fig.~\ref{fig:dphase-eobandpade} we also plot the phase differences
between the numerical and the untuned, original
E-approximants~\cite{BuonannoDamour1999,BuonannoDamour2000,DamourJaranowski2000} $F^m_n/H_p$.  At
3.5PN order the phase difference at $M \PsiFourFreq=0.1$ is $0.50$
radians. We also computed the phase differences at $M
\PsiFourFreq=0.1$ of the E-approximants ${}^{\rm nK} F^3_4/H_7$, ${}^{\rm nK} F_7/H_7$ and
$F_7/H_7$ and found 0.45, 2.56 and $2.7$ radians, respectively.  Thus,
for untuned EOB models it is crucial to have introduced the \pade
flux.  When contrasting the original E-approximants with generic
Taylor approximants, we find that the phase differences are less
scattered as the PN order is increased. However, despite the fact that
the \pade-based EOB flux is closer to the numerical flux (see
Figs.~\ref{CompFomega2} and \ref{CompFomegaNorm}), untuned, original
E-approximants accumulate more phase difference than
P-approximants. This could be a consequence of the fact that
independently of the flux and the energy functions, what seems to
matter is the way the equations of motions are solved to get the
phasing.

\begin{figure}
  %diffEPphases_new
  \includegraphics[width=\linewidth]{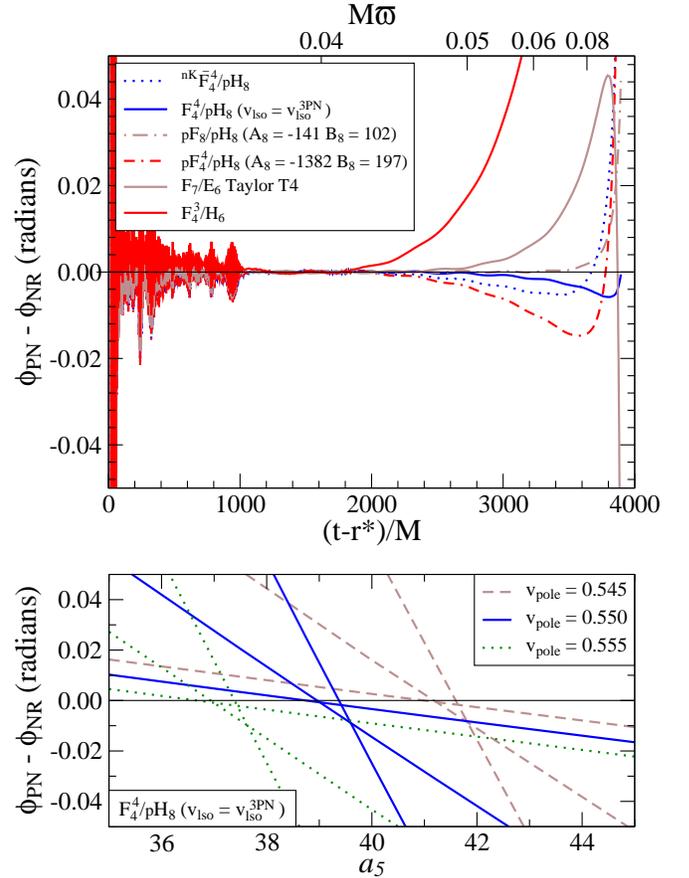}
  \caption{\label{fig:dphase-eob-time} The {\bf upper panel} shows
    phase differences versus time (lower $x$-axis) and versus GW
    frequency (upper $x$-axis) for several tuned and untuned
    E-approximants. For the tuned models, the optimal $a_5$ and
    $v_{\rm pole}$ values displayed in Table~\ref{tab:best_a5_vpole}.
    In the {\bf lower panel} we show phase differences between
    numerical and E-approximants computed at $t_{0.063}$, $t_{0.1}$,
    and the end of the numerical simulation $t_{0.16}$, as functions
    of $a_5$.  For the same color and style, the curve with the
    steepest slope corresponds to $t_{0.16}$ and the curve with the
    smallest slope corresponds to $t_{0.063}$ (For notation see
    Table~\ref{tab:approximants} and caption therein).}
\end{figure}

Because of the reduction of the dynamics to {\it a few} crucial
functions determining the inspiral
evolution~\cite{BuonannoDamour1999,BuonannoDamour2000,Damour2001}, notably $A$, $D$ and
${\cal F}$, and because of the rather simple procedure to match the
inspiral(-plunge) waveform to the ringdown waveform, the EOB model
turned out to be particularly suitable for matching the full numerical
waveforms~\cite{BuonannoCook2007,BuonannoPan2007,DamourNagar2007,
  DamourNagar2008,DamourNagar2008b}. In view of a future study which will include
merger and ringdown, we start here exploring the possibility of
improving the agreement with numerical waveforms by tuning the pseudo
4PN order coefficients $a_5$, $A_8$ and $B_8$ and/or, if present, the
pole location $v_{\rm pole}$.  In the lower panel of
Fig.~\ref{fig:dphase-eob-time}, using different $v_{\rm pole}$ values,
we show the phase differences computed at $t_{0.063}$ and $t_{0.1}$ as
functions of the unknown PN-expansion coefficient $a_5$ [see
Eq.~(\ref{eq:a5})]. As first pointed out and discussed 
in Ref.~\cite{DamourNagar2008} (see e.g., Fig. 3 therein), 
we find that there is a strong degeneracy between $a_5$ 
and $v_{\rm pole}$. In fact, for different $v_{\rm
  pole}$ values, the curves in Fig.~\ref{fig:dphase-eob-time} are
almost identical except for a shift in $a_5$. Although in this test we
use the E-approximant $F^4_4/\small{p}H_8 (v_{\rm lso}=v_{\rm
  lso}^{\rm 2PN})$, we find that this degeneracy appears in all
E-approximants considered.

To obtain the optimal $a_5$ and $v_{\rm pole}$ that minimize phase
differences during the entire numerical simulation, we first choose an
arbitrary $v_{\rm pole}$ in the range of degeneracy. Then, we
determine the $a_5$ value by minimizing the sum of the squares of the
phase difference at $t_{0.063}$ and $t_{0.1}$.  In the upper panel of
Fig.~\ref{fig:dphase-eob-time}, we show phase differences in time and
GW frequency for several E-approximants using those optimal $a_5$ and
$v_{\rm pole}$ values, which are given in
Table~\ref{tab:best_a5_vpole}. In Fig.~\ref{fig:dphase-eob-time}, we
also show phase differences for E-approximants with pseudo 4PN order
coefficients determined by the flux fit of Sec.~\ref{sec:FittingFlux}
(see Fig. \ref{fitflux}) and tunable $a_5$.  The optimal $a_5$ values
are shown in Table~\ref{tab:best_a5_vpole}.  The smaller phase
differences along the entire inspiral are obtained with the
E-approximants with \pade flux ${F}^4_4/\small{p}H_8$ ($v_{\rm lso} =
v_{\rm lso}^{\rm 2PN}$) and tunable $v_{\rm pole}, a_5$ and Taylor
flux $\small{p}F_8/\small{p}H_8$ with tunable $A_8, B_8, a_5$. We
notice that for $t > t_{0.1}$ the phase difference increases more
abruptly for the latter model. In the best case, the absolute phase
difference during the entire numerical simulation is within the
numerical error, i.e., within $0.02$ radians.  The choice of the
best tuned
E-approximant~\cite{PanBuonanno2008,BuonannoPan2007,DamourNagar2008,
DamourNagar2008a,DamourNagar2008b} will be determined once merger and
ringdown are included, and when long and accurate comparisons with
numerical simulations are extended to BBH with mass ratio different
from one.

\begin{table}\renewcommand{\arraystretch}{1.25}
  \begin{tabular}{|c|c|c|c|}
    \hline \multicolumn{2}{|c|}{\rm EOB model and fixed parameters}  & 
           {$a_5$} & {$v_{\rm pole}$} \\\hline
    $^{\rm nK}\bar{F}_4^4/\small{p}H_8$ & --- & 29.78 & 0.52 \\\hline
    $F_4^4/\small{p}H_8$ & $v_{\rm lso}=v_{\rm lso}^{\rm 2PN}$ & 
           39.35 & 0.55 \\\hline
    $\small{p}F_8/\small{p}H_8$ & $A_8=-141, B_8=102$ & 5.32 & N/A \\\hline
$\small{p}F_4^4/\small{p}H_8$ & $A_8=-1382, B_8=197,$
 & -3.10 & N/A \\
& $v_{\rm lso}=v_{\rm lso}^{\rm 2PN},
    v_{\rm pole}=v_{\rm pole}^{\rm 2PN}$ &&\\\hline

\hline
  \end{tabular}
  \caption{\label{tab:best_a5_vpole} Optimal $a_5$ and $v_{\rm pole}$
    that minimize phase differences between tuned EOB models and the
    numerical simulation.}
\end{table}

Finally, in Ref.~\cite{DamourNagar2008}, Damour and Nagar extracted the
data of the numerical simulation used in the present paper from one of
the figures of Ref.~\cite{BoyleBrown2007} and compared those data with the
EOB approach. They found for their ``non-tuned'' EOB model phase
differences $ \pm 0.05$ radians. This phase difference is 
smaller than the phase differences we discuss in this paper 
for untuned EOB models (see Fig.~\ref{fig:dphase-eobandpade} and 
discussion around it). However, we notice that $ \pm 0.05$ radians 
in Ref.~\cite{DamourNagar2008} refers to {\it half} the maximum phase 
difference accumulated over the entire evolution 
when matching the numerical and EOB phases at 
$M \omega = 0.1$. By contrast, in this paper, and 
in particular in Fig.~\ref{fig:dphase-eobandpade}, 
we match numerical and EOB phases in a time interval and compute 
the phase differences at $M \omega = 0.1$.

Moreover, we observe that their ``non-tuned'' EOB model is not 
really untuned, because it uses the \pade summation of the radial 
potential at 4PN order and {\it then} sets $a_5=0$. 
This is not equivalent to using the radial potential at 3.5PN
order with $a_5=0$. In fact, to recover the 3.5PN order \pade radial
potential from the 4PN order \pade potential one should use $a_5
=-17.16$.  They also use the non-Keplerian flux at 4PN order ${}^{\rm 
nK}\overline{F}_4^4$ which is different from the 3.5PN order one ${}^{\rm nK}F_4^3$. 
For our untuned EOB model at 3.5PN order which uses ${}^{\rm nK}F_4^3$ 
and the EOB dynamics at 3PN order, if we apply Ref.~\cite{DamourNagar2008} procedure 
and  compute {\it half} the maximum phase difference when matching 
the numerical and EOB phases at $M \omega = 0.1$, we find a
phase difference of $\pm 0.18 $ radians

\section{Conclusions}
\label{sec:Conclusions}

In this paper, using a highly accurate and long numerical
simulation~\cite{BoyleBrown2007} of a non-spinning equal-mass black hole
binary, we compute the gravitational waveform, GW energy flux, and GW
frequency derivative. Imposing the balance equation, we also estimate
the (derivative of) center-of-mass energy.  We compare these
quantities to those computed using adiabatic TaylorT4 and
\pade~\cite{DamourIyer1998,DamourJaranowski2000, BuonannoChen2003}, and non-adiabatic EOB PN
approximants~\cite{BuonannoDamour1999,BuonannoDamour2000,DamourJaranowski2000}.

We find that for the first 15 GW cycles, the 3.5PN order T-approximant
and the 3.5PN order untuned P- and E-approximants (see
Table~\ref{tab:approximants}) reproduce the numerical results for
energy flux, GW frequency derivative and (derivative of)
center-of-mass energy quite well (see Figs.  ~\ref{CompFomega2},
\ref{CompFomegaNorm}, \ref{omegadotnorm},
\ref{fig:dOmOrbNormSmoothed}, and \ref{dE}), but with interesting
differences.

We attempted to study the convergence of the PN expansion for the
energy flux.\footnote{We also tried to apply the criterion suggested
  in Ref.~\cite{YunesBerti2008} to assess the region of validity of the PN
  series for the flux in the equal-mass case.  Unfortunately, the
  numerical simulation starts at too high a frequency, when the Taylor
  series at 3.5PN order seems to already be outside the region of
  validity.}  We find that \pade approximants to the flux introduced
in Ref.~\cite{DamourIyer1998} do not accelerate the convergence of the
Taylor series, but are closer to the numerical flux than are the
T-approximants.  In particular, the Taylor flux at all orders through
3.5 PN is outside the numerical flux error bars even $\sim 25$ GW
cycles before merger (see Fig.~\ref{CompFomega2}).  We find that the
non-adiabatic non-Keplerian E-approximants to the flux at 3.5PN
order are  within $\sim 2\%$ of the numerical flux over the
entire frequency range we consider (see Fig.~\ref{CompFomegaNorm}).

Quite interestingly, in the equal-mass case the numerical normalized
energy flux $F/F_{\rm Newt}$ starts decreasing at high frequency
during the late part of the inspiral and blurred plunge (see
Fig.~\ref{CompFomegaNorm}).  This differs from the behavior of
$F/F_{\rm Newt}$ in the test-mass limit (see Fig.~\ref{flux-tml}).
Both the Taylor and \pade-based E-approximants with non-Keplerian
flux~\cite{DamourGopakumar2006} show a similar decreasing behavior at
high frequency.  This fact suggests that if a pole is present in the
energy flux of equal-mass binaries, it is located at a larger
frequency than that at which the common apparent horizon forms. As
seen in Sec.~\ref{sec:FittingFlux}, when fitting for $v_{\rm pole}$ we
obtain $v_{\rm pole}(\nu=1/4) = 0.74$, which is to be contrasted with
the test-mass case $v_{\rm pole}(\nu=0) = 1/\sqrt{3}\approx0.58$.
These values of $v_{\rm pole}$ correspond to orbital frequencies
$M\OrbitalFreq=0.405$ and $M\OrbitalFreq=0.192$, respectively.

For the GW frequency derivative $\dot{\hDotFreq}$, we find that at low
frequency the Taylor, \pade and EOB models at 3.5PN order are within
the numerical error (see Fig.~\ref{omegadotnorm}). At high frequency,
as already observed in Ref.~\cite{DamourNagar2008}, only the non-adiabatic
E-approximant has a GW frequency derivative that flattens out, as does
the numerical result. The non-Keplerian E-approximant at 3.5PN order
is closest to the numerical data (see
Fig.~\ref{fig:dOmOrbNormSmoothed}).

When estimating the derivative of center-of-mass energy
$dE/d\hDotFreq$, we expect the numerical result and adiabatic
PN-approximants to be close only in the range of frequencies over
which the balance equation and the adiabatic approximation are valid.
We find that this range of frequencies is $M \hDotFreq \lesssim 0.08$
(see Fig.~\ref{dE}) for the 2.5PN T-approximant and all the 3.5PN
approximants.\footnote{It is not clear whether the failure of the
  adiabatic models is a result of the assumption of adiabaticity, or
  if the accuracy of those models would continue to improve if terms
  at order higher than 3.5PN were known.}  At higher frequency, the
3.5PN order non-adiabatic E-approximants are closer to the numerical
$dE/d\hDotFreq$ than are the adiabatic approximants, and the
non-Keplerian E-approximant is the closest.

Applying a new matching procedure, we compared the numerical waveforms
with TaylorT4, \pade, and EOB waveforms.  We find that the accumulated
phase difference from the numerical solution at $M\PsiFourFreq = 0.1$
is $-0.12$ radians for the untuned 3.5PN
P-approximant~\cite{DamourIyer1998,DamourJaranowski2000, BuonannoChen2003}, $0.50$ radians
for the untuned, original 3.5PN
E-approximant~\cite{BuonannoDamour1999,BuonannoDamour2000,DamourJaranowski2000}, and 0.45 radians
for the untuned non-Keplerian~\cite{DamourGopakumar2006} 3.5PN E-approximant (see
Fig.~\ref{fig:dphase-eobandpade}).  Although those phase differences
are larger than for 3.5PN TaylorT4 ($-0.04$ radians), the phase
differences for the P-approximants are less scattered as a function of
PN order than are the phase differences for generic Taylor
approximants.

The analyses of the flux, GW frequency derivative and (derivative of
the) center-of-mass energy emphasize again the importance of including
non-adiabatic effects during the last stages of
inspiral~\cite{BuonannoDamour2000}. Roughly, we can say that non-adiabatic
effects are no longer negligible starting from a frequency $M
\hDotFreq\,\sim\, 0.08 \mbox{--} 0.12$, as can be seen in
Figs.~\ref{CompFomegaNorm}, \ref{omegadotnorm}, and \ref{dE}. As seen
in these figures, non-adiabatic E-approximants can capture some of the
relevant features of the late time evolution. We expect that by
further improving these models by fitting higher-order PN coefficients
to the numerical data, they will become excellent candidates for
developing an analytic template bank of coalescing
BBHs~\cite{BuonannoCook2007,BuonannoPan2007,DamourNagar2007,
  DamourNagar2008,DamourNagar2008b}.

In this paper we started to explore the possibility of reducing the
phase differences between numerical and E-approximant waveforms by
fitting the unknown parameters $a_5$, ${\cal F}_8$, and $v_{\rm pole}$
(see Fig.~\ref{fig:dphase-eob-time}).  As a first step, for several
E-approximants we searched for a local minimal phase difference by
varying $a_5$, ${\cal F}_8$, and $v_{\rm pole}$.  We found that we
were able to reduce phase differences to below the numerical
uncertainty.  In a future work which will include merger and ringdown,
we plan to determine the {\it region} of the parameter space ($a_5$,
${\cal F}_8$, $v_{\rm pole}$) in which the phase difference is within
the numerical uncertainty of the simulation.

\begin{acknowledgments}
  We thank Emanuele Berti, Lee Lindblom, Etienne Racine,
  Bangalore Sathyaprakash, Saul Teukolsky, and Kip Thorne for informative
  discussions. We also thank Emanuele Berti and Eric Poisson for
  providing us the numerical data of the GW flux in the test-mass
  limit case. We thank Thibault Damour and Alessandro Nagar for 
  clarifications on the ``non-tuned'' EOB model 
  used in Ref.~\cite{DamourNagar2008}.

  A.B. and Y.P. acknowledge support from NSF grant PHY-0603762, and
  A.B. also acknowledges support from the Alfred P Sloan Foundation.
  M.B., L.K., A.M., H.P. and M.S. are supported in part by grants from
  the Sherman Fairchild Foundation to Caltech and Cornell, and from
  the Brinson Foundation to Caltech; by NSF grants PHY-0601459,
  PHY-0652995, DMS-0553302 and NASA grant NNG05GG52G at Caltech; by
  NSF grants PHY-0652952, DMS-0553677, PHY-0652929, and NASA grant
  NNG05GG51G at Cornell.
\end{acknowledgments}

\begin{figure}
  %Comparison_tml1_2panels
  \includegraphics[width=0.9\linewidth]{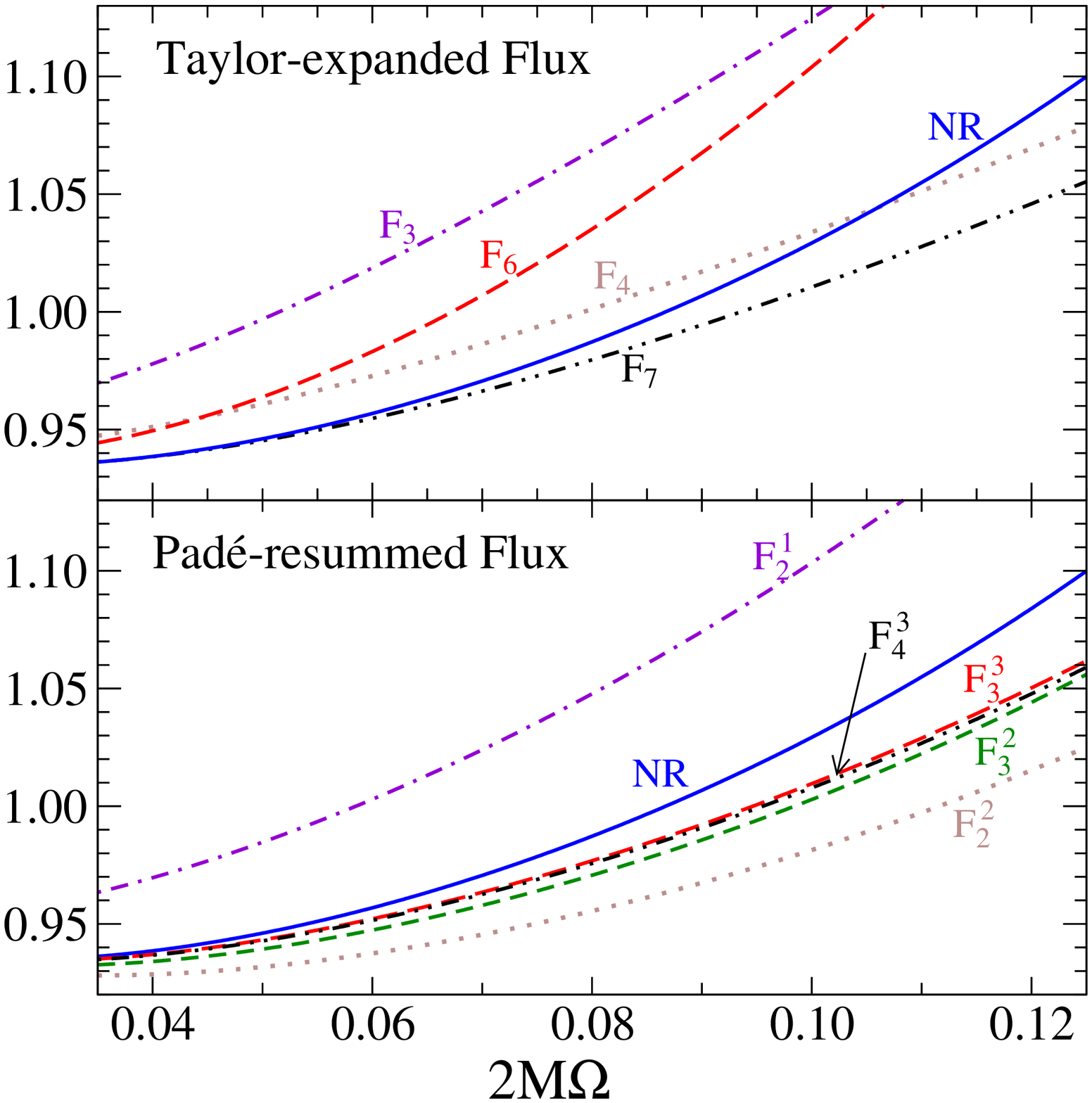}
  %Comparison_tml2_2panels
  \includegraphics[width=0.9\linewidth]{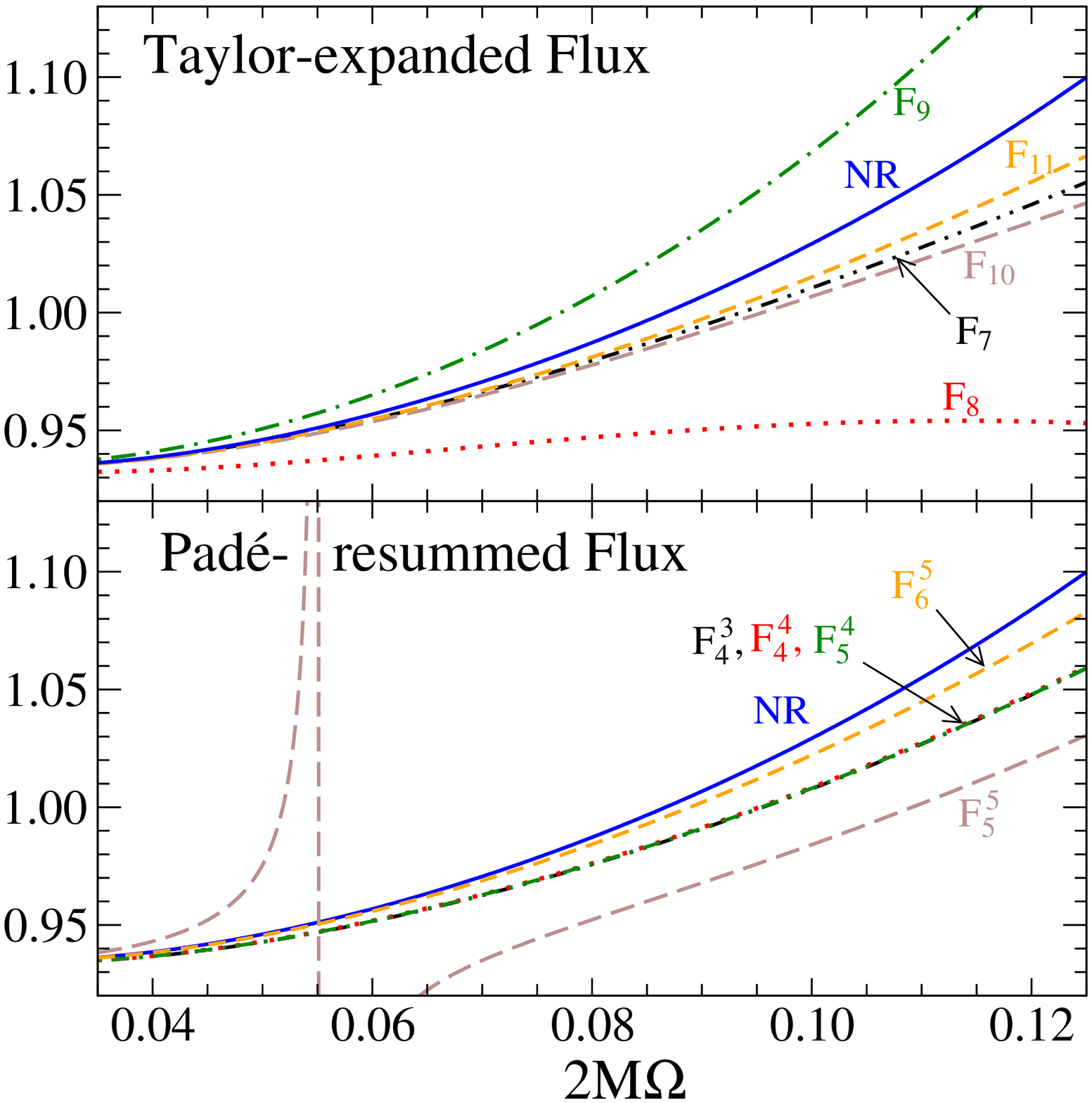}
  \caption{Normalized energy flux $F/F_{\rm Newt}$ versus GW frequency
    $2 \OrbitalFreq$ in the test-mass limit. For notation see
    Table~\ref{tab:approximants} and caption therein. For comparison,
    both panels also include the result of the numerical calculation
    of Poisson~\cite{Poisson1995}, labeled with `NR'.
    \label{flux-tml}}
\end{figure}

\begin{figure}
  %PaperPN-Convergence-Test-Test-Mass
  \includegraphics[width=0.9\linewidth]{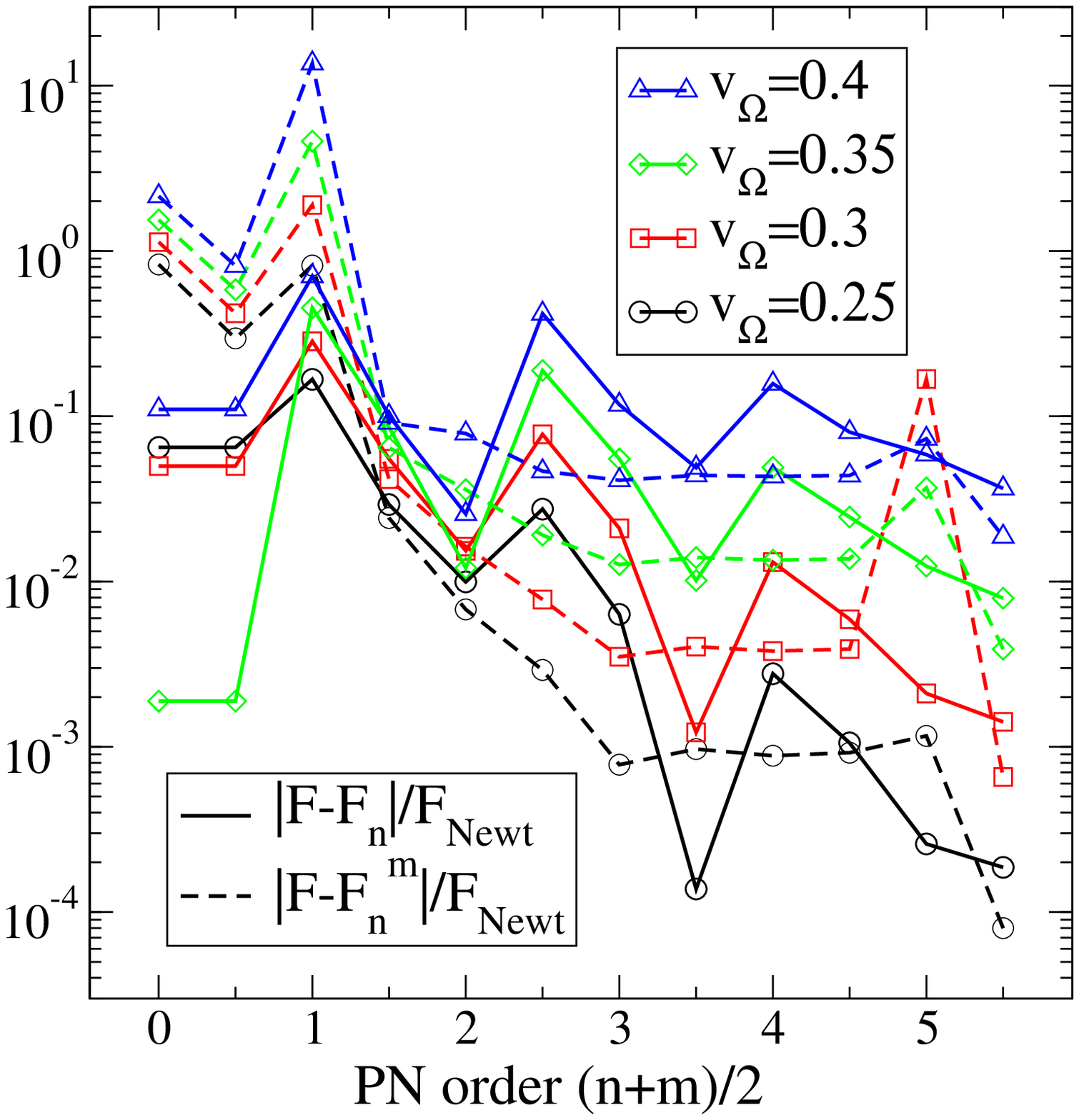}
  \caption{\label{fig:TestMass-FluxConvergence} Convergence of the
    PN-approximants in the test-mass limit. Plotted are differences of
    $F/F_{\rm Newt}$ from the numerical result.  The
    P-approximants do not converge faster than the Taylor series.  }
\end{figure}

\appendix*

\section{Pad\'e approximants to the energy flux in the test particle
  limit}
\label{sec:TestParticleFlux}

In the test-mass-limit case the GW energy flux is known through 5.5PN
order~\cite{TanakaTagoshi1996}.  The explicit coefficients entering
Eq.~(\ref{flux}) for $i \geq 8$ and $\nu =0$ can be read from
Eqs.~(4.1) and (4.2) of Ref.~\cite{DamourIyer1998}.

\begin{figure}
  %TestMassCauchyConvergence
  \includegraphics[width=0.9\linewidth]{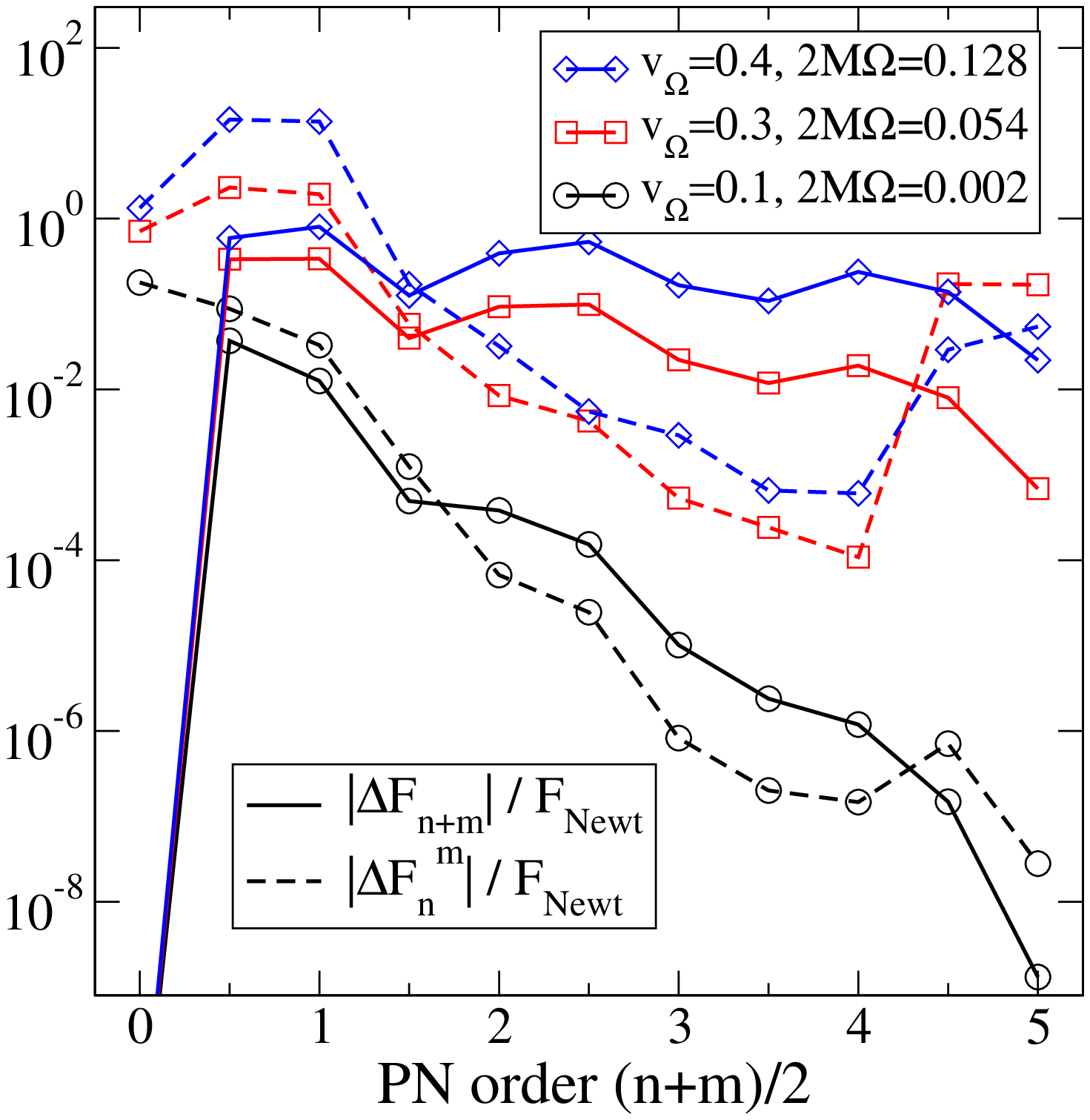}
  \caption{\label{fig:testmassCC}Cauchy convergence test of $F/F_{\rm
      Newt}$ in the test-mass limit for the T- and P-approximants. 
     We plot $\Delta{F}_{n+m} \equiv
    {F}_{n+m+1} - {F}_{n+m}$, and $\Delta{F}_n^m \equiv
    {F}_{n+1}^m - {F}_n^m$  at
    three different frequencies.  At high frequencies, the
    4.5 and 5 PN \pade approximants are contaminated by the extraneous
    pole of the 5PN \pade series; for low frequencies
    ($v_\Omega=0.1$), the pole is apparently irrelevant.}
\end{figure}

\begin{table*}
  \begin{tabular}{c|cc|cc|cc}
PN order & \multicolumn{2}{c|}{$v_\Omega = 0.1;\, 2M\Omega = 0.002$} & \multicolumn{2}{c|}{$v_\Omega= 0.3;\, 2M\Omega = 0.054 $}
& \multicolumn{2}{c}{$v_\Omega=0.4;\, 2M\Omega = 0.128$} \\
    (n+m)/2  & ${F}_{n+m}/F_{\rm Newt}$ & $F_n^m/F_{\rm Newt}$ & ${F}_{n+m}/F_{\rm Newt}$ & $F_n^m/F_{\rm Newt}$ &
    ${F}_{n+m}/F_{\rm Newt}$ & $F_n^m/F_{\rm Newt}$ \\\hline
    0.0&       1.0000000000&             1.20948977  &      1.0000&    2.0817   & 1.000 & 3.255\\
    0.5&       1.0000000000&             1.03092783  &      1.0000&    1.3699  & 1.000 & 1.923\\
    1.0&       {\bf 0.9}628869047&  {\bf 0.9}4287089 &      0.6660&   -0.9467  & 0.406&-12.52\\
    1.5&       {\bf 0.97}54532753&  {\bf 0.97}587569 &      1.0053&    0.9916 & 1.210& {\bf 1}.201\\
    2.0&       {\bf 0.974}9604292&  {\bf 0.974}62770 & {\bf 0.}9653&   0.9337 & 1.084& {\bf 1.0}31\\
    2.5&       {\bf 0.974}5775009&  {\bf 0.974}69475 & {\bf 0.}8723&   0.9422 & 0.692 &{\bf 1.0}63\\
    3.0&       {\bf 0.9747}307757&  {\bf 0.97471}937 & {\bf 0.9}710&   0.9465 & 1.227 &{\bf 1.0}69\\
    3.5&       {\bf 0.9747}206248&  {\bf 0.97471}854 & {\bf 0.9}488&   0.9460 & 1.061  &{\bf 1.0}66\\
    4.0&       {\bf 0.97471}82352&  {\bf 0.97471}874 & {\bf 0.9}369&   0.9462 & 0.952  &{\bf 1.0}67\\
    4.5&       {\bf 0.974719}4262&  {\bf 0.97471}859 & {\bf 0.9}559&   0.9461 & {\bf 1}.190  &{\bf 1.0}66\\
    5.0&       {\bf 0.97471927}76&  {\bf 0.974719}30 & {\bf 0.94}79&   1.1178 &   {\bf 1.0}51  &{\bf 1.0}37\\
    5.5&       {\bf 0.97471927}63&  {\bf 0.974719}28 & {\bf 0.94}85&   {\bf 0.94}93 & {\bf 1.0}73   &{\bf 1.0}91
  \end{tabular}
  \caption{\label{tab:testmass} Normalized energy flux $F/F_{\rm
      Newt}$ in the test-mass limit for the T- and P-approximants at
    different PN orders and at three different frequencies.  We use
    boldface to indicate the range of significant figures that do not
    change with increasing PN order.}
\end{table*}

In Fig.~\ref{flux-tml} we compare the normalized energy flux
function~\cite{Poisson1995} $F/F_{\rm Newt}$ to the T- and
P-approximants.  To easily compare Fig.~\ref{flux-tml} with the other
figures in the paper, we plot quantities as functions of the
approximate GW frequency defined by $2 M\OrbitalFreq$.  As noticed in
Ref.~\cite{DamourIyer1998}, the P-approximants approach the numerical data
more systematically.  The differences between different PN orders are
difficult to see in Fig.~\ref{flux-tml}.  To obtain a clearer view,
Fig.~\ref{fig:TestMass-FluxConvergence} plots the differences between
PN flux and numerical flux at four fixed frequencies.
Fig.~\ref{fig:TestMass-FluxConvergence} shows this somewhat better
behavior of \pade; however, the \pade-approximants show little
improvement between PN orders 3.5 and 4.5, and at order 5 there occurs
an extraneous pole.  At frequency $2M\OrbitalFreq=0.04$,
P-approximants with order $\ge 2.5$ are within 0.5 percent of the
numerical data, as are T-approximants with order $\ge 3.5$.  Good
agreement at low frequency is rather important because that is where
the majority of the waveform phasing accumulates.

Table~\ref{tab:testmass} and Fig.~\ref{fig:testmassCC} test the internal
convergence of T- and P-approximants {\em without} referring to a
numerical result.  Table~\ref{tab:testmass} displays the flux at all
known PN-orders at select frequencies, with boldface highlighting the
digits that have already converged.  Although the \pade summation does
not accelerate the convergence, the P-approximant at 5.5PN order is
closest to the numerical data (see Fig.~\ref{fig:TestMass-FluxConvergence}).

Comparing Table~\ref{tab:testmass} with Table~\ref{tab:equalmass}, and
Fig.~\ref{fig:testmassCC} with Fig.~\ref{fig:CauchyConvergence} we
observe that the P-approximants converge more systematically in the
equal-mass case than in the test-mass limit. This is also evident by
comparing Fig.~\ref{fig:TestMass-FluxConvergence} with
Fig.~\ref{CompFomega2}: We see that P-approximants at 3PN and 3.5PN
orders are inside the numerical flux error whereas T-approximants at
all orders through 3.5 PN are outside the numerical flux error bars
even $\sim 25$ GW cycles before merger.  However, as the \pade
approximant does not converge faster, it is not immediately clear
whether similar superior behavior of \pade can be expected for more
generic binary black holes.

\bibliography{References}

\begin{thebibliography}{53}
\expandafter\ifx\csname natexlab\endcsname\relax\def\natexlab#1{#1}\fi
\expandafter\ifx\csname bibnamefont\endcsname\relax
  \def\bibnamefont#1{#1}\fi
\expandafter\ifx\csname bibfnamefont\endcsname\relax
  \def\bibfnamefont#1{#1}\fi
\expandafter\ifx\csname citenamefont\endcsname\relax
  \def\citenamefont#1{#1}\fi
\expandafter\ifx\csname url\endcsname\relax
  \def\url#1{\texttt{#1}}\fi
\expandafter\ifx\csname urlprefix\endcsname\relax\def\urlprefix{URL }\fi
\providecommand{\bibinfo}[2]{#2}
\providecommand{\eprint}[2][]{\url{#2}}

\bibitem[{\citenamefont{Barish and Weiss}(1999)}]{BarishWeiss1999}
\bibinfo{author}{\bibfnamefont{B.~C.} \bibnamefont{Barish}} \bibnamefont{and}
  \bibinfo{author}{\bibfnamefont{R.}~\bibnamefont{Weiss}},
  \bibinfo{journal}{Phys. Today} \textbf{\bibinfo{volume}{52}},
  \bibinfo{pages}{44} (\bibinfo{year}{1999}).

\bibitem[{\citenamefont{Waldman}(2006)}]{Waldman2006}
\bibinfo{author}{\bibfnamefont{S.~J.} \bibnamefont{Waldman}},
  \bibinfo{journal}{Class. Quant. Grav.} \textbf{\bibinfo{volume}{23}},
  \bibinfo{pages}{S653} (\bibinfo{year}{2006}).

\bibitem[{\citenamefont{Hild}(2006)}]{Hild2006}
\bibinfo{author}{\bibfnamefont{S.}~\bibnamefont{Hild}},
  \bibinfo{journal}{Class. Quant. Grav.} \textbf{\bibinfo{volume}{23}},
  \bibinfo{pages}{S643} (\bibinfo{year}{2006}).

\bibitem[{\citenamefont{Acernese et~al.}(2002)\citenamefont{Acernese, Amico,
  Arnaud, Arnault, Babusci, Ballardin, Barone, Barsuglia, Bellachia, Beney
  et~al.}}]{AcerneseAmico2002}
\bibinfo{author}{\bibfnamefont{F.}~\bibnamefont{Acernese}},
  \bibinfo{author}{\bibfnamefont{P.}~\bibnamefont{Amico}},
  \bibinfo{author}{\bibfnamefont{N.}~\bibnamefont{Arnaud}},
  \bibinfo{author}{\bibfnamefont{C.}~\bibnamefont{Arnault}},
  \bibinfo{author}{\bibfnamefont{D.}~\bibnamefont{Babusci}},
  \bibinfo{author}{\bibfnamefont{G.}~\bibnamefont{Ballardin}},
  \bibinfo{author}{\bibfnamefont{F.}~\bibnamefont{Barone}},
  \bibinfo{author}{\bibfnamefont{M.}~\bibnamefont{Barsuglia}},
  \bibinfo{author}{\bibfnamefont{F.}~\bibnamefont{Bellachia}},
  \bibinfo{author}{\bibfnamefont{J.~L.} \bibnamefont{Beney}},
  \bibnamefont{et~al.}, \bibinfo{journal}{Class. Quant. Grav.}
  \textbf{\bibinfo{volume}{19}}, \bibinfo{pages}{1421} (\bibinfo{year}{2002}).

\bibitem[{\citenamefont{Acernese et~al.}(2006)\citenamefont{Acernese, Amico,
  Alshourbagy, Antonucci, Aoudia, Avino, Babusci, Ballardin, Barone, Barsotti
  et~al.}}]{AcerneseAmico2006}
\bibinfo{author}{\bibfnamefont{F.}~\bibnamefont{Acernese}},
  \bibinfo{author}{\bibfnamefont{P.}~\bibnamefont{Amico}},
  \bibinfo{author}{\bibfnamefont{M.}~\bibnamefont{Alshourbagy}},
  \bibinfo{author}{\bibfnamefont{F.}~\bibnamefont{Antonucci}},
  \bibinfo{author}{\bibfnamefont{S.}~\bibnamefont{Aoudia}},
  \bibinfo{author}{\bibfnamefont{S.}~\bibnamefont{Avino}},
  \bibinfo{author}{\bibfnamefont{D.}~\bibnamefont{Babusci}},
  \bibinfo{author}{\bibfnamefont{G.}~\bibnamefont{Ballardin}},
  \bibinfo{author}{\bibfnamefont{F.}~\bibnamefont{Barone}},
  \bibinfo{author}{\bibfnamefont{L.}~\bibnamefont{Barsotti}},
  \bibnamefont{et~al.}, \bibinfo{journal}{Class. Quant. Grav.}
  \textbf{\bibinfo{volume}{23}}, \bibinfo{pages}{S635} (\bibinfo{year}{2006}).

\bibitem[{\citenamefont{Flanagan and Hughes}(1998)}]{FlanaganHughes1998}
\bibinfo{author}{\bibfnamefont{E.~E.} \bibnamefont{Flanagan}} \bibnamefont{and}
  \bibinfo{author}{\bibfnamefont{S.~A.} \bibnamefont{Hughes}},
  \bibinfo{journal}{Phys. Rev. D} \textbf{\bibinfo{volume}{57}},
  \bibinfo{pages}{4535} (\bibinfo{year}{1998}).

\bibitem[{\citenamefont{Damour et~al.}(2000{\natexlab{a}})\citenamefont{Damour,
  Iyer, and Sathyaprakash}}]{DamourIyer2000}
\bibinfo{author}{\bibfnamefont{T.}~\bibnamefont{Damour}},
  \bibinfo{author}{\bibfnamefont{B.~R.} \bibnamefont{Iyer}}, \bibnamefont{and}
  \bibinfo{author}{\bibfnamefont{B.~S.} \bibnamefont{Sathyaprakash}},
  \bibinfo{journal}{Phys. Rev. D} \textbf{\bibinfo{volume}{62}},
  \bibinfo{pages}{084036} (\bibinfo{year}{2000}{\natexlab{a}}).

\bibitem[{\citenamefont{Blanchet}(2006)}]{Blanchet2006}
\bibinfo{author}{\bibfnamefont{L.}~\bibnamefont{Blanchet}},
  \bibinfo{journal}{Living Rev.~Rel.} \textbf{\bibinfo{volume}{9}},
  \bibinfo{pages}{4} (\bibinfo{year}{2006}).

\bibitem[{\citenamefont{Buonanno
  et~al.}(2007{\natexlab{a}})\citenamefont{Buonanno, Cook, and
  Pretorius}}]{BuonannoCook2007}
\bibinfo{author}{\bibfnamefont{A.}~\bibnamefont{Buonanno}},
  \bibinfo{author}{\bibfnamefont{G.~B.} \bibnamefont{Cook}}, \bibnamefont{and}
  \bibinfo{author}{\bibfnamefont{F.}~\bibnamefont{Pretorius}},
  \bibinfo{journal}{Phys. Rev. D} \textbf{\bibinfo{volume}{75}},
  \bibinfo{pages}{124018} (\bibinfo{year}{2007}{\natexlab{a}}).

\bibitem[{\citenamefont{Baker et~al.}(2007{\natexlab{a}})\citenamefont{Baker,
  van Meter, McWilliams, Centrella, and Kelly}}]{BakerVanMeter2007}
\bibinfo{author}{\bibfnamefont{J.~G.} \bibnamefont{Baker}},
  \bibinfo{author}{\bibfnamefont{J.~R.} \bibnamefont{van Meter}},
  \bibinfo{author}{\bibfnamefont{S.~T.} \bibnamefont{McWilliams}},
  \bibinfo{author}{\bibfnamefont{J.}~\bibnamefont{Centrella}},
  \bibnamefont{and} \bibinfo{author}{\bibfnamefont{B.~J.} \bibnamefont{Kelly}},
  \bibinfo{journal}{Phys. Rev. Lett.} \textbf{\bibinfo{volume}{99}},
  \bibinfo{pages}{181101} (\bibinfo{year}{2007}{\natexlab{a}}).

\bibitem[{\citenamefont{Hannam et~al.}(2008)\citenamefont{Hannam, Husa,
  Gonz{\'a}lez, Sperhake, and Br{\"u}gmann}}]{HannamHusa2008}
\bibinfo{author}{\bibfnamefont{M.}~\bibnamefont{Hannam}},
  \bibinfo{author}{\bibfnamefont{S.}~\bibnamefont{Husa}},
  \bibinfo{author}{\bibfnamefont{J.~A.} \bibnamefont{Gonz{\'a}lez}},
  \bibinfo{author}{\bibfnamefont{U.}~\bibnamefont{Sperhake}}, \bibnamefont{and}
  \bibinfo{author}{\bibfnamefont{B.}~\bibnamefont{Br{\"u}gmann}},
  \bibinfo{journal}{Phys. Rev. D} \textbf{\bibinfo{volume}{77}},
  \bibinfo{pages}{044020} (\bibinfo{year}{2008}).

\bibitem[{\citenamefont{Boyle et~al.}(2007)\citenamefont{Boyle, Brown, Kidder,
  Mrou{\'e}, Pfeiffer, Scheel, Cook, and Teukolsky}}]{BoyleBrown2007}
\bibinfo{author}{\bibfnamefont{M.}~\bibnamefont{Boyle}},
  \bibinfo{author}{\bibfnamefont{D.~A.} \bibnamefont{Brown}},
  \bibinfo{author}{\bibfnamefont{L.~E.} \bibnamefont{Kidder}},
  \bibinfo{author}{\bibfnamefont{A.~H.} \bibnamefont{Mrou{\'e}}},
  \bibinfo{author}{\bibfnamefont{H.~P.} \bibnamefont{Pfeiffer}},
  \bibinfo{author}{\bibfnamefont{M.~A.} \bibnamefont{Scheel}},
  \bibinfo{author}{\bibfnamefont{G.~B.} \bibnamefont{Cook}}, \bibnamefont{and}
  \bibinfo{author}{\bibfnamefont{S.~A.} \bibnamefont{Teukolsky}},
  \bibinfo{journal}{Phys. Rev. D} \textbf{\bibinfo{volume}{76}},
  \bibinfo{pages}{124038} (\bibinfo{year}{2007}).

\bibitem[{\citenamefont{Gopakumar et~al.}()\citenamefont{Gopakumar, Hannam,
  Husa, and Br{\"u}gmann}}]{GopakumarHannam0712.3737}
\bibinfo{author}{\bibfnamefont{A.}~\bibnamefont{Gopakumar}},
  \bibinfo{author}{\bibfnamefont{M.}~\bibnamefont{Hannam}},
  \bibinfo{author}{\bibfnamefont{S.}~\bibnamefont{Husa}}, \bibnamefont{and}
  \bibinfo{author}{\bibfnamefont{B.}~\bibnamefont{Br{\"u}gmann}},
  \eprint{arXiv:0712.3737}.

\bibitem[{\citenamefont{Hannam et~al.}()\citenamefont{Hannam, Husa,
  Br{\"u}gmann, and Gopakumar}}]{HannamHusa0712.3787}
\bibinfo{author}{\bibfnamefont{M.}~\bibnamefont{Hannam}},
  \bibinfo{author}{\bibfnamefont{S.}~\bibnamefont{Husa}},
  \bibinfo{author}{\bibfnamefont{B.}~\bibnamefont{Br{\"u}gmann}},
  \bibnamefont{and}
  \bibinfo{author}{\bibfnamefont{A.}~\bibnamefont{Gopakumar}},
  \eprint{arXiv:0712.3787}.

\bibitem[{\citenamefont{Pan et~al.}(2008)\citenamefont{Pan, Buonanno, Baker,
  Centrella, Kelly, McWilliams, Pretorius, and van Meter}}]{PanBuonanno2008}
\bibinfo{author}{\bibfnamefont{Y.}~\bibnamefont{Pan}},
  \bibinfo{author}{\bibfnamefont{A.}~\bibnamefont{Buonanno}},
  \bibinfo{author}{\bibfnamefont{J.~G.} \bibnamefont{Baker}},
  \bibinfo{author}{\bibfnamefont{J.}~\bibnamefont{Centrella}},
  \bibinfo{author}{\bibfnamefont{B.~J.} \bibnamefont{Kelly}},
  \bibinfo{author}{\bibfnamefont{S.~T.} \bibnamefont{McWilliams}},
  \bibinfo{author}{\bibfnamefont{F.}~\bibnamefont{Pretorius}},
  \bibnamefont{and} \bibinfo{author}{\bibfnamefont{J.~R.} \bibnamefont{van
  Meter}}, \bibinfo{journal}{Phys. Rev. D} \textbf{\bibinfo{volume}{77}},
  \bibinfo{pages}{024014} (\bibinfo{year}{2008}).

\bibitem[{\citenamefont{Buonanno
  et~al.}(2007{\natexlab{b}})\citenamefont{Buonanno, Pan, Baker, Centrella,
  Kelly, McWilliams, and van Meter}}]{BuonannoPan2007}
\bibinfo{author}{\bibfnamefont{A.}~\bibnamefont{Buonanno}},
  \bibinfo{author}{\bibfnamefont{Y.}~\bibnamefont{Pan}},
  \bibinfo{author}{\bibfnamefont{J.~G.} \bibnamefont{Baker}},
  \bibinfo{author}{\bibfnamefont{J.}~\bibnamefont{Centrella}},
  \bibinfo{author}{\bibfnamefont{B.~J.} \bibnamefont{Kelly}},
  \bibinfo{author}{\bibfnamefont{S.~T.} \bibnamefont{McWilliams}},
  \bibnamefont{and} \bibinfo{author}{\bibfnamefont{J.~R.} \bibnamefont{van
  Meter}}, \bibinfo{journal}{Phys. Rev. D} \textbf{\bibinfo{volume}{76}},
  \bibinfo{pages}{104049} (\bibinfo{year}{2007}{\natexlab{b}}).

\bibitem[{\citenamefont{Ajith et~al.}(2008)\citenamefont{Ajith, Babak, Chen,
  Hewitson, Krishnan, Sintes, Whelan, Br\"{u}gmann, Diener, Dorband
  et~al.}}]{AjithBabak2008}
\bibinfo{author}{\bibfnamefont{P.}~\bibnamefont{Ajith}},
  \bibinfo{author}{\bibfnamefont{S.}~\bibnamefont{Babak}},
  \bibinfo{author}{\bibfnamefont{Y.}~\bibnamefont{Chen}},
  \bibinfo{author}{\bibfnamefont{M.}~\bibnamefont{Hewitson}},
  \bibinfo{author}{\bibfnamefont{B.}~\bibnamefont{Krishnan}},
  \bibinfo{author}{\bibfnamefont{A.~M.} \bibnamefont{Sintes}},
  \bibinfo{author}{\bibfnamefont{J.~T.} \bibnamefont{Whelan}},
  \bibinfo{author}{\bibfnamefont{B.}~\bibnamefont{Br\"{u}gmann}},
  \bibinfo{author}{\bibfnamefont{P.}~\bibnamefont{Diener}},
  \bibinfo{author}{\bibfnamefont{N.}~\bibnamefont{Dorband}},
  \bibnamefont{et~al.}, \bibinfo{journal}{Phys. Rev. D}
  \textbf{\bibinfo{volume}{77}}, \bibinfo{pages}{104017}
  (\bibinfo{year}{2008}).

\bibitem[{\citenamefont{Damour and Nagar}(2008)}]{DamourNagar2008}
\bibinfo{author}{\bibfnamefont{T.}~\bibnamefont{Damour}} \bibnamefont{and}
  \bibinfo{author}{\bibfnamefont{A.}~\bibnamefont{Nagar}},
  \bibinfo{journal}{Phys. Rev. D} \textbf{\bibinfo{volume}{77}},
  \bibinfo{pages}{024043} (\bibinfo{year}{2008}).

\bibitem[{\citenamefont{Damour et~al.}(2008)\citenamefont{Damour, Nagar,
  Dorband, Pollney, and Rezzolla}}]{DamourNagar2008a}
\bibinfo{author}{\bibfnamefont{T.}~\bibnamefont{Damour}},
  \bibinfo{author}{\bibfnamefont{A.}~\bibnamefont{Nagar}},
  \bibinfo{author}{\bibfnamefont{E.~N.} \bibnamefont{Dorband}},
  \bibinfo{author}{\bibfnamefont{D.}~\bibnamefont{Pollney}}, \bibnamefont{and}
  \bibinfo{author}{\bibfnamefont{L.}~\bibnamefont{Rezzolla}},
  \bibinfo{journal}{Phys. Rev. D} \textbf{\bibinfo{volume}{77}},
  \bibinfo{pages}{084017} (\bibinfo{year}{2008}).

\bibitem[{\citenamefont{{Damour} et~al.}(2008)\citenamefont{{Damour}, {Nagar},
  {Hannam}, {Husa}, and {Br{\"u}gmann}}}]{DamourNagar2008b}
\bibinfo{author}{\bibfnamefont{T.}~\bibnamefont{{Damour}}},
  \bibinfo{author}{\bibfnamefont{A.}~\bibnamefont{{Nagar}}},
  \bibinfo{author}{\bibfnamefont{M.}~\bibnamefont{{Hannam}}},
  \bibinfo{author}{\bibfnamefont{S.}~\bibnamefont{{Husa}}}, \bibnamefont{and}
  \bibinfo{author}{\bibfnamefont{B.}~\bibnamefont{{Br{\"u}gmann}}},
  \bibinfo{journal}{Phys. Rev. D} \textbf{\bibinfo{volume}{78}},
  \bibinfo{pages}{044039} (\bibinfo{year}{2008}).

\bibitem[{\citenamefont{Damour et~al.}(1998)\citenamefont{Damour, Iyer, and
  Sathyaprakash}}]{DamourIyer1998}
\bibinfo{author}{\bibfnamefont{T.}~\bibnamefont{Damour}},
  \bibinfo{author}{\bibfnamefont{B.~R.} \bibnamefont{Iyer}}, \bibnamefont{and}
  \bibinfo{author}{\bibfnamefont{B.~S.} \bibnamefont{Sathyaprakash}},
  \bibinfo{journal}{Phys. Rev. D} \textbf{\bibinfo{volume}{57}},
  \bibinfo{pages}{885} (\bibinfo{year}{1998}).

\bibitem[{\citenamefont{Buonanno and Damour}(1999)}]{BuonannoDamour1999}
\bibinfo{author}{\bibfnamefont{A.}~\bibnamefont{Buonanno}} \bibnamefont{and}
  \bibinfo{author}{\bibfnamefont{T.}~\bibnamefont{Damour}},
  \bibinfo{journal}{Phys. Rev. D} \textbf{\bibinfo{volume}{59}},
  \bibinfo{pages}{084006} (\bibinfo{year}{1999}).

\bibitem[{\citenamefont{Buonanno and Damour}(2000)}]{BuonannoDamour2000}
\bibinfo{author}{\bibfnamefont{A.}~\bibnamefont{Buonanno}} \bibnamefont{and}
  \bibinfo{author}{\bibfnamefont{T.}~\bibnamefont{Damour}},
  \bibinfo{journal}{Phys. Rev. D} \textbf{\bibinfo{volume}{62}},
  \bibinfo{pages}{064015} (\bibinfo{year}{2000}).

\bibitem[{\citenamefont{Damour et~al.}(2000{\natexlab{b}})\citenamefont{Damour,
  Jaranowski, and Sch\"afer}}]{DamourJaranowski2000}
\bibinfo{author}{\bibfnamefont{T.}~\bibnamefont{Damour}},
  \bibinfo{author}{\bibfnamefont{P.}~\bibnamefont{Jaranowski}},
  \bibnamefont{and}
  \bibinfo{author}{\bibfnamefont{G.}~\bibnamefont{Sch\"afer}},
  \bibinfo{journal}{Phys. Rev. D} \textbf{\bibinfo{volume}{62}},
  \bibinfo{pages}{084011} (\bibinfo{year}{2000}{\natexlab{b}}).

\bibitem[{\citenamefont{Damour}(2001)}]{Damour2001}
\bibinfo{author}{\bibfnamefont{T.}~\bibnamefont{Damour}},
  \bibinfo{journal}{Phys. Rev. D} \textbf{\bibinfo{volume}{64}},
  \bibinfo{pages}{124013} (\bibinfo{year}{2001}).

\bibitem[{\citenamefont{Damour et~al.}(2003)\citenamefont{Damour, Iyer,
  Jaranowski, and Sathyaprakash}}]{DamourIyer2003}
\bibinfo{author}{\bibfnamefont{T.}~\bibnamefont{Damour}},
  \bibinfo{author}{\bibfnamefont{B.~R.} \bibnamefont{Iyer}},
  \bibinfo{author}{\bibfnamefont{P.}~\bibnamefont{Jaranowski}},
  \bibnamefont{and} \bibinfo{author}{\bibfnamefont{B.~S.}
  \bibnamefont{Sathyaprakash}}, \bibinfo{journal}{Phys. Rev. D}
  \textbf{\bibinfo{volume}{67}}, \bibinfo{pages}{064028}
  (\bibinfo{year}{2003}).

\bibitem[{\citenamefont{Damour and Nagar}(2007)}]{DamourNagar2007}
\bibinfo{author}{\bibfnamefont{T.}~\bibnamefont{Damour}} \bibnamefont{and}
  \bibinfo{author}{\bibfnamefont{A.}~\bibnamefont{Nagar}},
  \bibinfo{journal}{Phys. Rev. D} \textbf{\bibinfo{volume}{76}},
  \bibinfo{pages}{064028} (\bibinfo{year}{2007}).

\bibitem[{\citenamefont{Pfeiffer et~al.}(2007)\citenamefont{Pfeiffer, Brown,
  Kidder, Lindblom, Lovelace, and Scheel}}]{PfeifferBrown2007}
\bibinfo{author}{\bibfnamefont{H.~P.} \bibnamefont{Pfeiffer}},
  \bibinfo{author}{\bibfnamefont{D.~A.} \bibnamefont{Brown}},
  \bibinfo{author}{\bibfnamefont{L.~E.} \bibnamefont{Kidder}},
  \bibinfo{author}{\bibfnamefont{L.}~\bibnamefont{Lindblom}},
  \bibinfo{author}{\bibfnamefont{G.}~\bibnamefont{Lovelace}}, \bibnamefont{and}
  \bibinfo{author}{\bibfnamefont{M.~A.} \bibnamefont{Scheel}},
  \bibinfo{journal}{Class. Quant. Grav.} \textbf{\bibinfo{volume}{24}},
  \bibinfo{pages}{S59} (\bibinfo{year}{2007}).

\bibitem[{\citenamefont{Berti et~al.}(2007)\citenamefont{Berti, Cardoso,
  Gonzalez, Sperhake, Hannam, Husa, and Br{\"u}gmann}}]{BertiCardoso2007}
\bibinfo{author}{\bibfnamefont{E.}~\bibnamefont{Berti}},
  \bibinfo{author}{\bibfnamefont{V.}~\bibnamefont{Cardoso}},
  \bibinfo{author}{\bibfnamefont{J.~A.} \bibnamefont{Gonzalez}},
  \bibinfo{author}{\bibfnamefont{U.}~\bibnamefont{Sperhake}},
  \bibinfo{author}{\bibfnamefont{M.}~\bibnamefont{Hannam}},
  \bibinfo{author}{\bibfnamefont{S.}~\bibnamefont{Husa}}, \bibnamefont{and}
  \bibinfo{author}{\bibfnamefont{B.}~\bibnamefont{Br{\"u}gmann}},
  \bibinfo{journal}{Phys. Rev. D} \textbf{\bibinfo{volume}{76}},
  \bibinfo{pages}{064034} (\bibinfo{year}{2007}).

\bibitem[{\citenamefont{Pollney et~al.}(2007)\citenamefont{Pollney, Reisswig,
  Rezzolla, Szil{\'a}gyi, Ansorg, Deris, Diener, Dorband, Koppitz, Nagar
  et~al.}}]{PollneyReisswig2007}
\bibinfo{author}{\bibfnamefont{D.}~\bibnamefont{Pollney}},
  \bibinfo{author}{\bibfnamefont{C.}~\bibnamefont{Reisswig}},
  \bibinfo{author}{\bibfnamefont{L.}~\bibnamefont{Rezzolla}},
  \bibinfo{author}{\bibfnamefont{B.}~\bibnamefont{Szil{\'a}gyi}},
  \bibinfo{author}{\bibfnamefont{M.}~\bibnamefont{Ansorg}},
  \bibinfo{author}{\bibfnamefont{B.}~\bibnamefont{Deris}},
  \bibinfo{author}{\bibfnamefont{P.}~\bibnamefont{Diener}},
  \bibinfo{author}{\bibfnamefont{E.~N.} \bibnamefont{Dorband}},
  \bibinfo{author}{\bibfnamefont{M.}~\bibnamefont{Koppitz}},
  \bibinfo{author}{\bibfnamefont{A.}~\bibnamefont{Nagar}},
  \bibnamefont{et~al.}, \bibinfo{journal}{Phys. Rev. D}
  \textbf{\bibinfo{volume}{76}}, \bibinfo{pages}{124002}
  (\bibinfo{year}{2007}).

\bibitem[{\citenamefont{Schnittman et~al.}(2008)\citenamefont{Schnittman,
  Buonanno, van Meter, Baker, Boggs, Centrella, Kelly, , and
  McWilliams}}]{SchnittmanBuonanno2008}
\bibinfo{author}{\bibfnamefont{J.~D.} \bibnamefont{Schnittman}},
  \bibinfo{author}{\bibfnamefont{A.}~\bibnamefont{Buonanno}},
  \bibinfo{author}{\bibfnamefont{J.~R.} \bibnamefont{van Meter}},
  \bibinfo{author}{\bibfnamefont{J.~G.} \bibnamefont{Baker}},
  \bibinfo{author}{\bibfnamefont{W.~D.} \bibnamefont{Boggs}},
  \bibinfo{author}{\bibfnamefont{J.}~\bibnamefont{Centrella}},
  \bibinfo{author}{\bibfnamefont{B.~J.} \bibnamefont{Kelly}}, ,
  \bibnamefont{and} \bibinfo{author}{\bibfnamefont{S.~T.}
  \bibnamefont{McWilliams}}, \bibinfo{journal}{Phys. Rev. D}
  \textbf{\bibinfo{volume}{77}}, \bibinfo{pages}{044031}
  (\bibinfo{year}{2008}).

\bibitem[{\citenamefont{Buonanno et~al.}(2003)\citenamefont{Buonanno, Chen, and
  Vallisneri}}]{BuonannoChen2003}
\bibinfo{author}{\bibfnamefont{A.}~\bibnamefont{Buonanno}},
  \bibinfo{author}{\bibfnamefont{Y.}~\bibnamefont{Chen}}, \bibnamefont{and}
  \bibinfo{author}{\bibfnamefont{M.}~\bibnamefont{Vallisneri}},
  \bibinfo{journal}{Phys. Rev. D} \textbf{\bibinfo{volume}{67}},
  \bibinfo{pages}{024016} (\bibinfo{year}{2003}), \bibinfo{note}{{\bf 74},
  029903(E) (2006)}.

\bibitem[{\citenamefont{Buonanno et~al.}(2006)\citenamefont{Buonanno, Chen, and
  Damour}}]{BuonannoChen2006}
\bibinfo{author}{\bibfnamefont{A.}~\bibnamefont{Buonanno}},
  \bibinfo{author}{\bibfnamefont{Y.}~\bibnamefont{Chen}}, \bibnamefont{and}
  \bibinfo{author}{\bibfnamefont{T.}~\bibnamefont{Damour}},
  \bibinfo{journal}{Phys. Rev. D} \textbf{\bibinfo{volume}{74}},
  \bibinfo{pages}{104005} (\bibinfo{year}{2006}).

\bibitem[{\citenamefont{Damour and Gopakumar}(2006)}]{DamourGopakumar2006}
\bibinfo{author}{\bibfnamefont{T.}~\bibnamefont{Damour}} \bibnamefont{and}
  \bibinfo{author}{\bibfnamefont{A.}~\bibnamefont{Gopakumar}},
  \bibinfo{journal}{Phys. Rev. D} \textbf{\bibinfo{volume}{73}},
  \bibinfo{pages}{124006} (\bibinfo{year}{2006}).

\bibitem[{\citenamefont{Jaranowski and
  Sch\"afer}(1998)}]{JaranowskiSchafer1998}
\bibinfo{author}{\bibfnamefont{P.}~\bibnamefont{Jaranowski}} \bibnamefont{and}
  \bibinfo{author}{\bibfnamefont{G.}~\bibnamefont{Sch\"afer}},
  \bibinfo{journal}{Phys. Rev. D} \textbf{\bibinfo{volume}{57}},
  \bibinfo{pages}{7274} (\bibinfo{year}{1998}), \bibinfo{note}{{\bf 63},
  029902(E) (2000)}.

\bibitem[{\citenamefont{de~Andrade et~al.}(2001)\citenamefont{de~Andrade,
  Blanchet, and Faye}}]{AndradeBlanchet2001}
\bibinfo{author}{\bibfnamefont{V.~C.} \bibnamefont{de~Andrade}},
  \bibinfo{author}{\bibfnamefont{L.}~\bibnamefont{Blanchet}}, \bibnamefont{and}
  \bibinfo{author}{\bibfnamefont{G.}~\bibnamefont{Faye}},
  \bibinfo{journal}{Class. Quant. Grav.} \textbf{\bibinfo{volume}{18}},
  \bibinfo{pages}{753} (\bibinfo{year}{2001}).

\bibitem[{\citenamefont{{Damour} et~al.}(2000)\citenamefont{{Damour},
  {Jaranowski}, and {Sch{\"a}fer}}}]{DamourJaranowski2000a}
\bibinfo{author}{\bibfnamefont{T.}~\bibnamefont{{Damour}}},
  \bibinfo{author}{\bibfnamefont{P.}~\bibnamefont{{Jaranowski}}},
  \bibnamefont{and}
  \bibinfo{author}{\bibfnamefont{G.}~\bibnamefont{{Sch{\"a}fer}}},
  \bibinfo{journal}{Phys. Rev. D} \textbf{\bibinfo{volume}{62}},
  \bibinfo{pages}{044024} (\bibinfo{year}{2000}).

\bibitem[{\citenamefont{Blanchet and Faye}(2001)}]{BlanchetFaye2001}
\bibinfo{author}{\bibfnamefont{L.}~\bibnamefont{Blanchet}} \bibnamefont{and}
  \bibinfo{author}{\bibfnamefont{G.}~\bibnamefont{Faye}},
  \bibinfo{journal}{Phys. Rev. D} \textbf{\bibinfo{volume}{63}},
  \bibinfo{pages}{062005} (\bibinfo{year}{2001}).

\bibitem[{\citenamefont{Damour et~al.}(2001)\citenamefont{Damour, Jaranowski,
  and Sch\"afer}}]{DamourJaranowski2001}
\bibinfo{author}{\bibfnamefont{T.}~\bibnamefont{Damour}},
  \bibinfo{author}{\bibfnamefont{P.}~\bibnamefont{Jaranowski}},
  \bibnamefont{and}
  \bibinfo{author}{\bibfnamefont{G.}~\bibnamefont{Sch\"afer}},
  \bibinfo{journal}{Phys. Rev. D} \textbf{\bibinfo{volume}{63}},
  \bibinfo{pages}{044021} (\bibinfo{year}{2001}), \bibinfo{note}{{\bf 66},
  029901(E) (2002)}.

\bibitem[{\citenamefont{Blanchet et~al.}(2002)\citenamefont{Blanchet, Faye,
  Iyer, and Joguet}}]{BlanchetFaye2002}
\bibinfo{author}{\bibfnamefont{L.}~\bibnamefont{Blanchet}},
  \bibinfo{author}{\bibfnamefont{G.}~\bibnamefont{Faye}},
  \bibinfo{author}{\bibfnamefont{B.~R.} \bibnamefont{Iyer}}, \bibnamefont{and}
  \bibinfo{author}{\bibfnamefont{B.}~\bibnamefont{Joguet}},
  \bibinfo{journal}{Phys. Rev. D} \textbf{\bibinfo{volume}{65}},
  \bibinfo{pages}{061501(R)} (\bibinfo{year}{2002}), \bibinfo{note}{{\bf 71},
  129902(E) (2005)}.

\bibitem[{\citenamefont{Blanchet et~al.}(2004)\citenamefont{Blanchet, Damour,
  and Esposito-Far\`{e}se}}]{BlanchetDamour2004}
\bibinfo{author}{\bibfnamefont{L.}~\bibnamefont{Blanchet}},
  \bibinfo{author}{\bibfnamefont{T.}~\bibnamefont{Damour}}, \bibnamefont{and}
  \bibinfo{author}{\bibfnamefont{G.}~\bibnamefont{Esposito-Far\`{e}se}},
  \bibinfo{journal}{Phys. Rev. D} \textbf{\bibinfo{volume}{69}},
  \bibinfo{pages}{124007} (\bibinfo{year}{2004}).

\bibitem[{\citenamefont{Bender and Orszag}(1978)}]{BenderOrszag}
\bibinfo{author}{\bibfnamefont{C.~M.} \bibnamefont{Bender}} \bibnamefont{and}
  \bibinfo{author}{\bibfnamefont{S.~A.} \bibnamefont{Orszag}},
  \emph{\bibinfo{title}{Advanced Mathematical Methods for Scientists and
  Engineers}} (\bibinfo{publisher}{McGraw-Hill}, \bibinfo{address}{New York},
  \bibinfo{year}{1978}).

\bibitem[{\citenamefont{Poisson}(1995)}]{Poisson1995}
\bibinfo{author}{\bibfnamefont{E.}~\bibnamefont{Poisson}},
  \bibinfo{journal}{Phys. Rev. D} \textbf{\bibinfo{volume}{52}},
  \bibinfo{pages}{5719} (\bibinfo{year}{1995}), \bibinfo{note}{{\bf 55}, 7980
  (1997)}.

\bibitem[{\citenamefont{Tanaka et~al.}(1996)\citenamefont{Tanaka, Tagoshi, and
  Sasaki}}]{TanakaTagoshi1996}
\bibinfo{author}{\bibfnamefont{T.}~\bibnamefont{Tanaka}},
  \bibinfo{author}{\bibfnamefont{H.}~\bibnamefont{Tagoshi}}, \bibnamefont{and}
  \bibinfo{author}{\bibfnamefont{M.}~\bibnamefont{Sasaki}},
  \bibinfo{journal}{Prog. Theor. Phys.} \textbf{\bibinfo{volume}{96}},
  \bibinfo{pages}{1087} (\bibinfo{year}{1996}).

\bibitem[{\citenamefont{Damour et~al.}(2002)\citenamefont{Damour, Iyer, and
  Sathyaprakash}}]{DamourIyer2002}
\bibinfo{author}{\bibfnamefont{T.}~\bibnamefont{Damour}},
  \bibinfo{author}{\bibfnamefont{B.~R.} \bibnamefont{Iyer}}, \bibnamefont{and}
  \bibinfo{author}{\bibfnamefont{B.~S.} \bibnamefont{Sathyaprakash}},
  \bibinfo{journal}{Phys. Rev. D} \textbf{\bibinfo{volume}{66}},
  \bibinfo{pages}{027502} (\bibinfo{year}{2002}), \bibinfo{note}{{\bf 72},
  029901(E) (2005)}.

\bibitem[{\citenamefont{{LIGO Scientific Collaboration}}()}]{LAL}
\bibinfo{author}{\bibnamefont{{LIGO Scientific Collaboration}}},
  \emph{\bibinfo{title}{{LSC} {A}lgorithm {L}ibrary software packages {\scshape
  lal}, {\scshape lalwrapper}, and {\scshape lalapps}}},
  \urlprefix\url{http://www.lsc-group.phys.uwm.edu/lal}.

\bibitem[{\citenamefont{Poisson}(1993)}]{Poisson1993}
\bibinfo{author}{\bibfnamefont{E.}~\bibnamefont{Poisson}},
  \bibinfo{journal}{Phys. Rev. D} \textbf{\bibinfo{volume}{47}},
  \bibinfo{pages}{1497} (\bibinfo{year}{1993}).

\bibitem[{\citenamefont{Cutler et~al.}(1993)\citenamefont{Cutler, Finn,
  Poisson, and Sussman}}]{CutlerFinn1993}
\bibinfo{author}{\bibfnamefont{C.}~\bibnamefont{Cutler}},
  \bibinfo{author}{\bibfnamefont{L.~S.} \bibnamefont{Finn}},
  \bibinfo{author}{\bibfnamefont{E.}~\bibnamefont{Poisson}}, \bibnamefont{and}
  \bibinfo{author}{\bibfnamefont{G.~J.} \bibnamefont{Sussman}},
  \bibinfo{journal}{Phys. Rev. D} \textbf{\bibinfo{volume}{47}},
  \bibinfo{pages}{1511} (\bibinfo{year}{1993}).

\bibitem[{\citenamefont{Kidder}(2008)}]{Kidder2008}
\bibinfo{author}{\bibfnamefont{L.~E.} \bibnamefont{Kidder}},
  \bibinfo{journal}{Phys. Rev. D} \textbf{\bibinfo{volume}{77}},
  \bibinfo{pages}{044016} (\bibinfo{year}{2008}).

\bibitem[{\citenamefont{Arun et~al.}(2004)\citenamefont{Arun, Blanchet, Iyer,
  and Qusailah}}]{ArunBlanchet2004}
\bibinfo{author}{\bibfnamefont{K.}~\bibnamefont{Arun}},
  \bibinfo{author}{\bibfnamefont{L.}~\bibnamefont{Blanchet}},
  \bibinfo{author}{\bibfnamefont{B.}~\bibnamefont{Iyer}}, \bibnamefont{and}
  \bibinfo{author}{\bibfnamefont{M.}~\bibnamefont{Qusailah}},
  \bibinfo{journal}{Class. Quant. Grav.} \textbf{\bibinfo{volume}{21}},
  \bibinfo{pages}{3771} (\bibinfo{year}{2004}), \bibinfo{note}{{\bf 22},
  3115--3117(E) (2005)}.

\bibitem[{\citenamefont{Baker et~al.}(2007{\natexlab{b}})\citenamefont{Baker,
  McWilliams, van Meter, Centrella, Choi, Kelly, and
  Koppitz}}]{BakerMcWilliams2007}
\bibinfo{author}{\bibfnamefont{J.~G.} \bibnamefont{Baker}},
  \bibinfo{author}{\bibfnamefont{S.~T.} \bibnamefont{McWilliams}},
  \bibinfo{author}{\bibfnamefont{J.~R.} \bibnamefont{van Meter}},
  \bibinfo{author}{\bibfnamefont{J.}~\bibnamefont{Centrella}},
  \bibinfo{author}{\bibfnamefont{D.-I.} \bibnamefont{Choi}},
  \bibinfo{author}{\bibfnamefont{B.~J.} \bibnamefont{Kelly}}, \bibnamefont{and}
  \bibinfo{author}{\bibfnamefont{M.}~\bibnamefont{Koppitz}},
  \bibinfo{journal}{Phys. Rev. D} \textbf{\bibinfo{volume}{75}},
  \bibinfo{pages}{124024} (\bibinfo{year}{2007}{\natexlab{b}}).

\bibitem[{\citenamefont{Lovelace et~al.}(2008)\citenamefont{Lovelace, Owen,
  Pfeiffer, and Chu}}]{LovelaceOwen2008}
\bibinfo{author}{\bibfnamefont{G.}~\bibnamefont{Lovelace}},
  \bibinfo{author}{\bibfnamefont{R.}~\bibnamefont{Owen}},
  \bibinfo{author}{\bibfnamefont{H.~P.} \bibnamefont{Pfeiffer}},
  \bibnamefont{and} \bibinfo{author}{\bibfnamefont{T.}~\bibnamefont{Chu}},
  \bibinfo{journal}{Phys. Rev. D}  (\bibinfo{year}{2008}), \bibinfo{note}{in
  press, arXiv:0805.4192}.

\bibitem[{\citenamefont{Yunes and Berti}(2008)}]{YunesBerti2008}
\bibinfo{author}{\bibfnamefont{N.}~\bibnamefont{Yunes}} \bibnamefont{and}
  \bibinfo{author}{\bibfnamefont{E.}~\bibnamefont{Berti}},
  \bibinfo{journal}{Phys. Rev. D} \textbf{\bibinfo{volume}{77}},
  \bibinfo{pages}{124006} (\bibinfo{year}{2008}).

\end{thebibliography}
\end{document}